%% file: 03_main_labor_family_economics.tex
\documentclass[11pt]{article}
\usepackage[utf8]{inputenc}
\usepackage{amsmath,setspace,geometry}
\usepackage{amsthm}
\usepackage{amsfonts}
\usepackage[shortlabels]{enumitem}
\usepackage{rotating}
\usepackage{pdflscape}
\usepackage{graphicx}
\usepackage{bbm}
\usepackage{comment}
\usepackage[dvipsnames]{xcolor}
\usepackage{hyperref}
\hypersetup{colorlinks=true, linkcolor= BrickRed, citecolor = BrickRed, filecolor = BrickRed, urlcolor = BrickRed, hypertexnames = true}
\usepackage[]{natbib} 
\bibpunct[:]{(}{)}{,}{a}{}{,}
\usepackage[english]{babel}
\usepackage{float}
\usepackage{caption}
\usepackage{subcaption}
\usepackage{booktabs}
\usepackage{pdfpages}
\usepackage{threeparttable}
\usepackage{lscape}
\usepackage{bm}
\usepackage{setspace}
\title{Marital Sorting on Pre-Marital Preferences for Household Behavior}
\author{Chihiro Inoue \thanks{\href{mailto:}{inoue@econ.kobe-u.ac.jp}, Graduate School of Economics, Kobe University} 
 \and
 Yusuke Ishihata \thanks{\href{mailto:}{yusuke.ishihata@duke.edu}, Department of Economics, Duke University} 
 \and
 Suguru Otani\thanks{\href{mailto:}{suguru.otani@e.u-tokyo.ac.jp}, Market Design Center, Department of Economics, The University of Tokyo  \\
 We thank Frederik Almar, Pierre-Andr{\'e} Chiappori, Matthias Doepke, Jun Goto, Fuhito Kojima, Kanato Nakakuni, Kosuke Shimamoto, and Kazuharu Yanagimoto for insightful discussions and suggestions.
 We also thank seminar and conference participants at GRIPS, Kobe University, Okayama University, %, EALE 2026, 
 and the Kansai and Tokyo Labor Study Group for their useful comments.
 We acknowledge Shuto Fukuda and Hirokazu Tsuchiya for sharing the data and their technical and institutional knowledge of the IBJ platform. This work was supported by JST ERATO Grant Number JPMJER2301, Japan. }
 }
 
\date{
First version: March 19, 2026\\
Current version: \today
}

\begin{document}

\maketitle

% \begin{center}
%     \textcolor{red}{Preliminary draft\\ 
%     Do not distribute this draft without authors' permission}
% \end{center}

\begin{abstract}
    We examine marital sorting using novel data from a marriage-matching platform that records both a dating-to-marriage pipeline and pre-marital attributes, including preferences for children and for the division of housework and childcare.
    Unlike census or post-marital surveys, characteristics are collected before matching, and objectively measurable attributes are verified using official documents, providing an ideal setting to study matching and sorting free from post-marital adjustment.
    Using a multidimensional matching framework across twelve attributes, we find assortative matching along all dimensions.
    Age is the most salient trait, while preferences for children are second—exceeding education—an economically important margin invisible in standard data.
    A low-dimensional factor representation shows that fertility preferences constitute a distinct sorting dimension.
    Exploiting the platform's dating-to-marriage pipeline, we show that sorting on fertility preferences emerges at later serious-relationship stages.
    A theoretical analysis suggests that the magnitude of sorting along this margin is consistent with the veto model of fertility decisions.
    %\fi
    %

    % 96 words
    \if0
    We examine marital sorting using a novel dataset from a marriage-matching platform that records pre-marital attributes, including preferences for children and for the division of housework and childcare. 
    Unlike census or post-marital surveys, all characteristics are collected prior to matching, and objectively measurable attributes are verified using official documents, providing a clean setting to study sorting on pre-marital attributes. 
    Using a multidimensional matching framework across twelve attributes, we find positive assortative matching along all dimensions. 
    Age is the most salient trait, while preferences for children are the second most important—exceeding education. 
    The magnitude of sorting along this margin is suggestive of a veto model of fertility decisions.
    \fi
    
    $\quad$ \\
    %100 words AER (now 236 words)
    \textbf{Keywords}: Marriage market, assortative matching, fertility preferences, multidimensional matching, dating \\
    \textbf{JEL code}: D13, J12, J13
\end{abstract}

%\tableofcontents

\newpage

\section{Introduction}
Understanding how individuals sort into marriage is central to household economics and to the distributional consequences of household formation. 
Prior work documents strong positive assortative matching on education \citep{greenwood2014marry, eika2019educational, chiappori2025changes}, age \citep{choo2006marries}, income \citep{chiappori2022assortative}, and other sociodemographic characteristics. 
However, much less is known about how individuals sort on preferences that directly govern household behavior—such as the desire to have children or the willingness to engage in housework and childcare—even though these preferences fundamentally shape subsequent household decisions, including fertility, labor supply, and within-household time allocation.
\footnote{A growing body of work emphasizes that household behavior and matching should be analyzed jointly, rather than treating family formation as exogenous to intra-household decisions. While empirical studies of household behavior have made substantial progress, they typically condition on existing couples, abstracting from the selection processes that determine who forms a household with whom. In contrast, matching models highlight that partner selection is itself an endogenous process that shapes subsequent household behavior, bargaining power, and specialization. See, for example, the surveys by \citet{chiappori2023mating}, \citet{chiappori2024frictionless}, and \citet{salanie2024matching} for discussions of this integration and its importance for household economics.}
This gap primarily reflects data limitations: these preferences are usually unavailable in registry and census data, and even when surveys collect information on preferences, observations prior to marriage are rarely available.
Existing evidence therefore provides limited guidance on whether and how pre-marital preferences contributes to observed assortative matching.
% \textcolor{black}{(We might preview potential problems here; the argument may be strengthen further than just saying our understanding is incomplete.)}

This paper brings new evidence to this question using a uniquely rich dataset with directly observed household-relevant preferences.
Specifically, we analyze a novel dataset from IBJ, Japan's largest structured marriage matching platform.
IBJ represents a non-negligible share of marriages in Japan: the data record over 10,000 confirmed engagements per year and about 3.3\% of all new marriages in 2024.
In addition to standard attributes used in the marriage market literature, such as sociodemographic characteristics (e.g., education, age, and income) and anthropometric measures (e.g., height and weight), the data include preferences for children and post-marital time allocation between market work and home. 
These preferences measures correspond directly to primitives that shape post-marital household decisions, letting us study marital sorting on theoretically first-order but rarely observed characteristics.
Notably, all attributes and preferences are collected prior to matching. 
As a result, they are not shaped by post-marital intra-household coordination, which can contaminate survey measures from already-married couples commonly used in the literature.
\footnote{For instance, \citet{chiappori2024analyzing}, who use survey data on couples in Italy to analyze matching patterns, note that ``an obvious caveat is that, since we observe couples a few years after marriage, some individual traits may have converged.''}
For example, spouses may adjust labor supply, occupation, and income to exploit household specialization, and their reported preferences over children may converge as part of joint life planning.
The pre-marital elicitation, therefore, makes our data an ideal setting for analyzing matching and sorting on a rich set of attributes, including preferences over fertility and time allocation.

\textcolor{black}{Beyond measuring these novel pre-marital preferences, the IBJ data allow us to characterize how they relate to conventional anthropometric and sociodemographic attributes. 
Preferences for children and for the division of childcare and housework vary systematically with age, income, and education, yet their pairwise correlations with standard sociodemographic and anthropometric characteristics are generally modest, suggesting that these measures capture a distinct margin of heterogeneity rather than simply proxying for conventional traits. 
At the same time, strong correlations among conventional sociodemographic attributes indicate that assortativeness is best interpreted in a multidimensional framework that jointly accounts for these interactions.}

Given these patterns, we apply the multidimensional matching framework of \citet{dupuy2014personality} to study sorting on pre-marital preferences together with sociodemographic and anthropometric variables.
With the rich set of variables, we do not restrict attention to a single attribute, unlike the approach in \citet{choo2006marries} and much of the subsequent literature \citep{chiappori2024frictionless}.
Specifically, we examine twelve observed or constructed attributes: education, age, income, occupational flexibility, height, weight, drinking, smoking, marital history, and preferences for housework, childcare, and children.
This framework quantifies how multiple attributes and their interactions shape matching while accounting for correlations across attributes.
Ignoring this multidimensional structure and focusing on a single trait can mislead inference when traits are strongly correlated.

The estimated affinity matrix, which summarizes whether each male–female attribute combination are complementary or substitutable, yields several findings.
First, we find strong positive assortative matching among couples formed on the platform.
Strikingly, all twelve attributes---sociodemographic, anthropometric, and preference measures---exhibit positive assortativeness, though this is not guaranteed ex ante.
Because attributes are measured pre-match, these are clean sorting patterns rather than post-marital coordination artifacts.
Age has the largest diagonal coefficient, consistent with prior work such as \citet{chiappori2024analyzing}.
Remarkably, the second-largest diagonal coefficient is preferences for children --- exceeding even education, which typically displays one of the strongest assortative patterns.
This highlights that our rich pre-marital data uncover key sorting dimensions mostly invisible in standard datasets.

Second, the off-diagonal elements of the affinity matrix highlight the importance of cross-attribute interactions across genders.
Among sociodemographic attributes (education, age, income) and anthropometric characteristics (height, weight), most interactions are statistically different from zero. 
Thus, a favorable trait in one dimension (e.g., higher income) tends to attract partners with stronger traits along other sociodemographic and anthropometric dimensions.

Occupational flexibility, constructed from occupations following \citet{goldin2014grand}, plays a smaller role in sorting than other sociodemographic characteristics.
Although sorting is positive on this dimension, the magnitude is small and interactions with other attributes are mostly absent.
This pattern is informative in light of theoretical mechanisms: while occupational flexibility is increasingly recognized as relevant for within-household specialization or coordination after matching \citep{bang2021job, erosa2022hours,calvo2024marriage}, it need not be a primary margin of the partner selection stage.

Finally, interactions involving pre-marital child preferences show distinct patterns.
Preferences for children are complements to childcare intentions: individuals willing to assume childcare responsibilities tend to match with partners who more strongly want children.
By contrast, interactions between child preferences and most sociodemographic or anthropometric traits are small and statistically indistinguishable from zero, with age as an important exception. 
In particular, the interaction between male child preference and female age is negative, indicating that stronger fertility preferences are associated with younger partners along the female age dimension. By contrast, child preferences exhibit little interaction with education or income, suggesting that, despite the common view that education or income matter for childbearing, a desire for children does not systematically attract higher-education or higher-income partners, and highly educated or high-income individuals do not systematically attract partners with stronger child preferences.
These striking patterns are only visible in a multidimensional matching framework with a rich set of attributes.
\textcolor{black}{A saliency decomposition analysis based on \citet{dupuy2014personality} also confirms that child preferences are an important sorting margin with a role distinct from sociodemographic and anthropometric attributes.: three indices explain almost 90\% of observable matching surplus, and child preferences dominate the third index (share 6\%, above 20\% of joint utility after accounting for cohort-restricted search).}

\textcolor{black}{We also evaluate whether standard unidimensional specifications provide reliable measures of assortative matching. 
We show that dimensionality matters: estimates derived from unidimensional specifications can diverge substantially from our fully multidimensional benchmark, particularly when attributes are correlated.
By contrast, specifications that jointly include key demographic variables approximate the multidimensional benchmark more closely. 
These findings clarify when parsimonious models are informative and when they may instead lead to incomplete or distorted inference.}

We further examine how preferences and sorting evolve over time using multi-year data.
Few multidimensional matching studies can do this because most datasets provide either rich attributes or long panels, but not both.
Using year-comparable normalization \citep{ciscato2020role}, we find that sorting on age, height, weight, and child preferences remains strongly positive and stable over the decade; education sorting declines, income sorting remains broadly stable, and child-preference sorting remains persistently strong.
These shifts suggest that the attributes valued in the marriage market have evolved over time, potentially reflecting changing social norms or economic conditions.
Despite such dynamics, sorting on preferences for children remains strong and persistent, highlighting the importance of pre-marital preferences for children as a fundamental dimension of the marriage market.

While the time-series analysis characterizes how assortative matching at the Proposal stage evolves across years, it does not reveal when these patterns arise during dating. 
Our unique data address this because they record not only final match outcomes but also interactions at each stage of relationship formation, including initial meeting applications and subsequent decisions to enter exclusive commitments. 
Using the many-to-many framework of \citet{fox2018estimating}, we estimate assortativeness from Application to Proposal: age and income are robust from Application onward, marital history and child preferences emerge later, and other attributes are not robust.
Overall, the stage-based analysis suggests that the assortative structure observed at the Proposal stage reflects selective continuation along a limited set of dimensions rather than the sudden emergence of sorting at the point of final commitment.

Finally, we discuss how our finding of strong sorting on pre-marital fertility preferences relates to standard frameworks in household economics.
We show that two leading models of household decision-making --- a unitary model in which the household jointly maximizes spousal utilities, and a veto model in which children require the consent of both spouses --- both predict positive assortative matching on fertility preferences.
The two models differ, however, in the predicted strength of sorting: the veto model predicts a strictly stronger degree of positive assortative matching on fertility preferences than the unitary alternative.
The strong sorting we document is therefore suggestive of the veto model, providing additional empirical support for a growing literature that adopts mutual consent as the preferred framework for modeling fertility decisions within the household.

\if0
Finally, we present a simple theoretical model of fertility and labor supply to illustrate why recognizing sorting on pre-marital preferences --- especially preferences for children --- is important.
Although conceptually natural, heterogeneous fertility preferences among households with the same observables are not commonly incorporated in theoretical or empirical structural models. 
This heterogeneity is often implicitly treated as inconsequential for average policy effects because aggregation is assumed to mask household-level differences.
%It is sometimes implicitly presumed that, largely to avoid added complexity and identification challenges, such heterogeneity is inconsequential for the purpose of estimating average policy effects because aggregation would mask household-level differences.
Our theoretical exercise shows otherwise: ignoring preference-based sorting generally overstates policy effects on fertility and labor supply.
%\textcolor{black}{The intuition is that fertility preferences are polarized due to sorting. Low-preference couples exhibit little policy response, while high-preference couples face diminishing marginal utility from additional children. Due to concavity, the latter responses cannot compensate for the former.}
Given our empirical finding of significant sorting on child preferences, analyses that abstract from this heterogeneity are likely to yield policy predictions biased in magnitude, typically away from zero.
\fi

\paragraph{Related Literature.}
This paper contributes to three strands of literature.
First and foremost, our work contributes to the literature on the marriage market.
Since the seminal contribution of \citet{choo2006marries}, there has been extensive structural empirical research on unidimensional sorting in the marriage market, primarily with respect to education \citep{chiappori2017partner}.
Given the importance of multiple attributes in shaping marital choices, we build on the quadratic affinity-matrix framework of \citet{dupuy2014personality}, which provides a flexible and empirically tractable model of multidimensional matching.\footnote{
The literature on multidimensional matching under transferable utility can be organized by modeling approach and the dimensions of sorting they emphasize.
Index-based models study assortative matching on anthropometric and socioeconomic characteristics such as age, height, body mass index, wages, and education \citep{chiappori2012fatter,chiappori2020erratum}, as well as on detailed educational programs and fields of study that proxy for earnings potential and work--life balance \citep{almar2025educational}.
Bidimensional matching models focus on trade-offs between human capital and specific traits or constraints, including smoking behavior \citep{chiappori2018bidimensional}, fertility \citep{low2024human}, migration status \citep{ahn2025matching}, race \citep{chiappori2016black}, sexual orientation \citep{ciscato2024matching}, and genetic characteristics \citep{zheng2025genetic}.
Quadratic affinity-matrix models based on \cite{dupuy2014personality} allow for richer interactions across multiple attributes, encompassing combinations of education, age, wages, anthropometric traits, health-related behaviors, personality traits, and household characteristics \citep{ciscato2020role,chiappori2024analyzing,ciscato2020like}.
Finally, a growing macro-oriented literature embeds multidimensional sorting into search-and-matching environments with frictions to study equilibrium implications for marriage and labor markets \citep{lindenlaub2017sorting,lindenlaub2023multidimensional,ciscato2024assessing,calvo2024marriage}.
}
Several papers apply this multidimensional method to analyze marriage market sorting with different substantive focuses, such as same-sex couples \citep{ciscato2020like} and changes in income inequality over time \citep{ciscato2020role}.\footnote{
\citet{adda2025there} use a related but distinct multidimensional matching framework to study how legal-status incentives affect the marriage decisions of natives and migrants in Italy.}

The closest related work is \citet{chiappori2024analyzing}, who use unique survey data from Italy to study multidimensional matching patterns with the model of \citet{dupuy2014personality}.
Their key finding is that multidimensional matching indexes have predictive power for children's outcomes, highlighting the role of sorting in shaping intergenerational inequality.
While our data do not include outcome variables, we observe a rich set of \textit{pre-marital} characteristics, including preference measures tied to household decisions, that are directly relevant to analyzing sorting at the time of marriage.
Our paper complements theirs by studying matching patterns using cleaner pre-marital attributes and by deriving implications for subsequent household decisions using these detailed preference measures.

In terms of data, our IBJ dataset is closest in spirit to datasets from online dating platforms.
Several papers use data from online dating apps \citep{hitsch2010makes, ong2015income, bapna2016one, egebark2021brains, buyukeren2025impact} and from speed dating events \citep{fisman2006gender, belot2013dating} to study matching patterns and to draw implications for the marriage market.
Compared with typical dating apps and speed dating environments, however, users on a marriage matching platform are substantially more selected into serious partnership formation due to screening processes and nontrivial financial costs.
This selection allows us to speak more directly to preferences relevant to marriage, rather than casual dating.
Moreover, unlike these datasets, user attributes in our data are verified through official documents, such as tax withholding forms and certified health checkup records, providing a high degree of reliability in analyzing matching patterns.\footnote{
\citet{lee2016effect} is an exception in that her dataset allows the observation of eventual marriage outcomes and includes legally verified key background variables. Our data further contain a larger number of couples and span longer periods, as well as information on preferences for children and time allocation.} Also, our data record all information about match formation processes with timestamps, enabling us to identify the dating status and stage.

Second, we contribute to the literature linking matching models with subsequent household decisions.
Recent surveys emphasize that understanding marriage market sorting and household behavior jointly is essential, as household formation and intra-household allocation are fundamentally interconnected processes \citep{chiappori2023mating, chiappori2024frictionless, salanie2024matching}.
In line with this view, several papers explicitly model the equilibrium marriage market together with life-cycle decisions to analyze these interactions \citep{chiappori2018marriage, gayle2019optimal, calvo2024marriage, reynoso2024impact, almar2025families, calvo2025effects, almar2026fertility}.
Although our dataset does not contain direct measures of post-marital household behavior, our evidence of sorting not only on sociodemographic characteristics but also on preferences has important implications for this literature.
In particular, models that link marriage market sorting to subsequent household decisions may need to incorporate such pre-marital preference-based sorting to generate accurate counterfactual policy predictions.
For example, our findings of sorting on pre-marital preferences for children suggest that households with identical observables may nevertheless differ in underlying preference parameters.
\footnote{A large share of empirical structural models of fertility and labor supply assume homogeneous preferences for children conditional on observables \citep[e.g.,][]{bick2016quantitative, garcia2017strings, yamaguchi2019effects, kim2024status, jakobsen2022fertility}.}
Incorporating such preference heterogeneity may therefore be important for accurately modeling household behavior in this literature.
%Our simple theoretical exercise shows that ignoring this heterogeneity and imposing preference homogeneity would overstate policy effects on fertility and labor supply.
%\footnote{See \citet{doepke2023economics} for a survey of models of fertility.}

Finally, we offer new insights into the literature on occupational flexibility and its implications for intra-household decisions.
\citet{goldin2014grand} shows that a large gender wage gap arises in occupations that reward long and inflexible hours, while \citet{cortes2019time} demonstrate that alleviating women's time constraints at home through greater availability of substitutes for household production reduces this gap.
Building on this work, recent studies emphasize that occupational choice and job flexibility shape intra-household specialization in labor supply, depending on the flexibility of each spouse, with important consequences for wages and career trajectories \citep{bang2021job,erosa2022hours}.
Because households adjust labor supply differently based on spouses' relative flexibility, occupational flexibility is a potentially relevant dimension of partner selection.
Despite this relevance, assortative matching based on occupational flexibility has received little empirical attention.\footnote{A related literature examines assortative matching by occupation or industry \citep[e.g.,][]{kalmijn_assortative_1994}. This work typically focuses on occupational titles rather than characteristics of occupations --- such as flexibility, schedules, or career dynamics --- that are central to intra-household coordination.}
A closely related exception is \citet{almar2025educational}, who study assortative matching in educational ambition, measured using expected initial wage levels and wage growth. They argue that collapsing labor market trajectories into a single earnings measure obscures heterogeneity in work--life balance that is relevant for family formation, and show that career paths with higher expected wage growth are negatively associated with various flexibility measures.
Our paper complements this literature by examining matching patterns along this dimension. 
Although we find that sorting on occupational flexibility is modest, documenting these patterns helps clarify the extent to which couples sort on flexibility prior to making joint intra-household decisions.

$\quad$

The remainder of the paper is organized as follows.
Section \ref{sec:data} describes our IBJ data and institutional setting.
Section \ref{sec:model} presents the multidimensional matching models and econometric approach.
Section \ref{sec:results} reports estimation results, Section \ref{sec:discussion} discusses implications for household-economic models, and Section \ref{sec:conclusion} concludes.

%%%%%%%%%%%%%%%%%%%%%%%%%%%%%%%%%%%%%%%%%%%%%%%%%%%
%%%%%%%%%%%%%%%%%%%%%%%%%%%%%%%%%%%%%%%%%%%%%%%%%%%

\section{Data}\label{sec:data}

This section describes the IBJ data and institutional environment and presents summary statistics. This section also serves two purposes: to establish external validity and to document novel descriptive patterns in pre-marital preferences that motivate our structural analysis.
Subsection \ref{subsec:ibj_institution} introduces the IBJ marriage agency platform and its institutional features. 
Subsection \ref{sec:ibj_user_behavior} outlines the matching process and dating stages through which engagements are formed. 
Subsection \ref{subsec:summary_stats} presents summary statistics for matched couples in 2024 used in the main estimation. 
Subsection \ref{subsec:correlation_preferences} reports correlations between preference measures and other observed attributes.
Subsection \ref{subsec:summary_stats_history} documents changes in the characteristics and preferences of matched users between 2015 and 2024.

\subsection{A Marriage Agency Platform}\label{subsec:ibj_institution}

IBJ is a structured marriage-matching platform designed for long-term partnerships through a highly intermediated process. Unlike typical dating apps with open-ended, user-initiated interaction, IBJ imposes strict entry requirements, including document checks, health certification, and upfront and ongoing fees. Users are matched through consultant-mediated proposals, and confirmed engagements require bilateral agreement. The platform follows a one-to-one rule and prohibits concurrent engagement. This design selects users actively seeking committed relationships and discourages casual participation, making the platform well suited for studying search and matching dynamics in a controlled, high-stakes environment.\footnote{Appendix \ref{sec:related_literature_data_comparison} provides a detailed comparison of the IBJ platform data with datasets used in recent related literature, highlighting the unique advantages of observing rich pre-marital attributes, unmatched users, and stage-by-stage matching outcomes.}

\paragraph{Macro Trends}

\begin{figure}[!htbp]
  \begin{center}
  \subfloat[Female $F$, Male $M$, and tightness ($\frac{F}{M}$)]{\includegraphics[width = 0.33\textwidth]{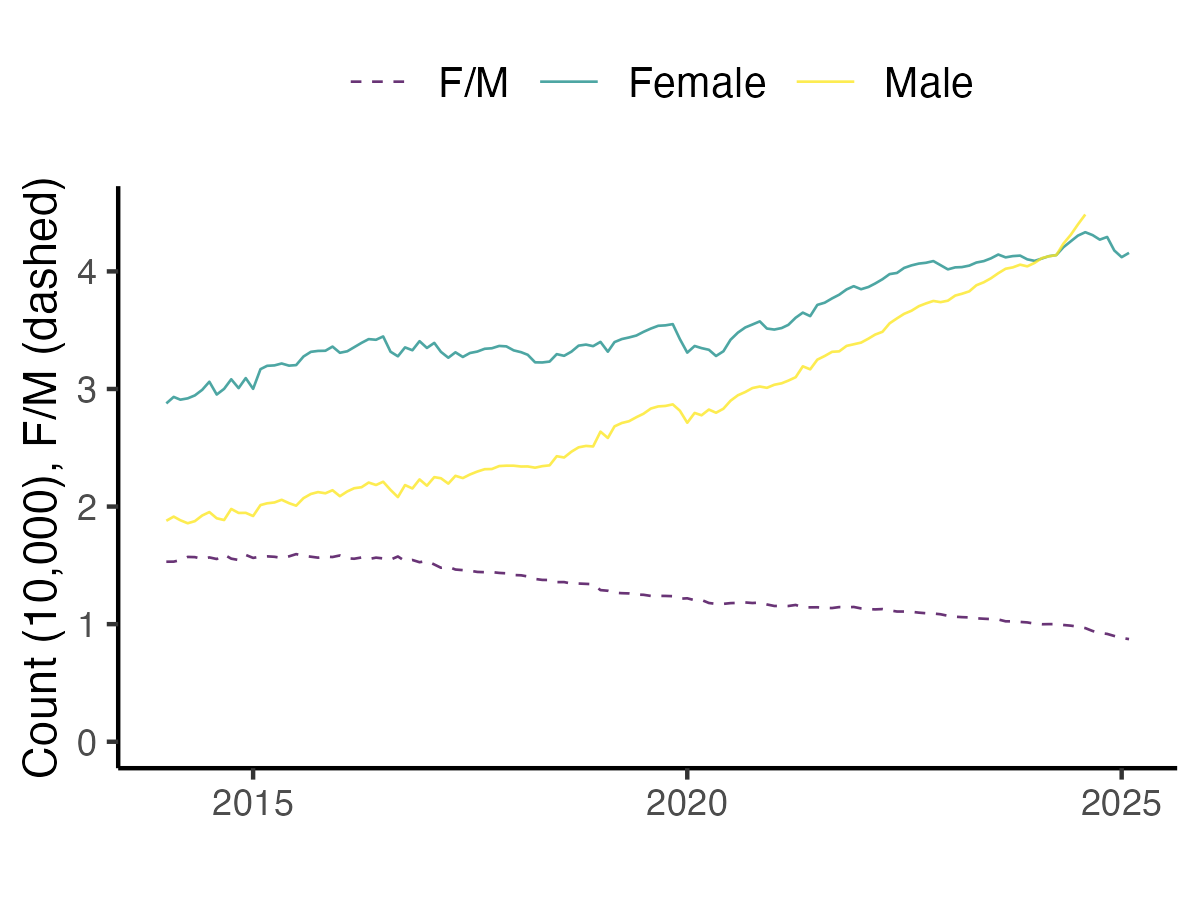}}
  \subfloat[Engagement $E$]{\includegraphics[width = 0.33\textwidth]{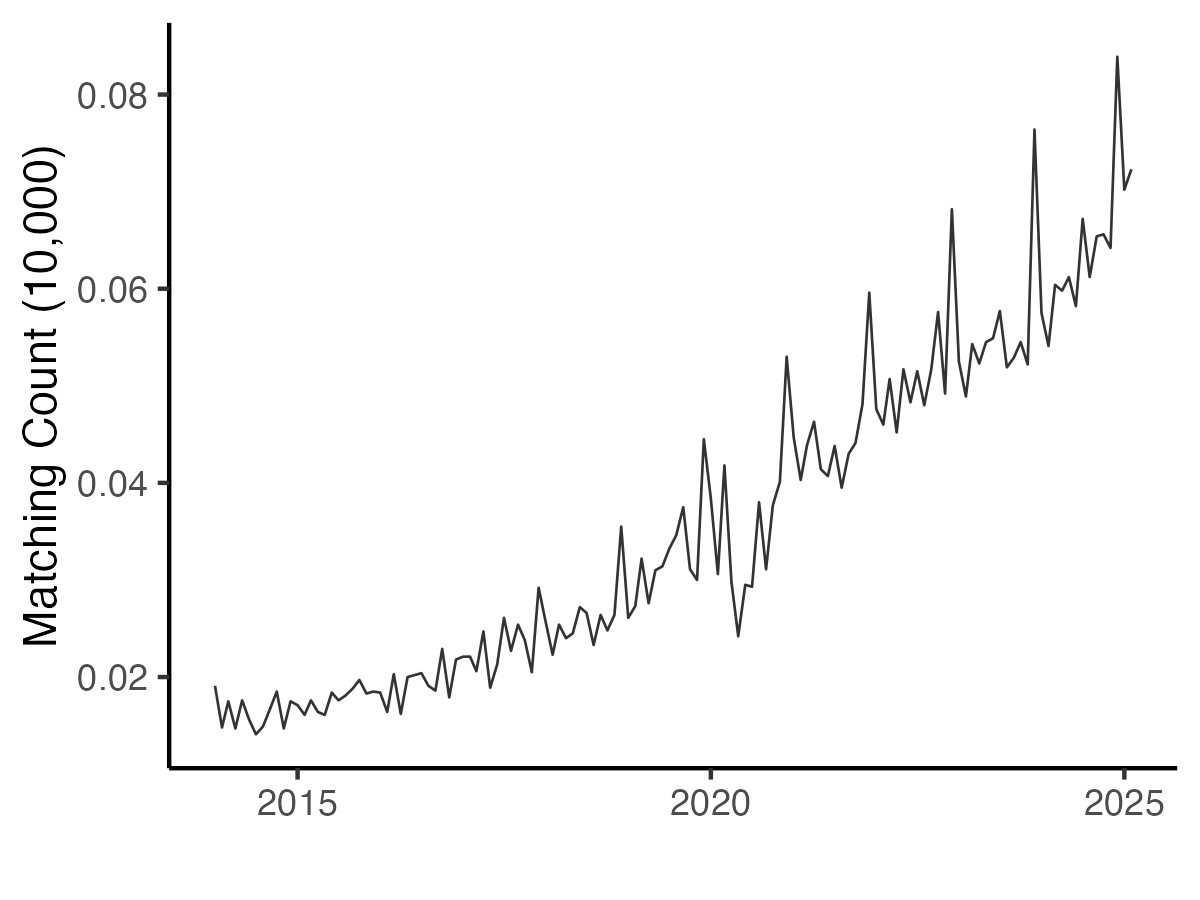}}
  \subfloat[Partner Finding Rate ($\frac{E}{F}$,$\frac{E}{M}$)]{\includegraphics[width = 0.33\textwidth]{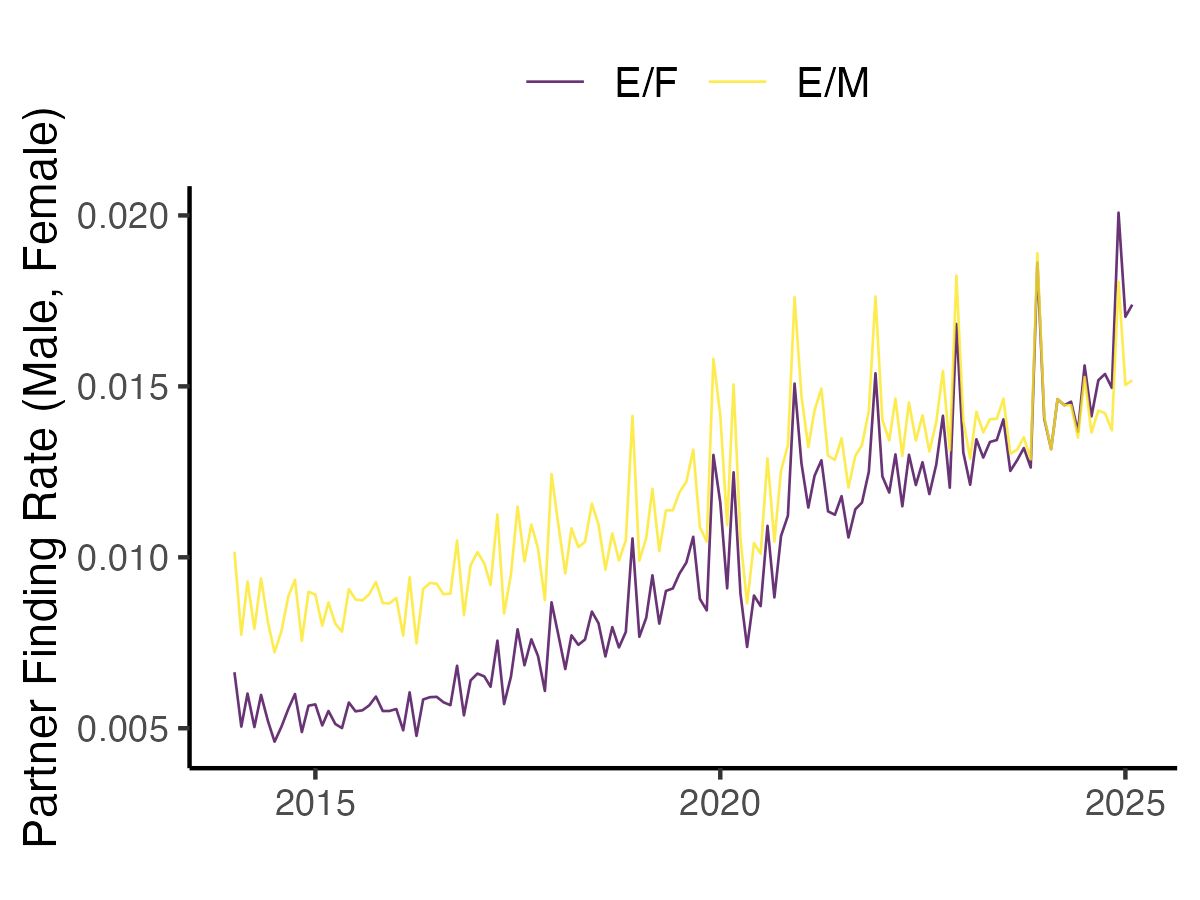}}
  \caption{Trends in key variables, 2014--2025}
  \label{fg:country_month_male_female_partner_finding_rate} 
  \end{center}
  \footnotesize
  Note: See \cite{otani2025nonparametric}. Also, see \cite{imamura2025note} for a model-free axiomatic assortativeness measurement and its application to the IBJ data.
\end{figure}

The IBJ platform has grown substantially in both user participation and matches. As Figure \ref{fg:country_month_male_female_partner_finding_rate} shows, active users rose steadily from 2014 to 2025, with especially sharp male-side growth after 2020. This gender-specific expansion lowered market tightness (the female-to-male ratio), changing the matching environment. Engagements also rose markedly, especially after 2020. In 2024, IBJ facilitated over 10,000 engagements, accounting for 3.3\% of marriages in Japan.\footnote{Approximately 50\% of engagements among IBJ users occur within the platform, while the remaining 50\% result from relationships formed outside the platform.} This coverage---especially for a high-verification, high-intent platform---offers an unprecedented view of modern partner search and formation.

\begin{figure}[!htbp]
  \begin{center}
  \subfloat[Age (Proportion)]{\includegraphics[width = 0.9\textwidth]{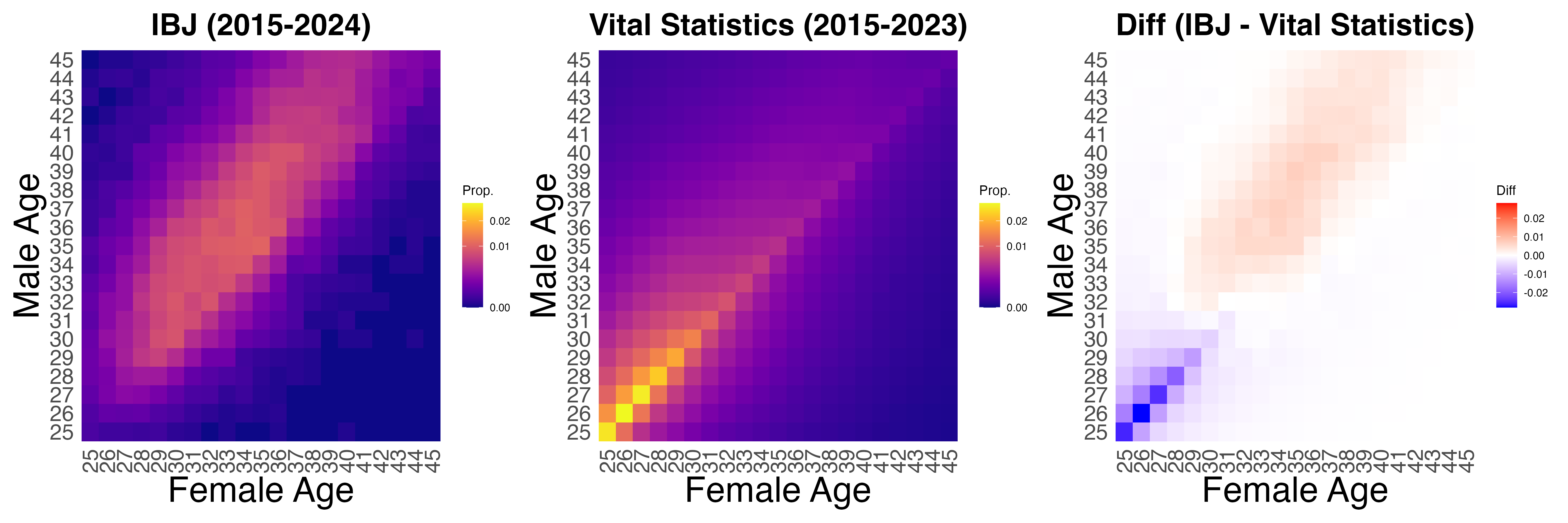}}\\
  \vspace{6mm}
  \subfloat[Education (Proportion)]{\includegraphics[width = 0.9\textwidth]{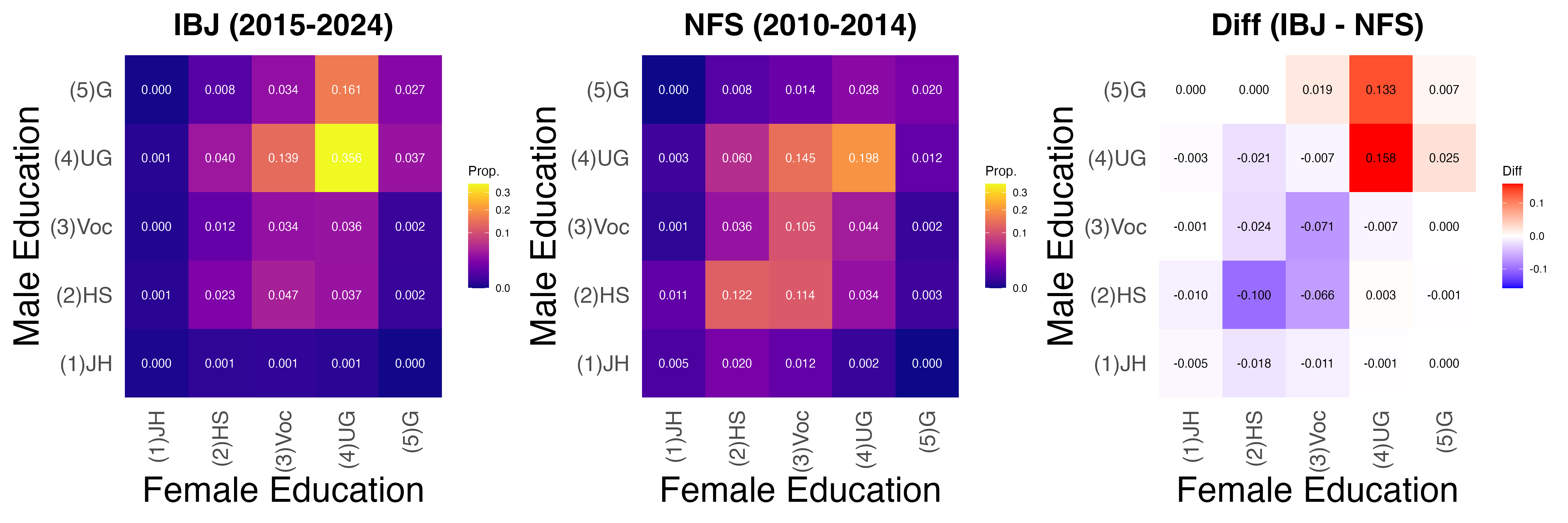}}\\
  \vspace{6mm}
  \subfloat[Income (Proportion)]{\includegraphics[width = 0.9\textwidth]{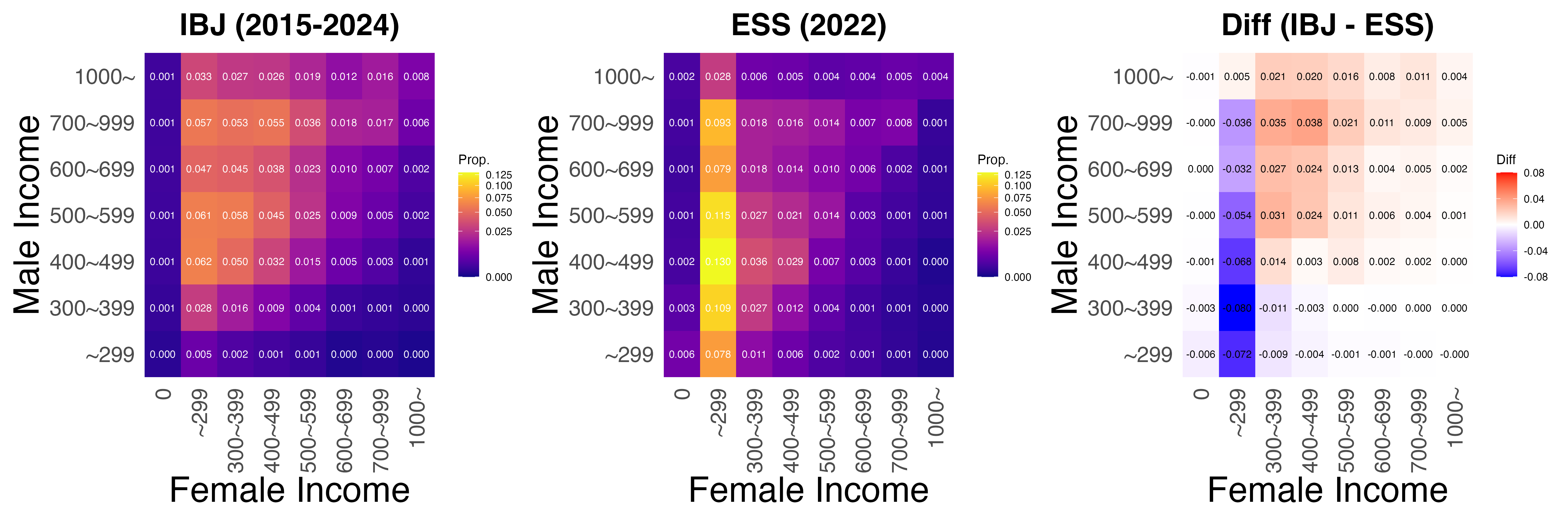}}
  \caption{IBJ Market Coverage Compared with National Statistics (NFS, VS, and ES): Proportion}
  \label{fg:representativeness_education_ibj_vs_national}
  \end{center}
  \footnotesize
  Note: Panels display the joint proportion $P(i,j) = n_{ij} / N$ for each cell $(i,j)$, where $n_{ij}$ is the number of couples in cell $(i,j)$ and $N$ is the total number of couples. IBJ age data cover 2015--2024 matched with Vital Statistics 2015--2023; education data compare IBJ 2015--2024 with the 15th National Fertility Survey (marriage cohort 2010--2014; the latest publicly available edition); income data compare IBJ 2015--2024 with the Employment Status Survey (2022, husband age $\leq$ 49).Income categories include a zero-income bin: IBJ women with occupation ``helping with housework'' are assigned zero income; ESS family workers, the non-employed, and wives not in the labor force are classified as zero income. Income represents the upper limit of the income category.
\end{figure}

\paragraph{Market Coverage}
Figure \ref{fg:representativeness_education_ibj_vs_national} compares the joint distribution of matched IBJ couples with national benchmarks and shows systematic coverage differences. Relative to Vital Statistics (VS, \textit{Jinko Dotai Chosa}), in Panel (a), IBJ engagements are concentrated among older couples, with greater mass in higher-age cells. This matches the platform’s positioning toward later entrants to the marriage market rather than the full population of newly married couples. 

Similarly, in Panel (b), compared with the National Fertility Survey (NFS, \textit{Shussho Doko Kihon Chosa}), IBJ couples are more highly educated: the education-by-education matrix has larger shares in higher education categories for both genders and smaller shares in lower categories.

Panel (c) compares the income matching matrix of IBJ couples with that of married households in the Employment Status Survey (ESS, \textit{Shugyo Kozo Kihon Chosa}, 2022, husband age $\leq$ 49).
Income categories include an explicit zero-income bin: on the IBJ side, women recorded as ``helping with housework'' (\textit{kajitetsudai}, a traditional label typically referring to unmarried women living with their parents without formal employment) are assigned zero income; on the ESS side, family workers (\textit{kazoku jugyosha}) and the non-employed (\textit{mugyosha}) are classified as zero income, and wives who are not in the labor force are also assigned to this category.
The income-by-income matrix shows that IBJ couples are distributed more broadly across higher income categories for both men and women, while the ESS data exhibit a clear concentration in cells with low-income women. 

The upward shift in the income distribution is partly mechanical: IBJ participants are older on average and thus observed at higher points of the life-cycle earnings profile, and institutional screening rules further select individuals with relatively higher earnings.
Beyond these compositional differences, a more fundamental distinction concerns the \emph{timing} at which income is recorded. IBJ captures income at the time of matching---that is, prior to marriage---whereas the ESS records income at the time of the survey, potentially years after marriage. This difference is likely to matter particularly for women: in nationally representative data collected after marriage, employment status and earnings typically adjust substantially around childbirth and early childrearing, with many women reducing labor supply or exiting the labor force. As a result, the ESS income matrix reflects not only pre-marital sorting but also subsequent intra-household specialization. 
In particular, the larger mass of women with zero income in the ESS relative to IBJ is consistent with post-marital labor supply reductions among wives. This comparison underscores that income matrices constructed solely from post-marital data may conflate assortative matching with post-marital labor supply adjustments, producing a biased picture of who sorts with whom at the point of marriage. IBJ's pre-marital income measures thus provide a cleaner lens for studying marital sorting on income.

\begin{figure}[!htbp]
  \begin{center}
  \subfloat[Age (Likelihood Ratio)]{\includegraphics[width = 0.9\textwidth]{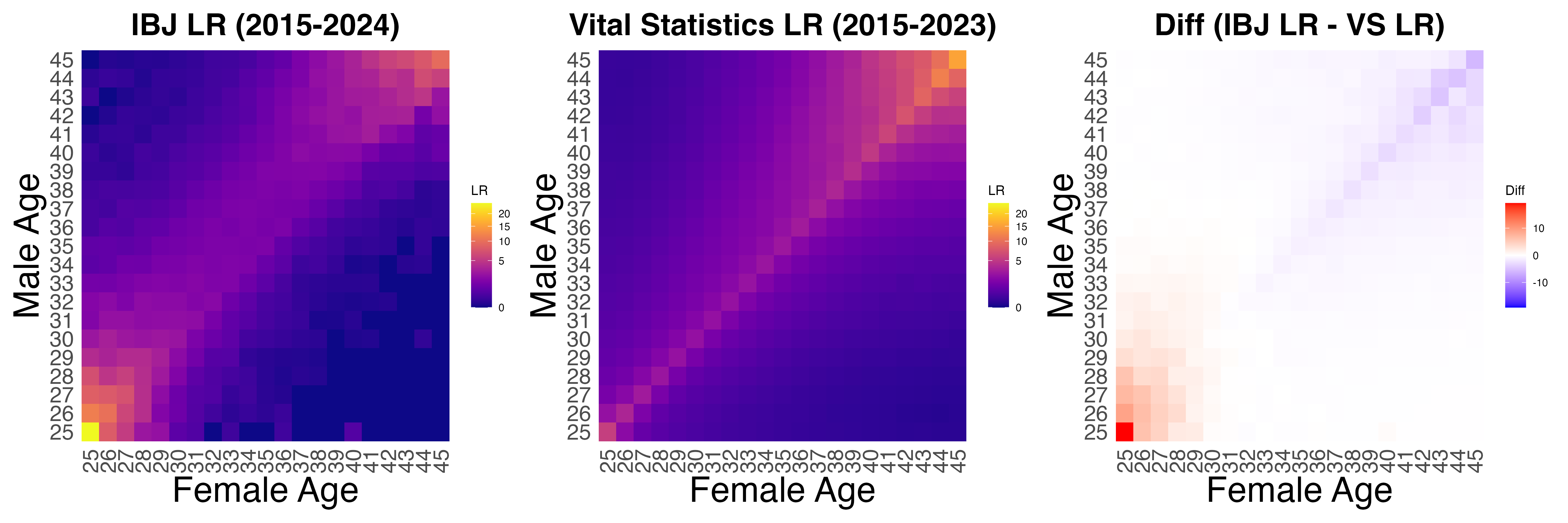}}\\
  \vspace{6mm}
  \subfloat[Education (Likelihood Ratio)]{\includegraphics[width = 0.9\textwidth]{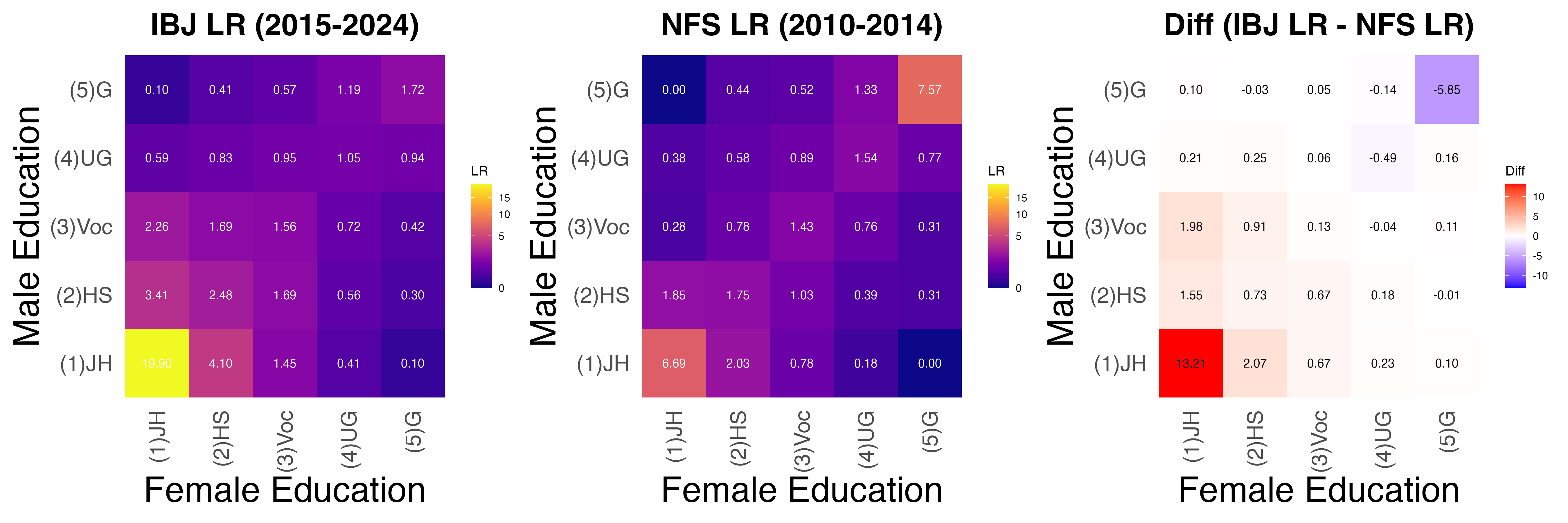}}\\
  \vspace{6mm}
  \subfloat[Income (Likelihood Ratio)]{\includegraphics[width = 0.9\textwidth]{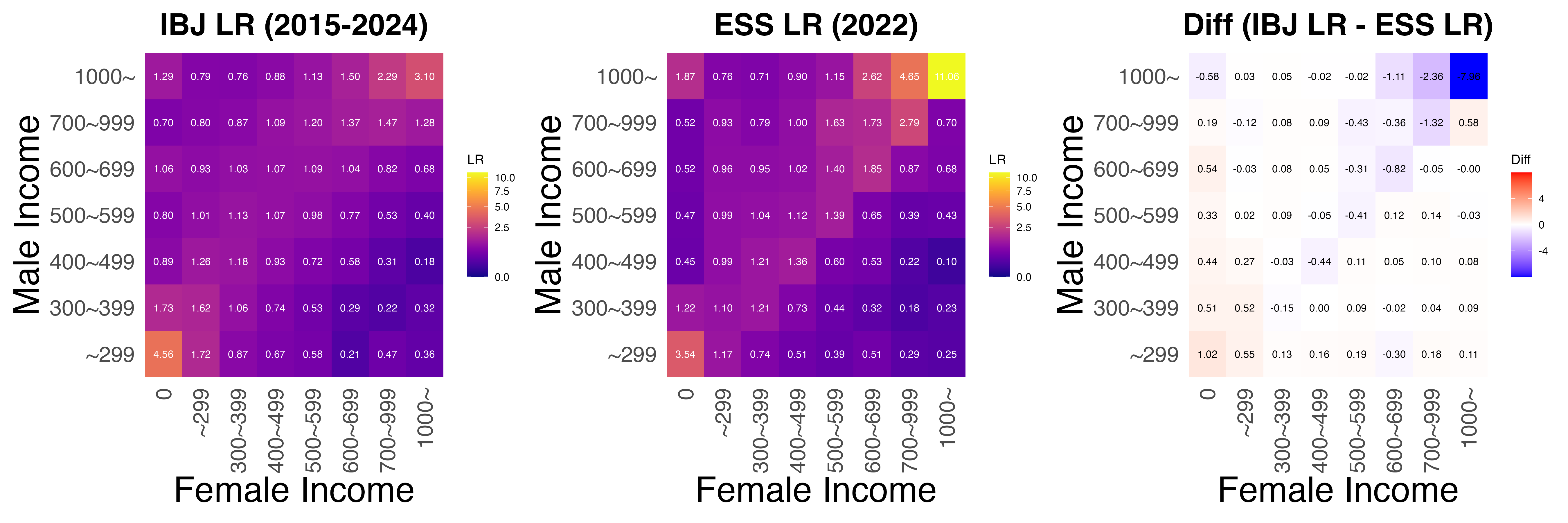}}
  \caption{IBJ Market Coverage Compared with National Statistics (NFS, VS, and ESS): Likelihood Ratio}
  \label{fg:representativeness_education_lr_ibj_vs_national}
  \end{center}
  \footnotesize
  Note: Panels report the likelihood ratio (LR), defined as $\text{LR}_{ij} = P(i,j) / [P_{\text{male}}(i) \times P_{\text{female}}(j)]$, where $P_{\text{male}}(i) = \sum_j P(i,j)$ and $P_{\text{female}}(j) = \sum_i P(i,j)$ are the marginal distributions because proportions are influenced by marginal distributions (e.g., the overall age or education composition of men and women). $\text{LR}_{ij} = 1$ indicates that the frequency of pairing $(i,j)$ is consistent with random matching; $\text{LR}_{ij} > 1$ indicates over-representation (positive sorting) and $\text{LR}_{ij} < 1$ indicates under-representation relative to what marginal frequencies alone would predict. The LR thus isolates the sorting pattern net of compositional differences between the two populations. IBJ age data cover 2015--2024 matched with Vital Statistics 2015--2023; education data compare IBJ 2015--2024 with the 15th National Fertility Survey (marriage cohort 2010--2014; the latest publicly available edition); income data compare IBJ 2015--2024 with the Employment Status Survey (2022, husband age $\leq$ 49). See notes to Figure \ref{fg:representativeness_education_ibj_vs_national} for the definition of zero-income categories.
\end{figure}

Figure \ref{fg:representativeness_education_lr_ibj_vs_national} presents the same comparisons using likelihood ratios (LR), which net out marginal-distribution differences and isolate assortative-matching strength. The LR heatmaps show that, after controlling for composition, positive age assortative matching is broadly similar between IBJ and Vital Statistics. The main-diagonal concentration persists in both datasets, indicating comparable age sorting once composition is controlled for. For education, LR comparisons with the National Fertility Survey likewise show positive within-category assortative matching in both samples, though the relative intensity differs across cells: IBJ exhibits comparatively stronger concentration in some lower- and middle-education pairings, whereas the National Fertility Survey shows stronger concentration in the highest education cells. For income, the LR comparison with the ESS also indicates positive assortative matching in both datasets. The broad diagonal structure is similar, but notable differences appear in the zero-income column: the ESS exhibits a large mass of wives with zero income across all husband income levels, consistent with post-marital labor force exit, whereas IBJ shows almost no zero-income wives. At the upper extreme, the ESS shows stronger concentration in the highest income cell, while IBJ exhibits relatively more mass in middle-income pairings. These LR comparisons suggest that IBJ-national differences mainly reflect composition shifts toward older and more educated users, while the underlying sorting structure remains broadly comparable.

Taken together, these comparisons show that IBJ is a sizable, policy-relevant segment of the Japanese marriage market, while attracting an older and more educated population than the national average. Once marginal-distribution differences are accounted for, the underlying assortative-matching structure---especially for age and education---is broadly comparable to representative data. For income, the comparison between pre-marital (IBJ) and post-marital (ESS) income matrices is itself informative: the shift in female income toward zero between the two datasets is consistent with post-marital reductions in female labor supply, suggesting that analyses relying solely on post-marital income data may conflate sorting with subsequent household specialization. We therefore interpret IBJ patterns as reflecting both compositional selection into the platform and genuine matching behavior, and account for both in later analyses.

\paragraph{Verified Matches.}
In this study, a match is a confirmed engagement formed on IBJ through mutual agreement. Because the platform enforces strict rules---including a ban on concurrent engagements and consultant-mediated confirmation---observed engagements are highly credible indicators of successful match formation. Engagements are also time-stamped and linked to full behavioral histories, allowing us to trace when and how matches form (Section \ref{sec:ibj_user_behavior}). This design minimizes false positives and makes matching operationally clear and empirically reliable.

\paragraph{Verified Variables}
IBJ data include highly verified demographic and behavioral variables. Key covariates---age, education, income, health information, and marital history---are cross-checked with government documents, tax records, and certified medical exams. User behavior is tracked at high frequency with timestamps, covering searches, proposals, messages, and mutual agreement. This data integrity contrasts with typical self-reported online-dating \citep{hitsch2010matching} and survey datasets \citep{dupuy2014personality, chiappori2024analyzing} and supports rigorous analysis of matching outcomes and behavioral determinants.
For the main empirical exercise, we restrict to the balanced sample with non-missing values for all variables used in the affinity matrix.

The data also contain occupation categories. 
Because employer information is observed during income verification, recorded occupations are highly reliable. 
Following \citet{goldin2014grand}, we build an occupational-flexibility index by mapping occupations to characteristics that capture how strongly jobs reward long, inflexible hours. 
Characteristics come from Japan's Occupational Information Network (O-NET), developed by the Ministry of Health, Labour and Welfare and modeled on U.S. O*NET.
We use five occupation-level characteristics from O-NET corresponding the U.S. O*NET characteristics used by \citet{goldin2014grand}: time pressure, contact with others, interpersonal relationships, structured work, and freedom to make decisions. 
Each is standardized to mean zero and unit variance at the O-NET occupation level and averaged within each IBJ occupation category. 
The final flexibility index averages the five standardized components, with all five component signs flipped so that higher values indicate greater flexibility.
Although occupational flexibility is not a standard marriage-market attribute, it captures time-use constraints and career dynamics central to intra-household allocation, so it is plausibly relevant for partner selection.

\subsection{IBJ Users' Dating Process and Stage}\label{sec:ibj_user_behavior}

Table \ref{tb:dating_process} summarizes IBJ's sequential partner-formation process and the distinction between non-exclusive exploration and exclusive commitment.\footnote{This paper focuses on year-level matching outcomes; the underlying IBJ data also contain time-stamped click impressions and action logs at the individual user level, as well as both user-stated and revealed partner-search filters (covering age, income, education, location, and other dimensions). These micro-level traces support ongoing companion work on individual-level search dynamics, partner-search filters, and the role of marriage agencies and consultants.} The process starts at Application, where users browse profiles and can initiate contact by sending an \textit{omiai} request. Initiation is not gender-specific: both men and women can apply. A request advances to Omiai Meeting only if accepted, after which the pair meets in person or online and independently decides whether to continue.

If both parties approve after the first meeting, they enter Pre-relationship, a sequence of non-exclusive dates. At this stage, users may interact with multiple partners and repeatedly decide whether to continue or exit each match. If both then choose to deepen the relationship, they move to the exclusive Serious relationship stage. This stage involves repeated committed dates with the same partner, though either party can still exit. The process ends at Proposal (engagement/matching), which requires mutual confirmation of intent to marry. Overall, the platform moves users gradually from open search to exclusive commitment through mutually agreed steps.

\begin{table}[!htbp]
\caption{Dating Process by Stage, Agent, and Action}
\label{tb:dating_process}
\begin{center}\footnotesize
    \begin{tabular}{@{}llll@{}}
    \toprule
    \textbf{Stage} & \textbf{Agent} & \textbf{Action} & \textbf{Note} \\
    \midrule
    1. Application & Proposer & Browse candidate profiles & Set filter conditions \\
     & Proposer & Click on candidate profiles & Expressing interest \\
     & Proposer & Send omiai request & Initiates non-exclusive interaction \\
     & Receiver & Accept or reject omiai request & \\
    \midrule
    2. Omiai Meeting & Both & Conduct omiai meeting & Face-to-face or virtual meeting \\
     & Both & Accept or reject trial dating & Decision to continue or exit \\
    \midrule
    3. Pre-relationship & Both & Begin first trial date & Non-exclusive\\
     & Both & Accept or reject second trial date & Decision to continue or exit \\
     & Both & Proceed with additional trial dates & Repeated decision process \\
     & Both & Accept or reject committed dating & Transition to exclusive dating \\
    \midrule
    4. Serious relationship & Both & Begin first date & Exclusive\\
     & Both & Accept or reject second date & Decision to continue \\
     & Both & Proceed with additional dates & Repeated decision process \\
    \midrule
    5. Proposal & Both & Accept marriage engagement & \\
    \bottomrule
\end{tabular}
\end{center}
\footnotesize
%Note: See Section \ref{sec:ibj_user_behavior} for definitions.

\end{table}

\begin{table}[!htbp]
  \begin{center}
      \caption{Summary Statistics of Action Counts Per Active User in 2024 by Gender}
      \label{tb:summary_statistics_action_male_female_2024}
      \input{figuretable/labor_family_economics_project/summary_statistics_action_male_female_2024}
  \end{center}\footnotesize
  \textit{Note}: The table reports stage-specific action counts per user in 2024 for users with at least one Application action. As a result, counts at later stages may be zero for users who exited earlier. The statistics pool actions across proposer and receiver roles and summarize total stage-level participation rather than role-specific behavior.
\end{table} 

Table \ref{tb:summary_statistics_action_male_female_2024} reports stage-specific activity counts per user in 2024, by gender, for users with at least one Application action. Because the sample is conditioned on initial Application activity, later-stage counts---Pre-relationship, Serious relationship, and Proposal---can be zero for users who exit earlier. Counts pool proposer and receiver roles and capture total stage-level engagement. The table shows a sharp funnel: Application activity is very intensive and dispersed for both genders (mean counts above 100 with large standard deviations) and exhibits pronounced right-skew---the male Application median of 55 sits well below its mean of 120 with a maximum of 15{,}788, indicating a thin tail of hyper-active senders---while participation drops sharply later. Average activity falls to just above two at Pre-relationship, around 0.2 at Serious relationship, and about 0.1 at Proposal, indicating substantial attrition as interactions move toward exclusivity and commitment. This implies that Proposal-stage assortativeness can reflect cumulative selection across stages, not only final-agreement preferences. 

\begin{figure}[!htbp]
  \begin{center}
  \subfloat[Application]{\includegraphics[width = 0.45\textwidth]{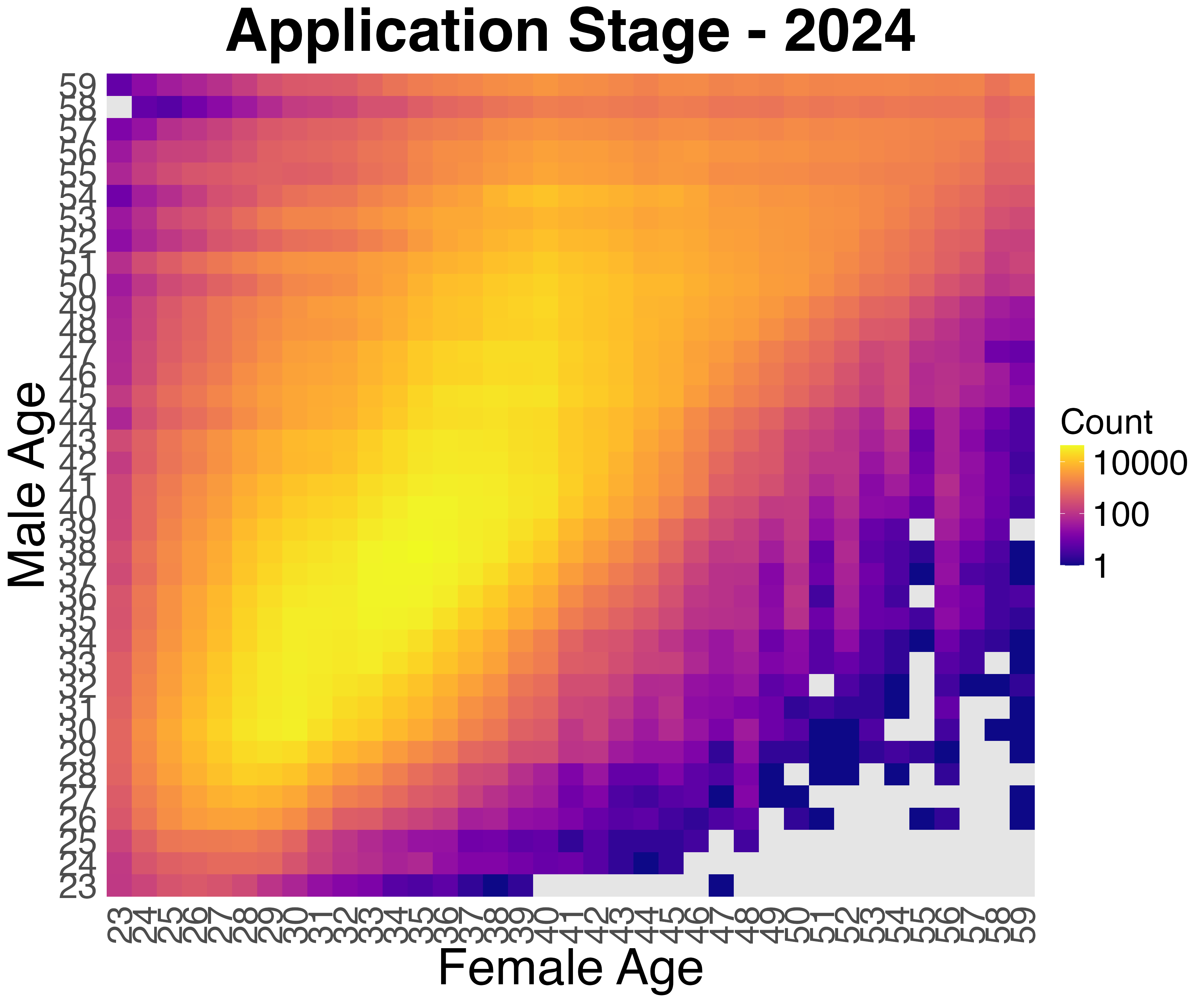}}
  \subfloat[Pre-relationship]{\includegraphics[width = 0.45\textwidth]{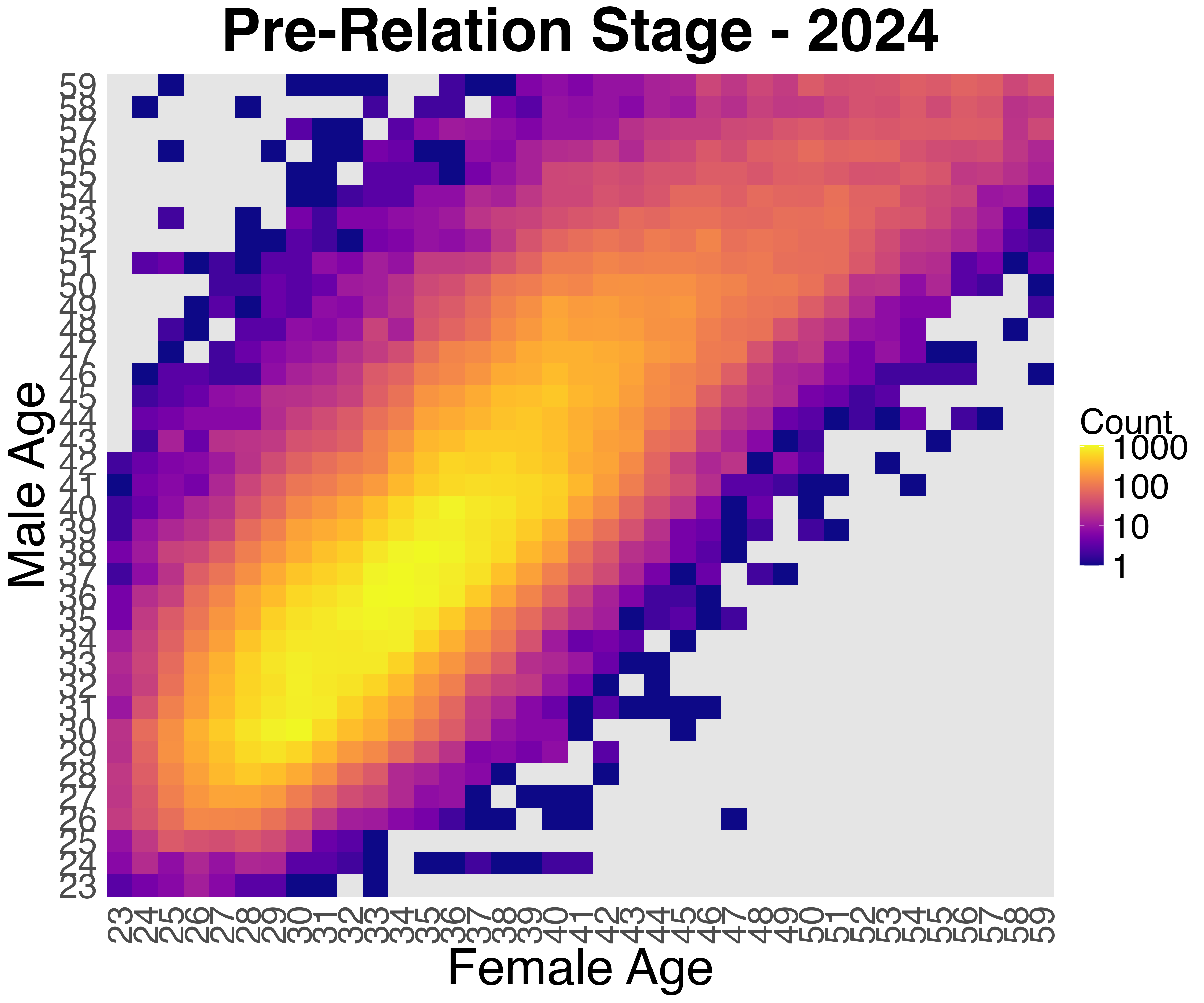}}\\
  \vspace{6mm}
  \subfloat[Serious relationship]{\includegraphics[width = 0.45\textwidth]{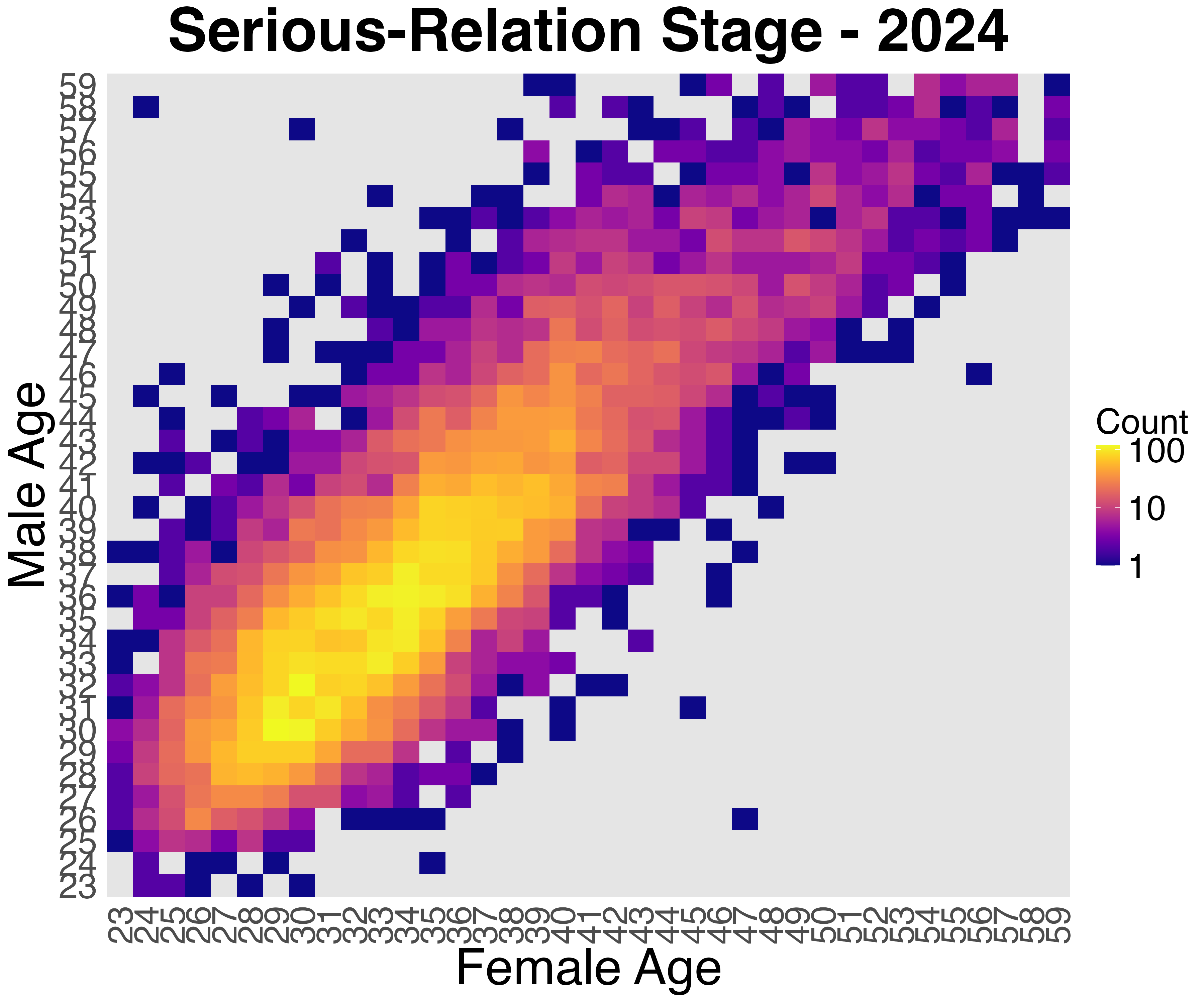}}
  \subfloat[Proposal]{\includegraphics[width = 0.45\textwidth]{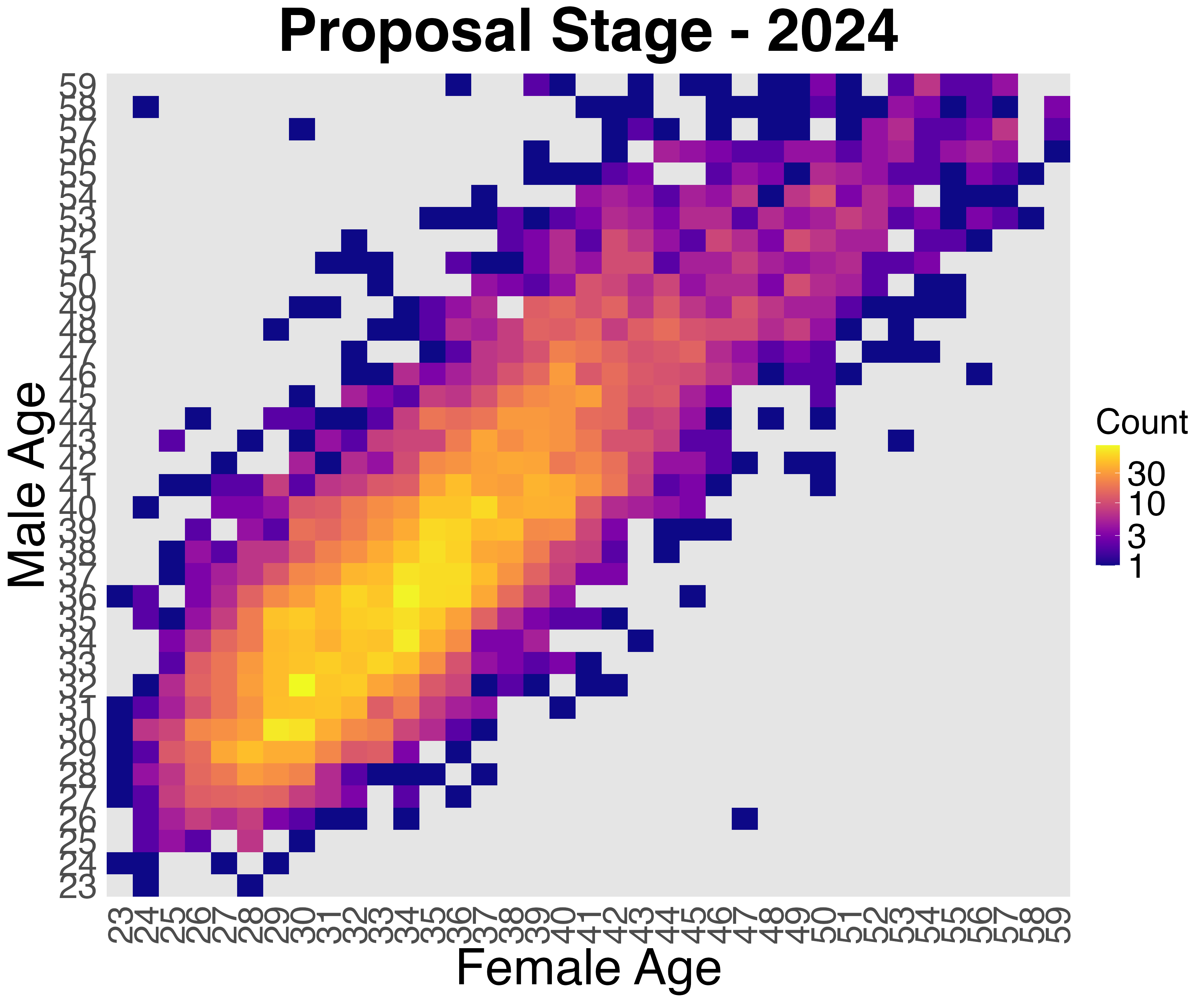}}
  \caption{Matching Matrix per Stage in 2024: Based on Age}
  \label{fg:age_matching_heatmap_proposal_2024} 
  \end{center}
  \footnotesize
  Note: Application stage includes all couples who applied for meetings in 2024. Pre-relationship, Serious relationship, and Proposal stages include couples who progressed to each respective stage in 2024.  
\end{figure}

Figure \ref{fg:age_matching_heatmap_proposal_2024} provides a complementary view of age matching across stages in 2024. In this one-dimensional age view, heatmaps show clear diagonal concentration even at Application, suggesting age-assortative interactions from search onset. But because the figure conditions only on age, apparent assortativeness may partly reflect selection on correlated attributes---such as education, income, or family preferences---that users jointly consider when initiating and continuing interactions. One-dimensional age patterns therefore cannot distinguish whether assortative matching originates early or is induced by multidimensional sorting on richer covariates.

These patterns motivate our subsequent analysis, which examines when assortative matching along different dimensions first emerges and how it evolves as matches are selectively retained or discarded throughout the dating process.

\subsection{Summary Statistics of Matched Users in 2024}\label{subsec:summary_stats}

Table \ref{tb:summary_statistics_matched_continuous_2024} reports the summary statistics of matched couples.
In 2024, men in our matched sample are on average approximately 3 years older than women (39.06 vs. 36.01), taller by about 13 cm (171.65 cm vs. 158.67 cm), and heavier by roughly 16 kg (65.19 kg vs. 49.40 kg), resulting in a higher BMI on average. 
Men also report higher annual income, with a mean upper-limit income category of 782.91 (in 10,000 JPY) compared to 503.62 for women.
These differences reflect well-established gender patterns in age, physical attributes, and earning potential in the marriage market.
The distributions of these variables are tightly clustered around the means, as indicated by relatively small standard deviations in age (7.15 for women, 8.02 for men) and income (208.45 for women, 347.35 for men, in 10,000 JPY). 
While the mean flexibility index is identical for men and women at -0.21, the distribution differs markedly across genders: the standard deviation is substantially larger for women (0.49) than for men (0.35), indicating greater dispersion in occupational flexibility among women.

In terms of discrete attributes, educational attainment differs by category across genders. Women are more concentrated in vocational and undergraduate categories (22.6\% and 62.0\%, respectively), whereas men are much more likely to hold graduate degrees (24.2\% vs. 6.7\% for women). Drinking and smoking habits also differ sharply by gender: 26.2\% of men are regular drinkers compared to 15.5\% of women, and 3.4\% of men regularly smoke versus 0.3\% of women. Meanwhile, the majority of both genders report never having been married before (88.4\% of women and 84.8\% of men). Regarding household preferences, most respondents in both groups favor either discussing housework with a partner or sharing it equally, but women are more likely to report taking primary responsibility themselves (9.1\% vs. 4.6\%), whereas men are more likely to favor equal sharing (70.3\% vs. 56.1\%). Childcare preferences are similarly centered on equal sharing, while stated desire for children is high for both genders: 71.5\% of women and 70.7\% of men report wanting children, with men somewhat more likely to report no clear preference (25.7\% vs. 21.8\%).

\begin{table}[!htbp]
  \begin{center}\footnotesize
      \caption{Summary Statistics in 2024 by Gender: Matched Couples}
      \label{tb:summary_statistics_matched_continuous_2024}
      \subfloat[Continuous]{\input{figuretable/labor_family_economics_project/summary_statistics_matched_continuous_2024}}\\
      \subfloat[Discrete]{\input{figuretable/labor_family_economics_project/summary_statistics_matched_discrete_2024}}
  \end{center}\footnotesize
  \textit{Note}: Income represents the upper limit of the income category. Occupational Flexibility follows \cite{goldin2014grand}.
\end{table}

\subsection{Preferences and Their Socioeconomic Correlates}\label{subsec:correlation_preferences}

Because these preference measures are rarely observed before marriage and are mostly absent from administrative data, it is informative to examine how they co-vary with anthropometric and sociodemographic characteristics used in matching research. Their empirical relationship with age, income, education, and occupation provides a first indication of whether preferences form an independent heterogeneity margin or merely proxy for existing socioeconomic traits. We therefore first describe the distribution of these variables across standard demographic dimensions before presenting structural assortative-matching estimates.

\paragraph{Preference Distribution across Age, Income, Education}

\begin{figure}[!htbp]
  \begin{center}
  \subfloat[Preference for Children]{\includegraphics[width = 0.82\textwidth]{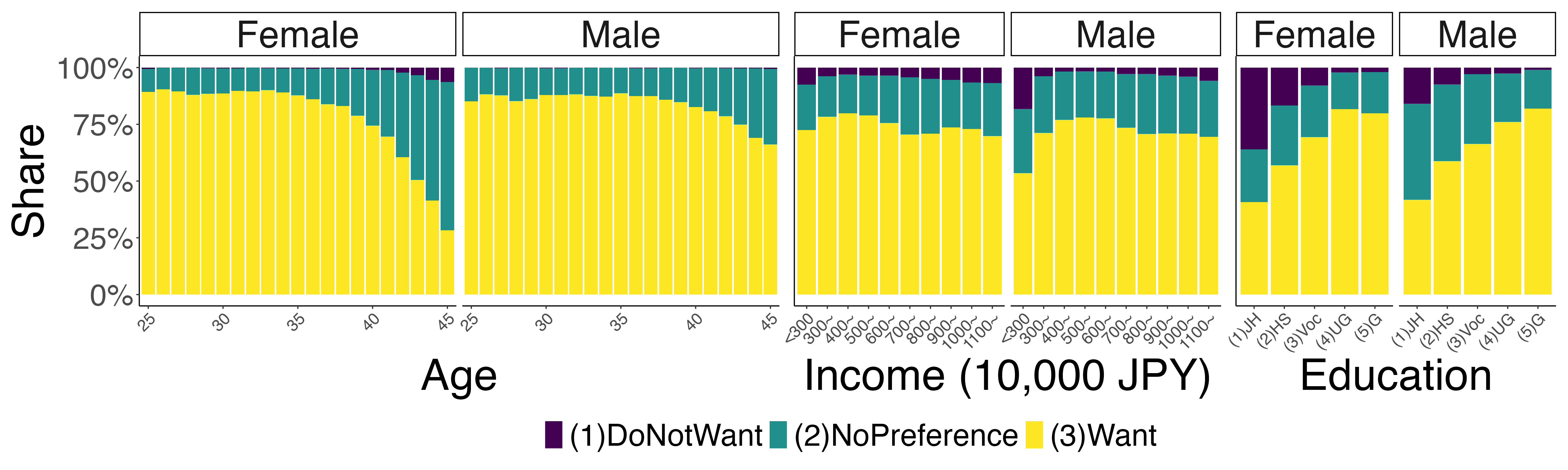}}\\
  \vspace{6mm}
  \subfloat[Childcare]
  {\includegraphics[width = 0.82\textwidth]{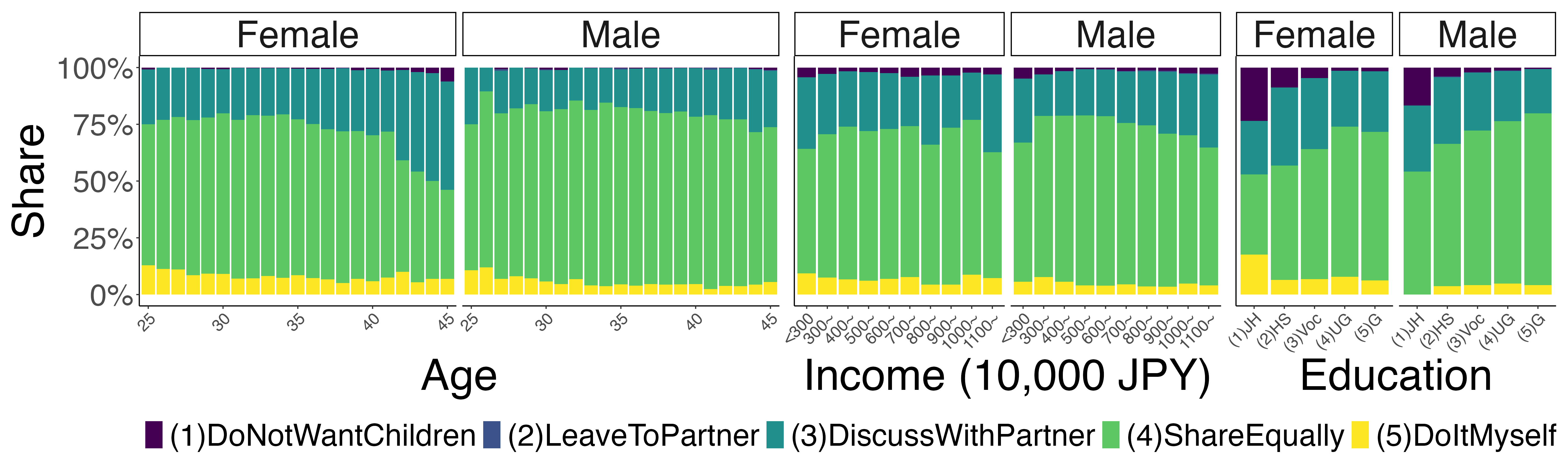}}\\
  \vspace{6mm}
  \subfloat[Housework]{\includegraphics[width = 0.82\textwidth]{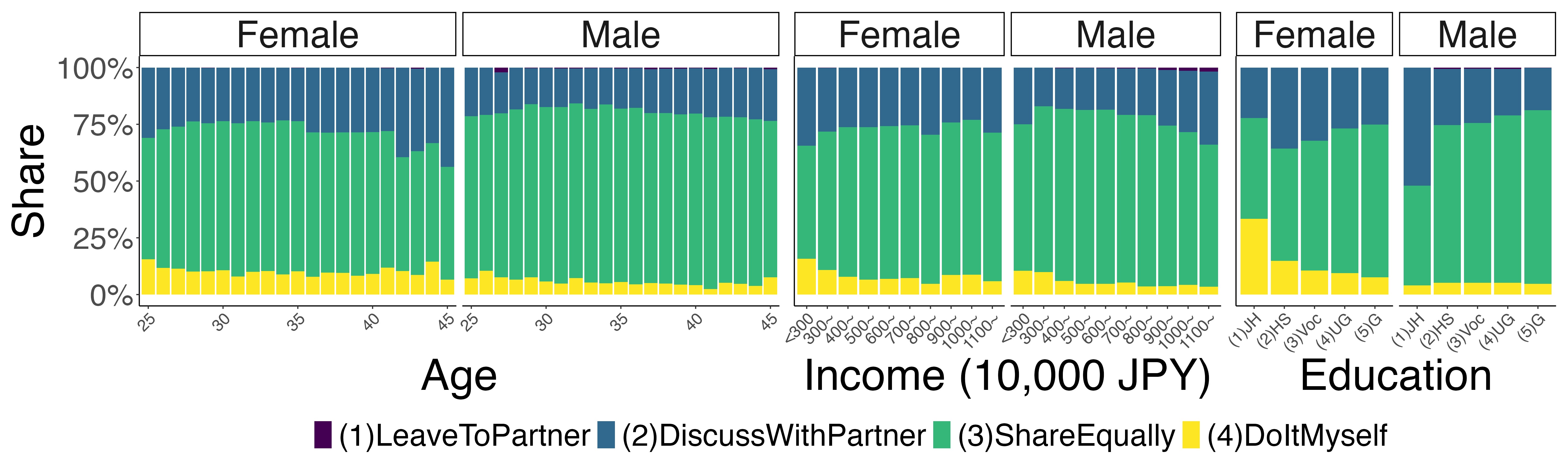}}\\
  \vspace{6mm}
  \subfloat[Occupational Flexibility]{\includegraphics[width = 0.82\textwidth]{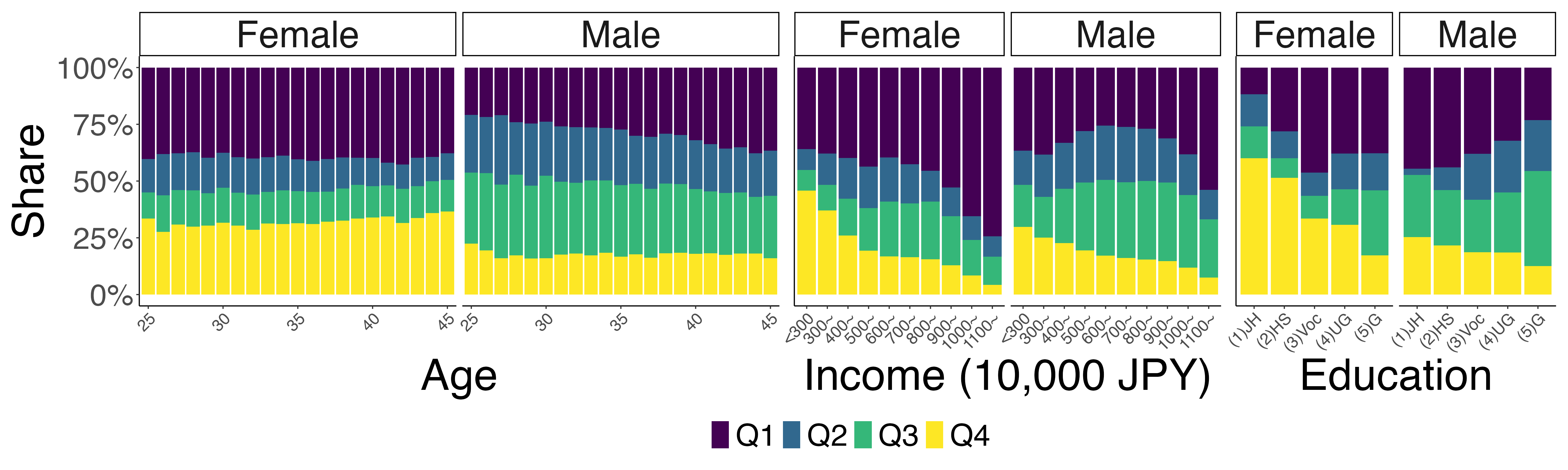}}
  \caption{Preference Distribution across Age, Income, Education for Matched Users in 2024}
  \label{fg:share_across_axes_flexibility} 
  \end{center}
  \footnotesize
  %Note: 
\end{figure} 

Panel (a) in Figure \ref{fg:share_across_axes_flexibility} shows clear age gradients in fertility preferences for both genders, with a steeper age-related decline in desire for children among women. Younger women and men mostly report wanting children, but at older ages the share of women reporting no desire or no clear preference rises more sharply. Income gradients exist for both genders and appear steeper for men: higher-income men are more likely to want children than lower-income men, while women’s fertility preferences vary less across income groups. 
Along education, the share wanting children generally rises with attainment, consistent with the fertility-education gradient in \citet{doepke2023economics}. The distribution is broadly similar across genders, though lower-educated women appear somewhat more polarized than lower-educated men. Overall, fertility preferences show demographic gradients and gender asymmetries, especially by age.

In Panel (b), “share equally” dominates childcare preferences in almost all demographic groups. %, but gender asymmetries remain. Women are more likely to report active childcare involvement, while men are somewhat more likely to favor discussion or delegation. 
Age differences are modest among men, whereas younger women are more likely than older women to prefer an equal division of childcare.
Income gradients are generally limited, although lower-income men appear slightly more supportive of equal sharing than higher-income men, while women’s responses remain relatively stable across income groups.
By education, higher-educated individuals---especially women---tend to cluster more around equal sharing.

A similar but stronger asymmetry appears in housework preferences in Panel (c). Although “share equally” is still the modal response for both genders, women are less likely than men to choose unilateral non-involvement and more likely to report sharing or taking responsibility themselves. Dispersion across alternatives is larger for men with lower income and education, indicating greater heterogeneity in male housework attitudes in those groups. Age gradients are weak for both genders, but gender gaps persist across demographics. Compared with fertility preferences, housework attitudes show stronger gender asymmetry and somewhat weaker socioeconomic stratification.

In Panel (d), occupational flexibility shows substantial heterogeneity across demographic groups and noticeable gender differences in dispersion.
Conditional on age, income, and education, women are substantially more likely than men to be in more flexible occupations.
Age patterns are relatively modest overall, although older women tend to be more concentrated in flexible occupations while older men tend to be more concentrated in less flexible occupations.
Higher-income groups tend to exhibit larger shares in the least flexible quartile (Q1) and smaller shares in the most flexible quartile (Q4), particularly among women.
Education gradients differ notably by gender.
Among women, higher education is associated with greater concentration in lower-flexibility occupations, whereas among men, higher education is associated with lower shares in both the least flexible (Q1) and most flexible (Q4) occupations and greater concentration in intermediate flexibility quartiles (Q2 and Q3).
Overall, the patterns suggest that occupational flexibility varies systematically with labor-market position and differs markedly by gender.

\paragraph{Correlation Matrix}
To assess whether preference measures capture an independent heterogeneity margin or proxy for standard socioeconomic traits, we examine the full correlation structure among attributes for matched users. Table \ref{tb:correlation_matrix_matched_male} reports pairwise correlations by gender in 2024. This complements the distributional evidence above and shows how strongly fertility and household preferences align statistically with age, education, income, and anthropometric characteristics.

%\begin{landscape}
\begin{table}[!htbp]
  \begin{center}\footnotesize
      \caption{Correlation Matrix by Gender: Matched Users in 2024}
      \label{tb:correlation_matrix_matched_male}
      \subfloat[Male]{\input{figuretable/labor_family_economics_project/correlation_matrix_matched_male}}\\
      \vspace{6mm}
      \subfloat[Female]{\input{figuretable/labor_family_economics_project/correlation_matrix_matched_female}}
  \end{center}\footnotesize
  \textit{Note}: Income represents the upper limit of the income category. Occupational Flexibility follows \cite{goldin2014grand}.
\end{table}
%\end{landscape}

Overall, preference variables are only weakly correlated with most sociodemographic and anthropometric characteristics, suggesting a distinct heterogeneity dimension. For both men and women, child preference is strongly negatively correlated with age, consistent with declining fertility intentions at older ages, but only modestly correlated with education and income. Household-responsibility preferences (housework and childcare) are positively correlated with child preference, especially among women, yet have limited association with education, income, height, or weight. By contrast, standard sociodemographic variables are more tightly interrelated: education and income are positively correlated for both genders, and height and weight are strongly correlated within gender. Gender asymmetries also appear: age is more strongly negatively correlated with fertility preferences among women, and childcare preferences are more internally coherent among women. Taken together, the correlations reinforce that fertility and household preferences form a relatively independent sorting margin rather than simply reflecting socioeconomic gradients.

\subsection{Trends in Covariate Distributions of Matched Users, 2015--2024}\label{subsec:summary_stats_history}

\begin{figure}[!htbp]
  \begin{center}
  \subfloat[Age]{\includegraphics[width = 0.42\textwidth]{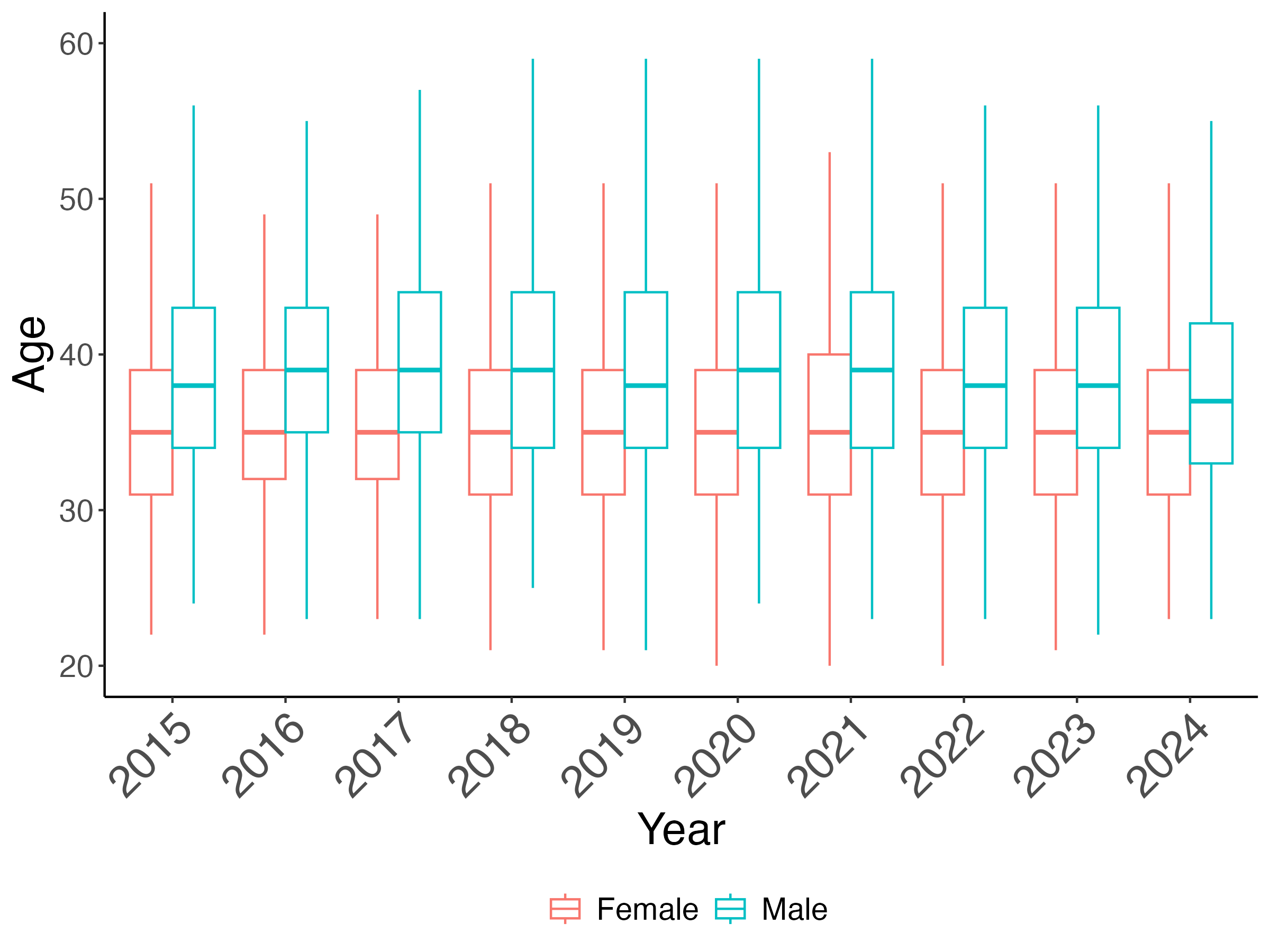}}
  \subfloat[Height]{\includegraphics[width = 0.42\textwidth]{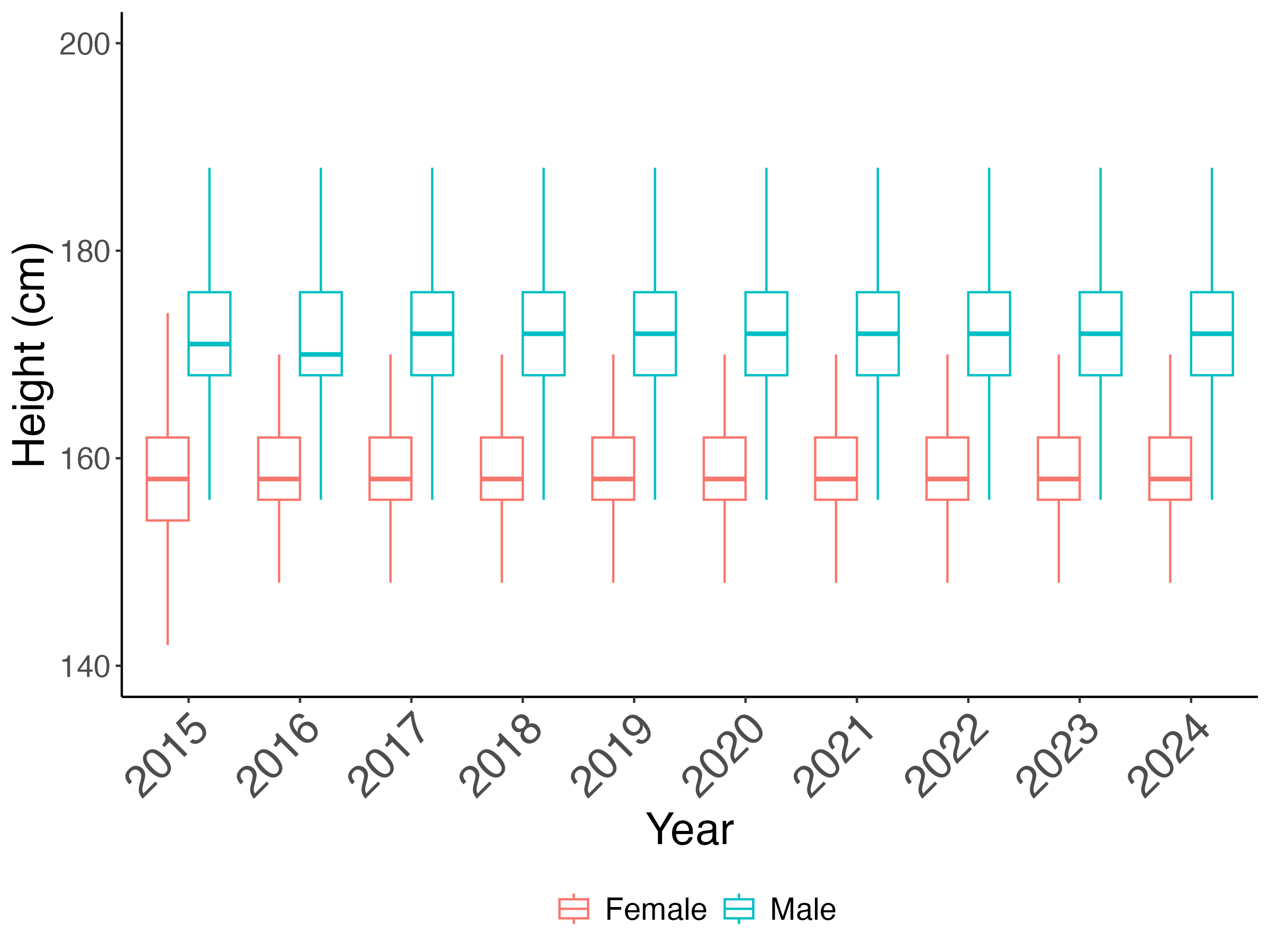}}\\
  \vspace{6mm}
  \subfloat[Weight]{\includegraphics[width = 0.42\textwidth]{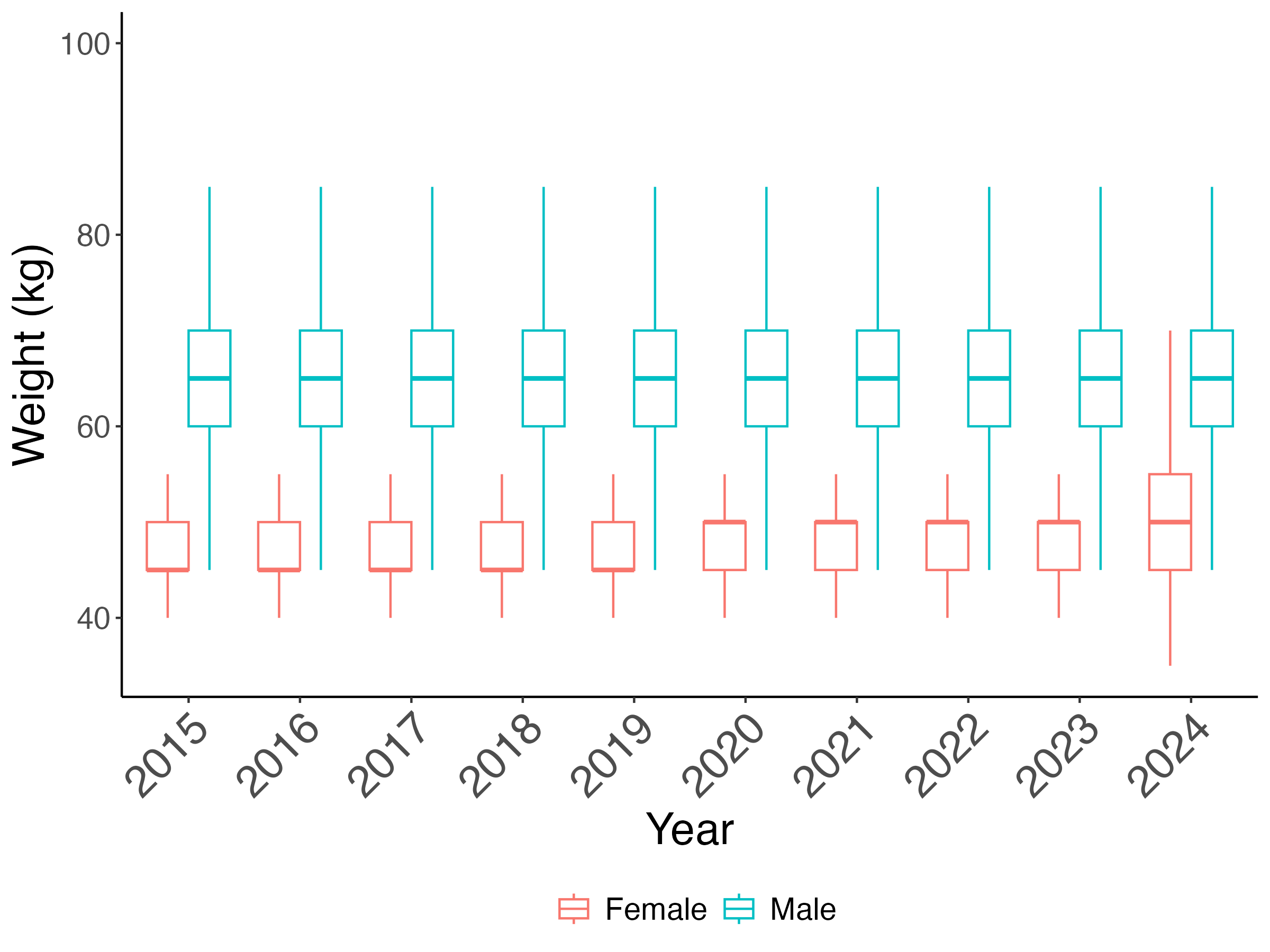}}
  \subfloat[BMI]{\includegraphics[width = 0.42\textwidth]{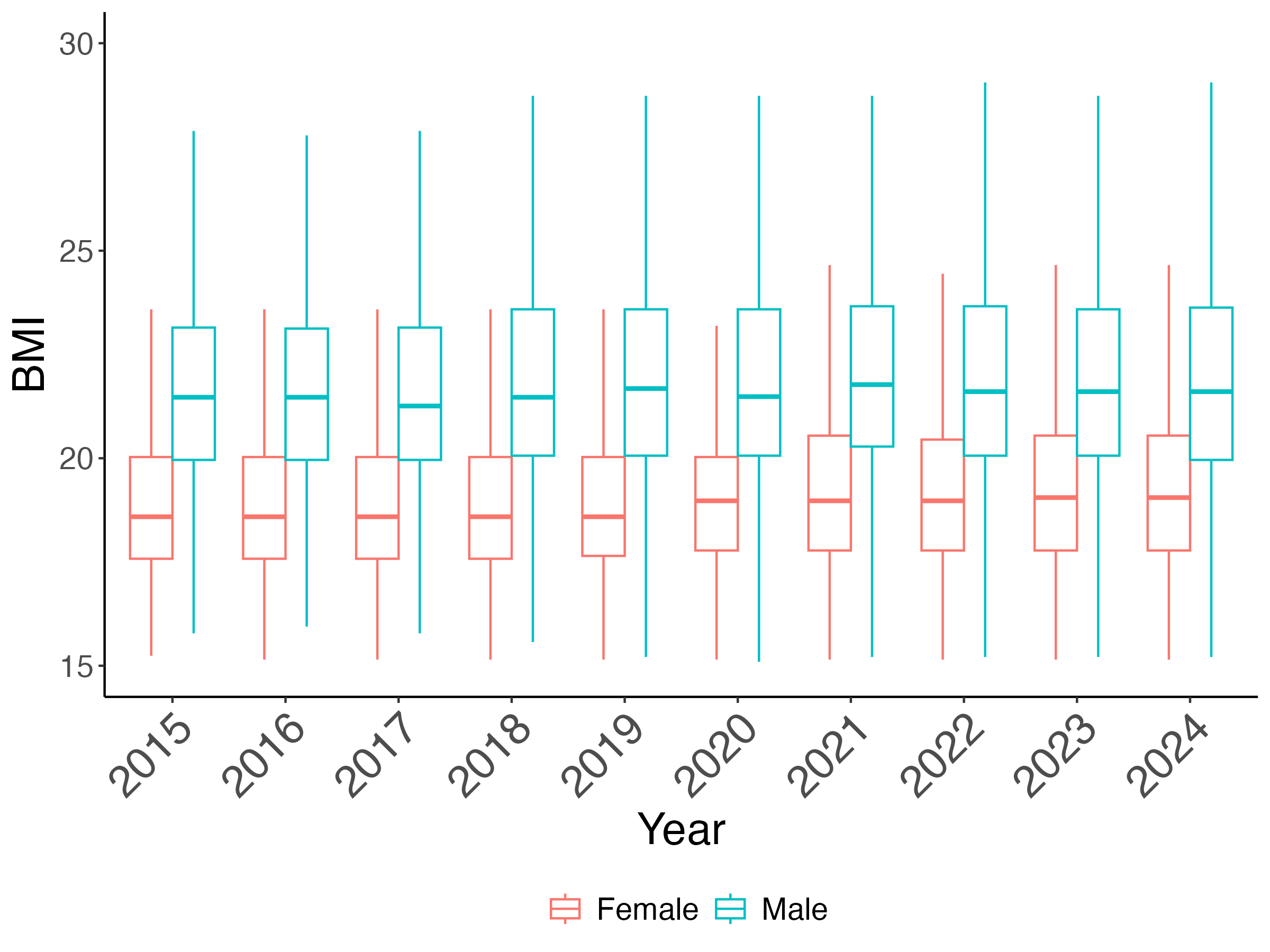}}\\
  \vspace{6mm}
  \subfloat[Income Category]{\includegraphics[width = 0.42\textwidth]{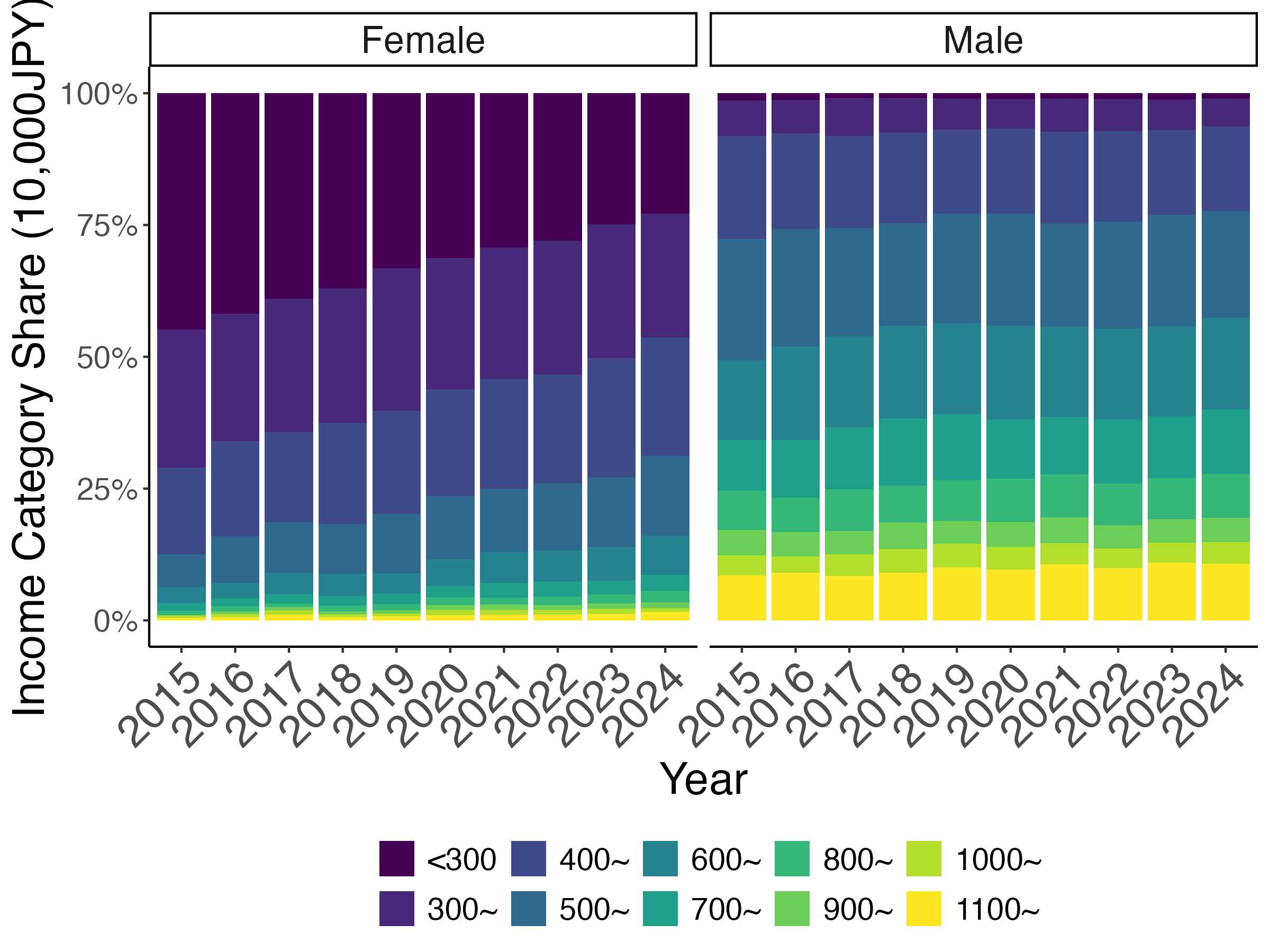}}
  \subfloat[Education]{\includegraphics[width = 0.42\textwidth]{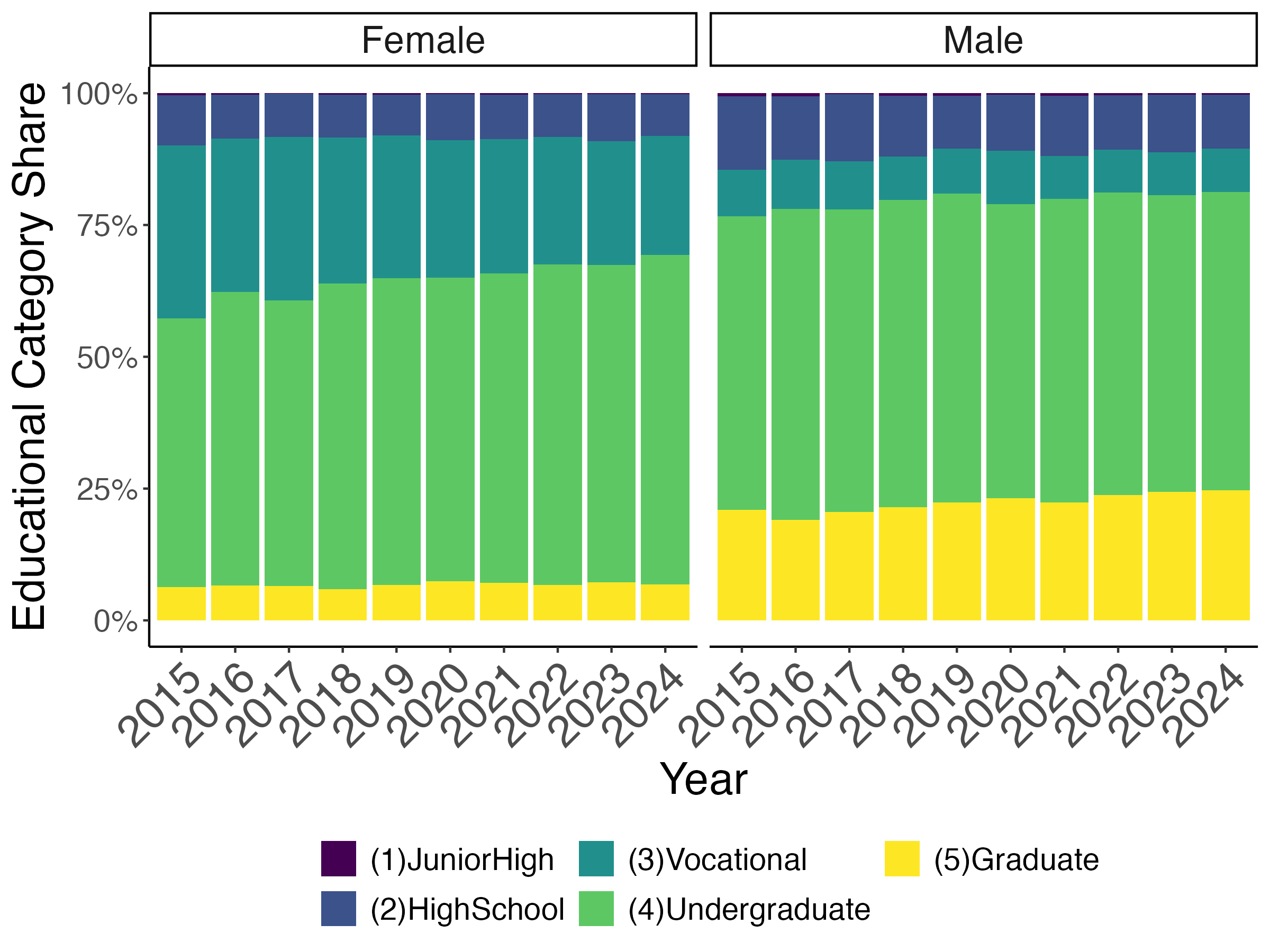}}
  \caption{Distribution and Share of Anthropometric and Sociodemographic Attractiveness in Married Couples}
  \label{fg:boxplot_education_2015_2024} 
  \end{center}
  \footnotesize
  Note: In each boxplot, the central box represents the interquartile range (IQR), spanning from the 25th to the 75th percentile of the distribution, with the horizontal line inside indicating the median (50th percentile). The vertical lines ("whiskers") extend to the most extreme values within 1.5 times the IQR from the box; values beyond this range are considered outliers and are not shown in the plot.
\end{figure}

Figure \ref{fg:boxplot_education_2015_2024} presents trends in anthropometric and sociodemographic characteristics of individuals in married couples over the 2015–2024 period. Panels (a)–(d) show boxplots of age, height, weight, and BMI by gender, revealing broadly stable distributions and persistent gender gaps across years. In contrast, panels (e) and (f) depict notable shifts in the composition of income and educational attainment. Among women, the income distribution shifts steadily toward higher categories over time, with declining shares in the lower bins and rising shares in the 5-million-JPY-and-above categories. This upward shift suggests that women’s economic attractiveness—traditionally less emphasized in the marriage market—has become increasingly important. Educational attainment also moves modestly upward, particularly among women, with a gradual rise in the undergraduate share and a corresponding decline in vocational categories.

\begin{figure}[!htbp]
  \begin{center}
  \subfloat[Drinking]{\includegraphics[width = 0.42\textwidth]{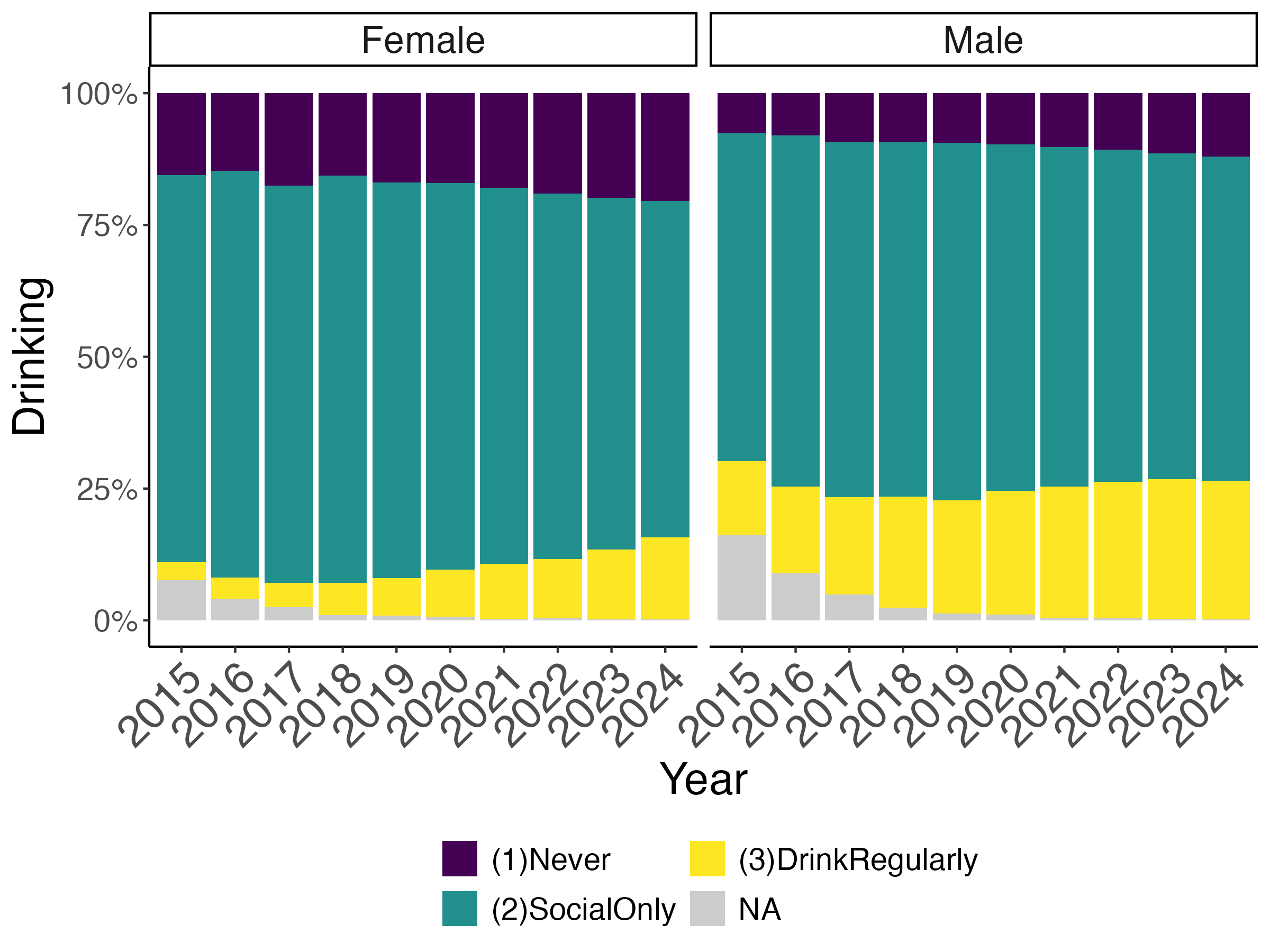}}
  \subfloat[Smoking]{\includegraphics[width = 0.42\textwidth]{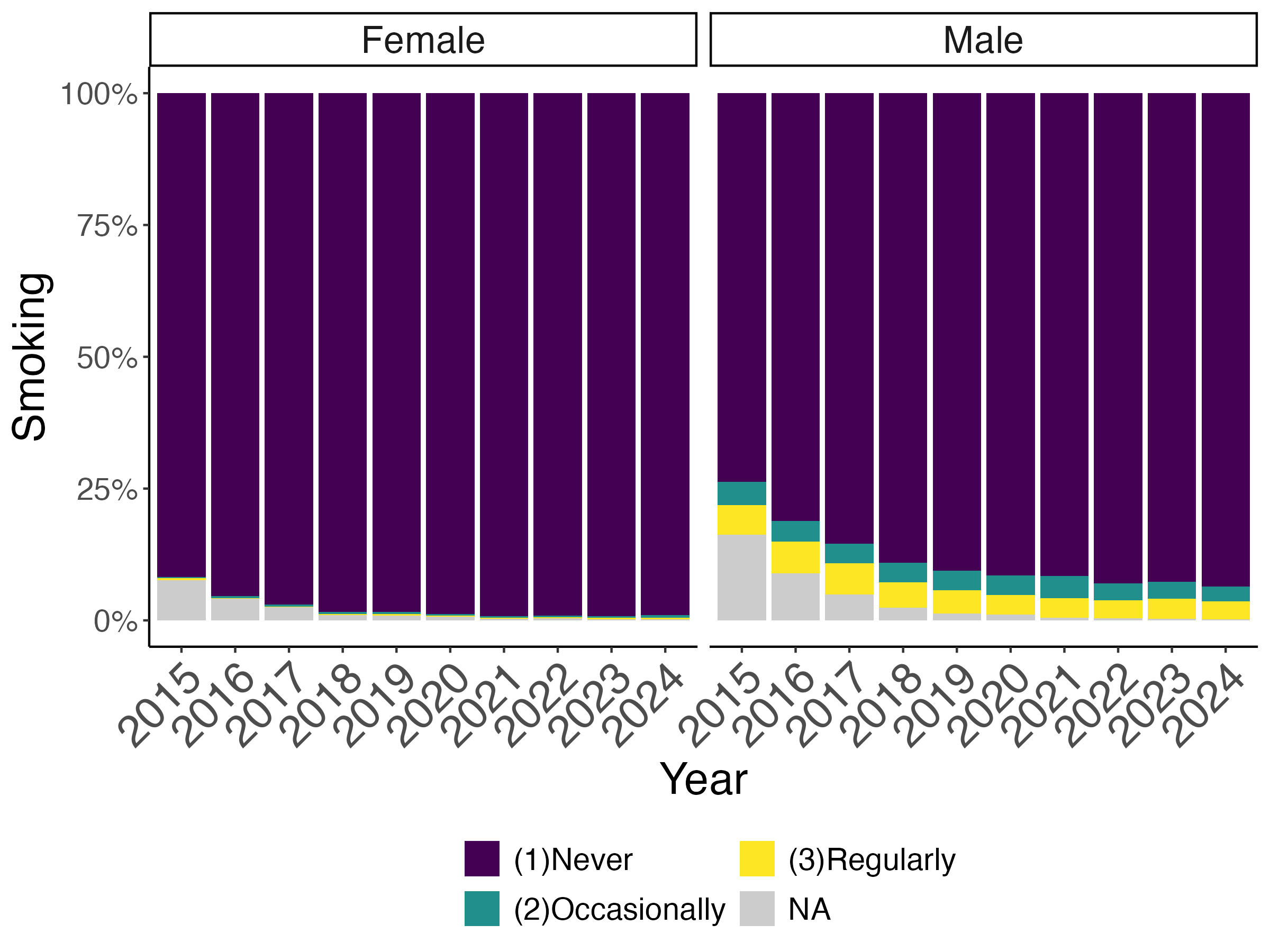}}\\
  \vspace{6mm}
  \subfloat[Marital History]{\includegraphics[width = 0.42\textwidth]{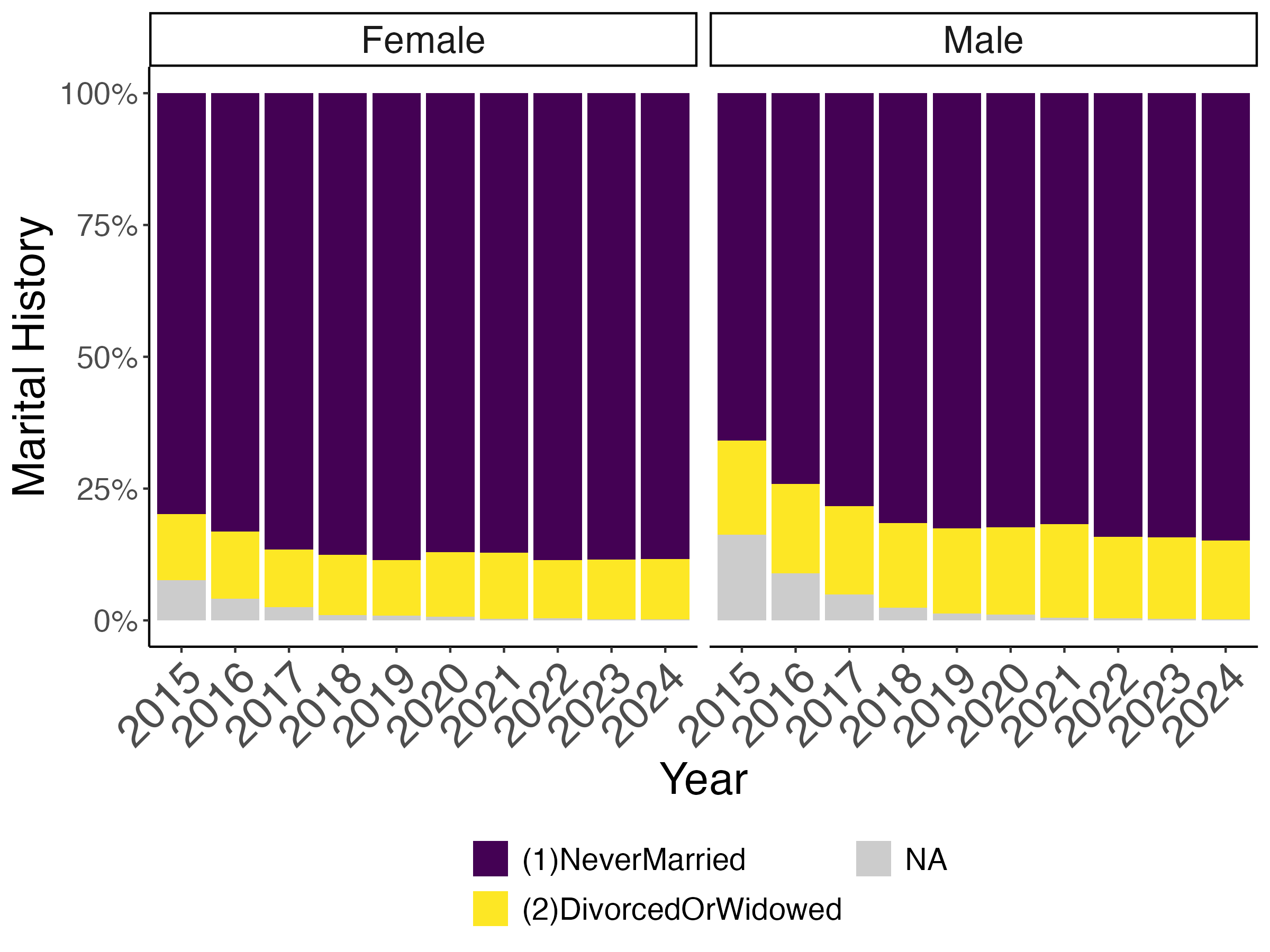}}
  \subfloat[Preference for Children]{\includegraphics[width = 0.42\textwidth]{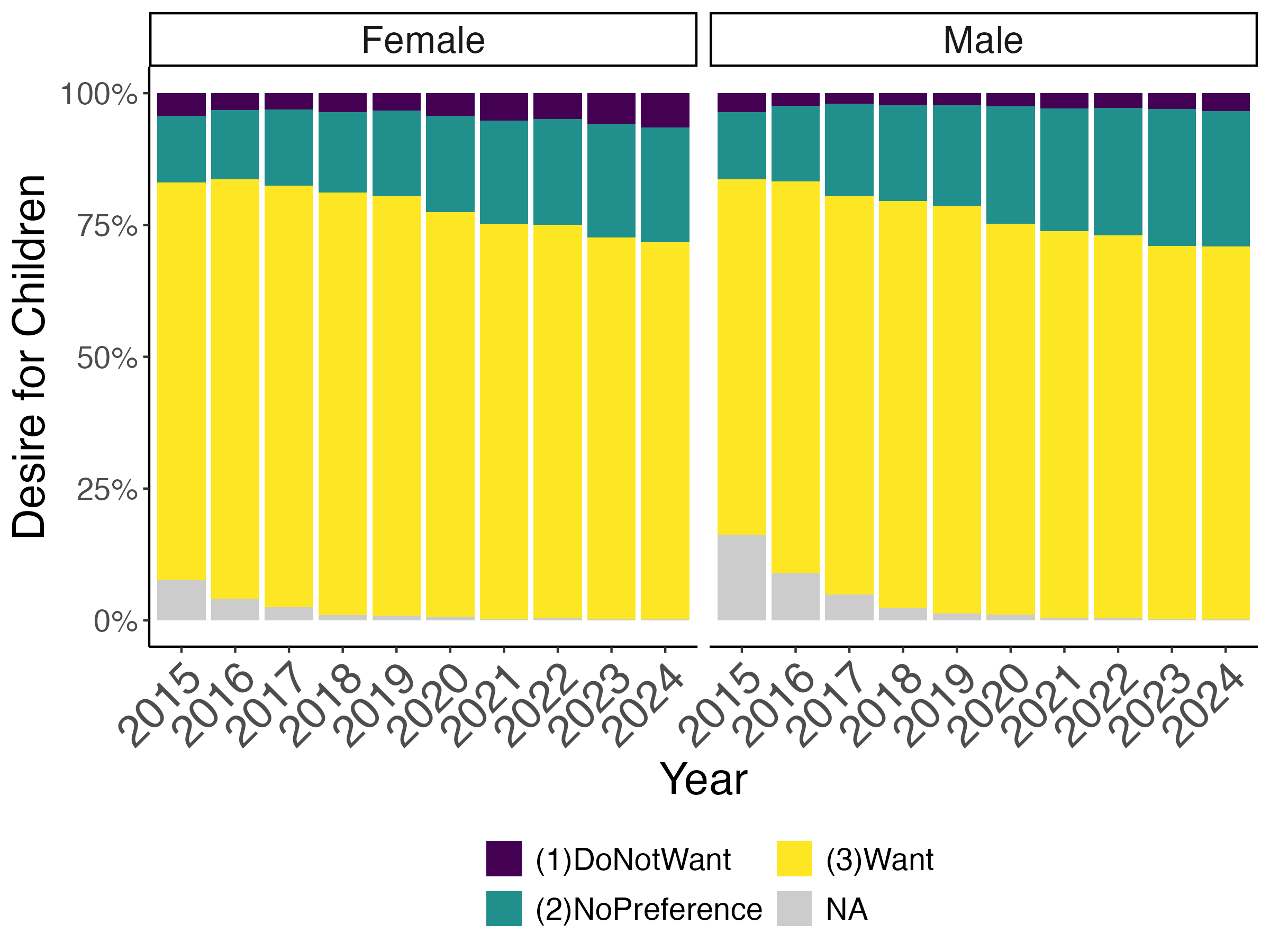}}\\
  \vspace{6mm}
  \subfloat[Childcare]{\includegraphics[width = 0.42\textwidth]{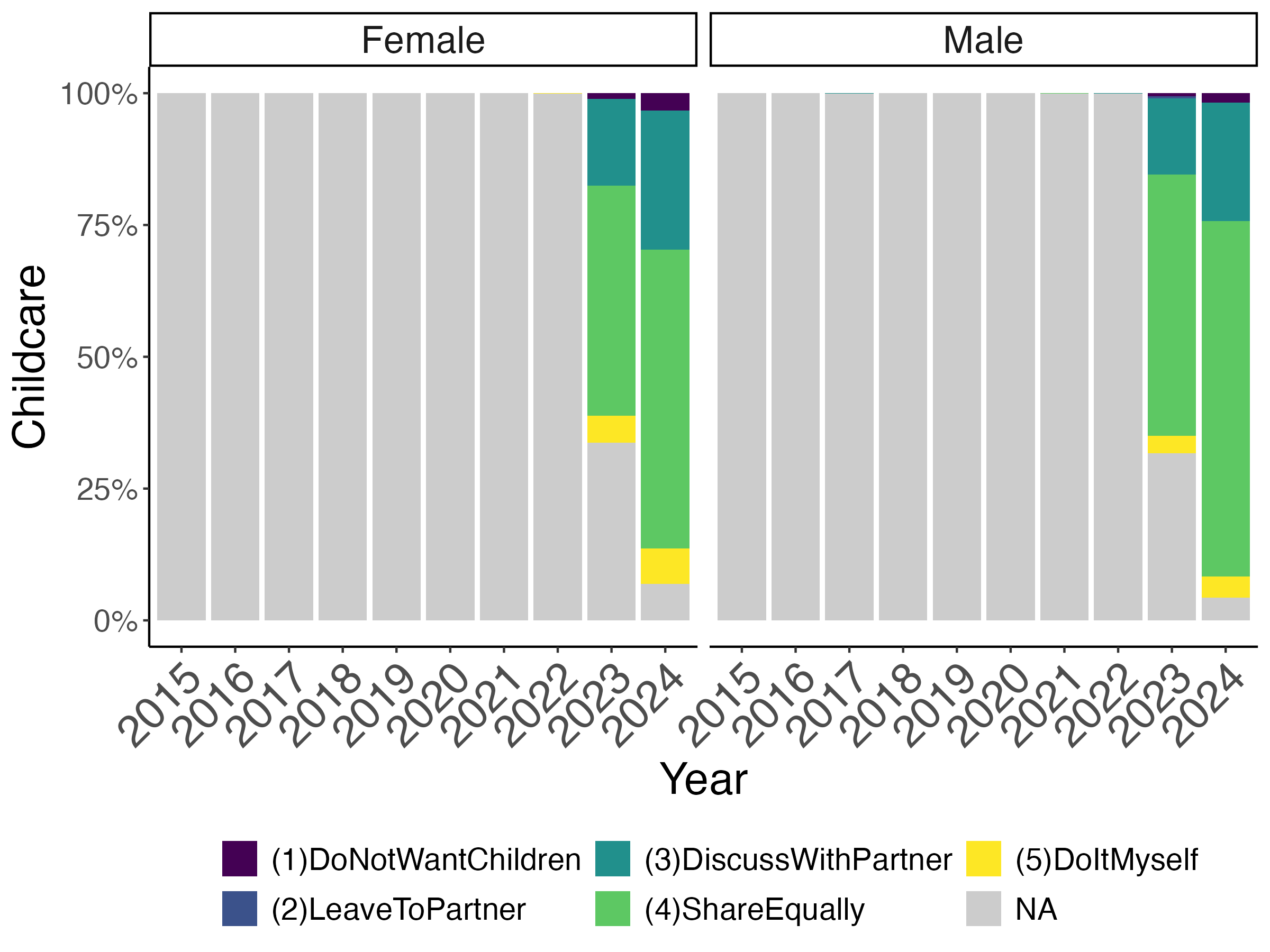}}
  \subfloat[Housework]{\includegraphics[width = 0.42\textwidth]{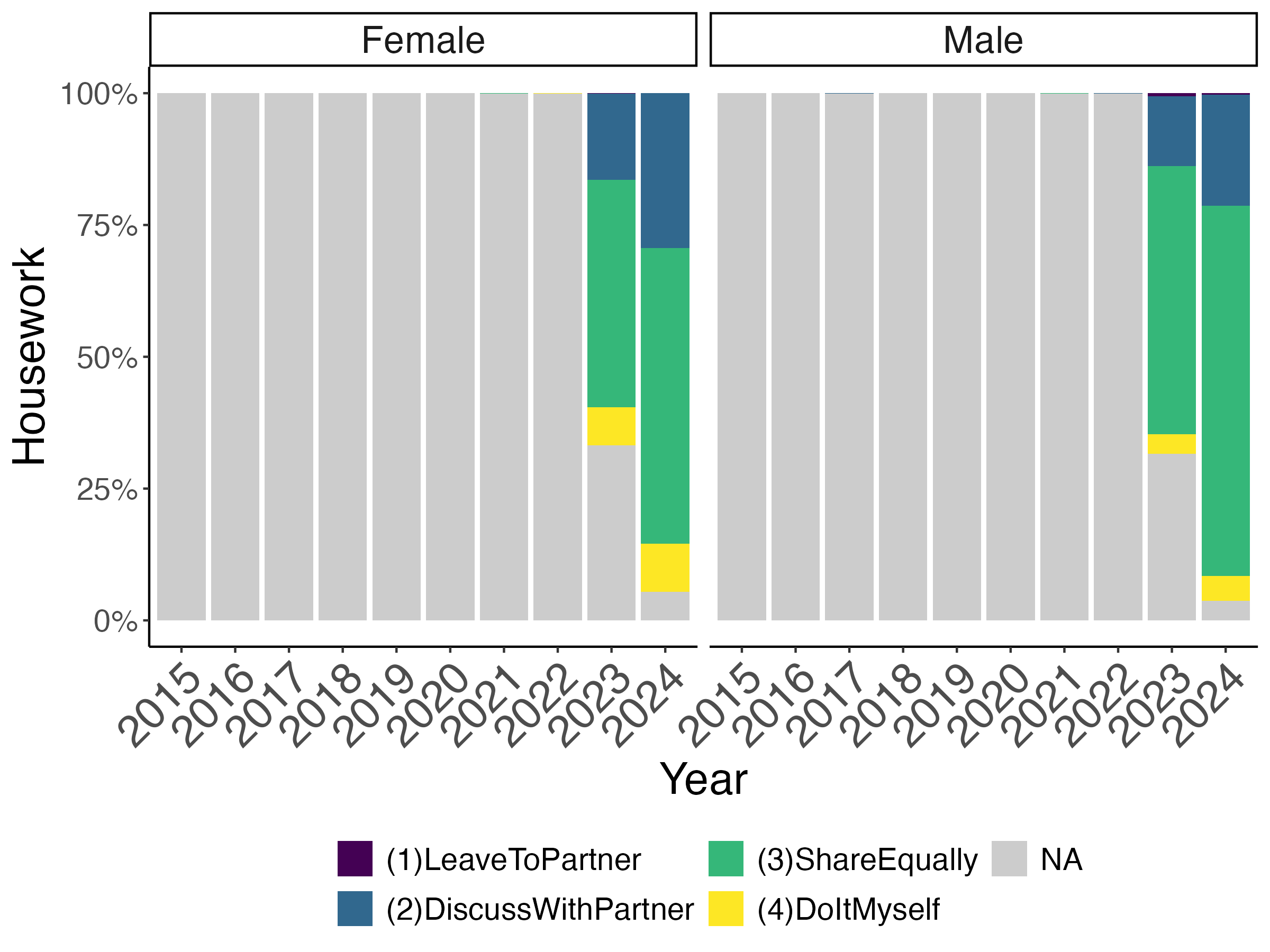}}
  \caption{Share of Lifestyle, Marital History, and Family Preferences in Married Couples}
  \label{fg:boxplot_housework_2015_2024} 
  \end{center}
  \footnotesize
  Note: 
  See \cite{chiappori2018bidimensional} and \cite{chiappori2024analyzing} for discussion of the use of smoking variables. 
\end{figure}

Figure \ref{fg:boxplot_housework_2015_2024} illustrates the evolution of lifestyle habits, marital history, and family preferences among married individuals from 2015 to 2024. Panels (a) and (b) show that smoking is rare while drinking is dominated by social drinking, and gender gaps persist: women are consistently less likely to drink regularly or smoke, though the share of regular female drinkers increased modestly over time. Panel (c) reveals a stable yet notable gender difference in marital history, with remarriage being more common among men than women. Panel (d) shows that the desire to have children remains widespread, but the share expressing no preference has grown gradually for both genders, indicating a slight softening of parenthood norms. Panels (e) and (f), which become informative only in the most recent years and are effectively observed from 2023 onward, highlight significant gender asymmetries in stated preferences for the division of childcare and housework. Men are somewhat more likely to prefer discussion or delegation, whereas women more frequently endorse shared or self-led responsibilities. These gender gaps in family role preferences have remained sizable and persistent since their introduction in the survey, with only limited signs of convergence.

\begin{figure}[!htbp]
  \begin{center}
  \includegraphics[width = 0.65\textwidth]{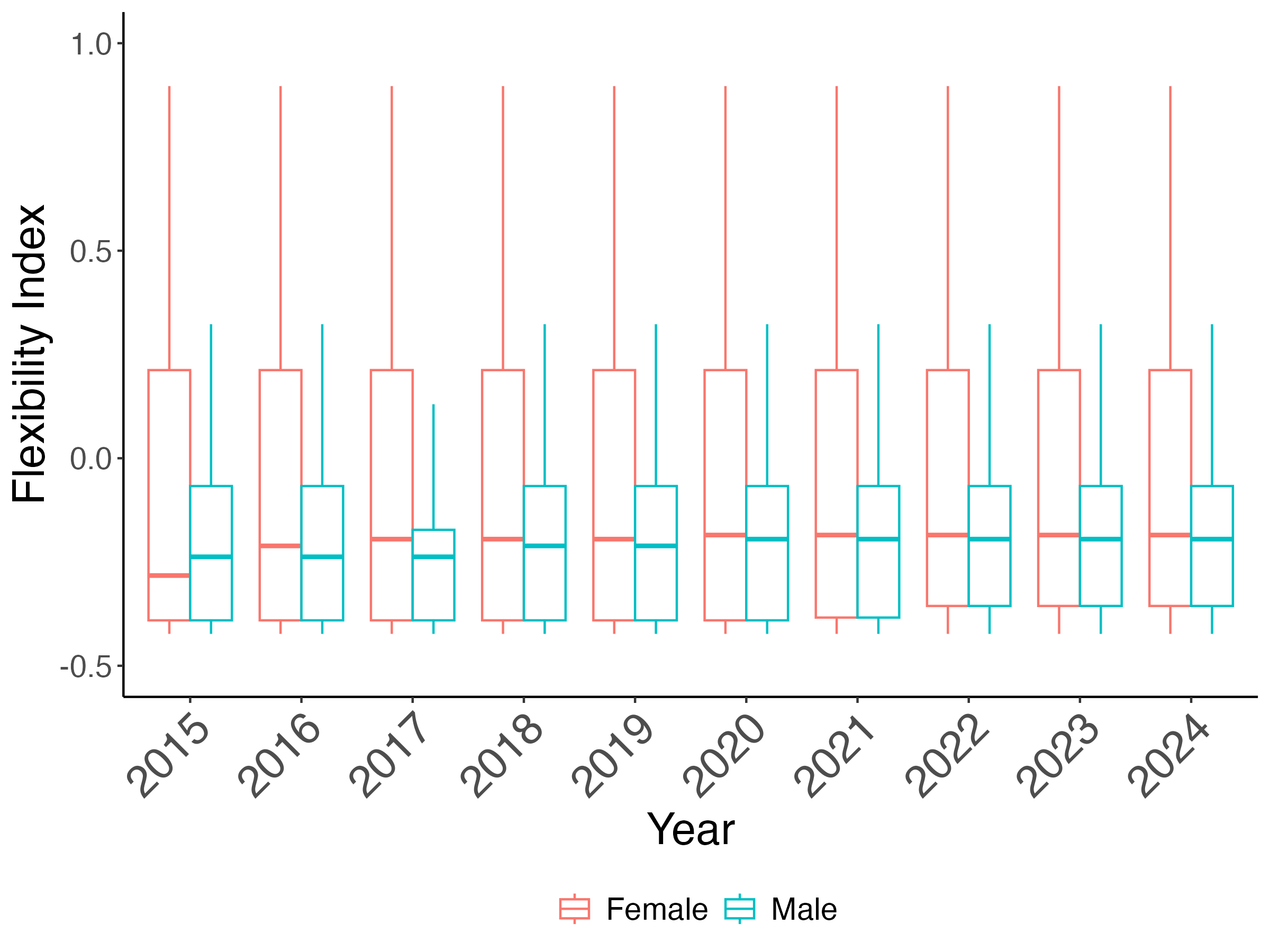}
  \caption{Distribution of Occupational Flexibility in Married Couples}
  \label{fg:boxplot_flexibility_2015_2024} 
  \end{center}
  \footnotesize
  Note: In each boxplot, the central box represents the interquartile range (IQR), spanning from the 25th to the 75th percentile of the distribution, with the horizontal line inside indicating the median (50th percentile). The vertical lines ("whiskers") extend to the most extreme values within 1.5 times the IQR from the box; values beyond this range are considered outliers and are not shown in the plot.
\end{figure}

Figure \ref{fg:boxplot_flexibility_2015_2024} shows that the median occupational flexibility index is very similar for men and women throughout the period. In contrast, the distribution is consistently more dispersed among women, with a wider interquartile range and longer tails, indicating greater heterogeneity in occupational flexibility. There is no clear time trend in the distribution for either gender over the sample period.

\section{Model}\label{sec:model}
This section outlines empirical multidimensional matching frameworks with transferable utility that are used to document sorting patterns in the marriage market.
Subsection \ref{subsec:dupuy_galichon} reviews the flexible and tractable multidimensional one-to-one matching model of \citet{dupuy2014personality}, which is well suited to describing final matching patterns in the Proposal stage.
Subsection \ref{subsec:fox_many_to_many_ibj} discusses the matching-with-trading-networks framework of \citet{fox2018estimating}.
Although this model requires an additional normalization and cross-market assumptions, which limit coefficient-level interpretability, it accommodates many-to-many matching and is therefore appropriate for dating stages where agents can interact with multiple potential partners.

\subsection{TU Multidimensional Continuous-Type Quadratic Matching Model}\label{subsec:dupuy_galichon}

We consider a frictionless transferable-utility marriage market in which each man and woman is described by observable characteristics, denoted $x \in \mathcal{X}$ for men and $y \in \mathcal{Y}$ for women.
The matching outcome is governed by a probability measure $\pi(x, y)$ giving the likelihood that type-$x$ men and type-$y$ women match. Upon matching, partners receive deterministic utility components $U(x, y)$ and $V(x, y)$ plus individual-specific stochastic shocks.

Following \citet{dupuy2014personality}, each agent draws an infinite sequence of potential partners characterized by observables and idiosyncratic preference shocks. For men, matches are evaluated over $\{(y^k, \varepsilon^k)\}_{k=1}^\infty$, where each $\varepsilon^k$ is i.i.d. type I extreme value. This yields a continuous-logit formulation in which utility from matching with $y^k$ is $U(x, y^k) + \frac{\sigma}{2}\varepsilon^k$, with $\sigma$ governing the relative importance of idiosyncratic versus systematic preferences. The same formulation applies for women.

Under this assumption, the equilibrium matching probabilities and surplus shares satisfy the following structure. The joint matching density $\pi(x, y)$ is characterized by
\[
\pi(x, y) = \exp\left( \frac{\Phi(x, y) - a(x) - b(y)}{\sigma} \right),
\]
where $\Phi(x, y)$ is total systematic surplus and $a(x)$, $b(y)$ are type-specific adjustment terms ensuring that matches do not exceed type supplies. Surplus shares are then:
\[
\begin{aligned}
U(x, y) &= \frac{\Phi(x, y) + a(x) - b(y)}{2}, \\
V(x, y) &= \frac{\Phi(x, y) + b(y) - a(x)}{2}.
\end{aligned}
\]

\paragraph{Parameterization and estimation.}
To make the model empirically implementable, we rely on a quadratic specification of the systematic surplus function, following \citet{dupuy2014personality}:
\[
\Phi(x, y)=x^{\prime} A y=\sum_{i, j \in \{1, \cdots, O\}} x_i A_{i j} y_j.
\]
The $O \times O$ affinity matrix $A$ summarizes how traits interact in matching and is the central object of interest.
Its elements encode both magnitude and sign: positive $a_{ij}$ indicates complementarity between traits $x_i$ and $y_j$, while negative values indicate substitutability.

Estimation exploits the equilibrium structure.
As shown by \citet{dupuy2014personality}, $B = A / \sigma$ is recovered as the solution to
\[
    \min_B \left\{\mathcal{W}(B, 1) - E_{\hat{\pi}}\left[  x' B y \right]\right\}
\]
where $\mathcal{W}(A, \sigma)$ denotes the social gain maximized at the equilibrium matching $\pi$ \citep{shapley1971assignment}:
\[
    \mathcal{W}(A, \sigma) \equiv \max_{\pi \in \mathcal{M}} \left\{ E_\pi[x'Ay] - \sigma E_\pi[ \log\:\pi(x,y) ] \right\},
\]
where expectations indexed by $\hat{\pi}$ are computed from the empirical joint distribution of observed matches.

Optimality conditions yield moment-matching restrictions.
Intuitively, the estimated model must reproduce the joint characteristic distribution observed in matches.
In particular, the implied equilibrium matching distribution equates predicted and empirical cross-moments of male and female traits:
\[
    E_{\pi} [X_i Y_j] = E_{\hat{\pi}} [X_i Y_j]
\]
for every pair of characteristics $(i,j)$.
In practice, estimation selects $B$ so model-implied assortative patterns align with the empirical ones.

\paragraph{Saliency analysis.}
An important implication of the quadratic specification is that it admits a low-dimensional representation of assortative forces.
In particular, $A$ can be factorized by singular value decomposition (SVD), which transparently characterizes both the number and nature of economically relevant sorting dimensions.
Formally, $A$ can be written as
\[
A = U' \Lambda V
\]
where $\Lambda$ is a diagonal matrix of positive singular values (largest to smallest), and $U$ and $V$ contain the associated loading vectors.
Each singular value captures the relative importance of a sorting dimension, while loadings describe how observables map to that dimension on each market side.

This decomposition rewrites surplus as a sum of independent components. Defining $\tilde{x} = Ux$ and $\tilde{y} = Vy$, surplus is
\[
\Phi(x,y) = x'Ay = \sum_{k=1}^K \lambda_k \tilde{x}_k \tilde{y}_k
\]
where each term captures the contribution of one assortative dimension.
In this representation, sorting operates through orthogonal indices, with larger singular values indicating more salient dimensions.

In the empirical analysis, we apply this decomposition to $\hat{A}$ to recover $U$, $V$, and $\Lambda$.
This allows us to assess both relative salience across sorting dimensions and the trait combinations defining each dimension.
Sampling uncertainty is quantified by bootstrap procedures.

A remaining issue is dimensionality. Following \citet{dupuy2014personality}, we conduct rank-based tests on the estimated affinity matrix.
The null imposes $rank(A) = k$; rejection implies sorting along more than $k$ economically relevant dimensions.

\paragraph{Identification with multiple years.}
Because the data span 2015--2024, one can estimate yearly affinity matrices and study changes in sorting forces.
However, $A$ is identified only up to scale (only $B = A / \sigma$ is identified), so additional restrictions are needed for cross-year comparisons.

To enable such comparisons, we follow \cite{ciscato2020role} and normalize each year’s affinity matrix to unit Frobenius norm, $ \lVert A^t \rVert = 1$.
\footnote{\citet{ciscato2020like} propose an alternative normalization in which $\mathcal{W}(A,\sigma)=1$. Note that the optimal matching function $\pi(x,y)$ depends only on $B = A / \sigma$, and is therefore invariant to the choice of normalization.}
Under this normalization, we have $\frac{B^t}{\lVert B^t \rVert} = A^t$ and $\sigma^t = \frac{1}{\lVert B^t \rVert}$.
After estimating $B^t$, we recover $A^t$, which is directly comparable across years and allows analysis of time trends in sorting.
The time series of $\sigma^t$ then tracks changes in the relative importance of unobserved heterogeneity.

\subsection{A Many-to-Many TU Matching Model for Multi-Stage Partner Formation}
\label{subsec:fox_many_to_many_ibj}
This subsection adapts the transferable-utility (TU) matching-with-trading-networks framework of \citet{fox2018estimating} to IBJ’s multi-stage partner-formation pipeline (Section \ref{sec:ibj_user_behavior}) and defines stage-specific assortativeness measures using continuous attributes.\footnote{Our use of ``trade'' follows \citet{fox2018estimating}: a trade is a relationship instance encoding observable characteristics of both participants and, optionally, additional observable contract/stage features. See \citet{fox2018estimating} for trading-network equilibrium and matching maximum score estimation.} 
The goal is to identify at which stage of the matching process assortative patterns along different dimensions first emerge, rather than to compare the magnitude of assortativeness across stages.

At each stage, agents can form multiple relationships, and equilibrium outcomes are characterized as competitive allocations over trade portfolios \citep{azevedo2018existence}, not single matches. Preferences depend on observables, with deterministic surplus parameterized linearly by stage-specific coefficients. Restricting to bilinear interactions between male and female continuous attributes yields a stage-specific affinity matrix: diagonal elements capture assortative matching, off-diagonals capture cross-attribute complementarities. Identification uses the rank-order property under exchangeable unobserved heterogeneity, and estimation uses matching maximum score, delivering set-consistent stage-specific affinity estimates up to scale.

\subsection{Technical Distinction Between \cite{dupuy2014personality} and \cite{fox2018estimating}}

Although both this framework and \citet{dupuy2014personality} use bilinear surplus specifications, they differ fundamentally in equilibrium concept and identifying variation. Dupuy--Galichon is a one-to-one continuous-logit model where entropic regularization yields a unique matching density and point identification of the scaled affinity matrix $B=A/\sigma$ (or, equivalently, a normalized version of $A$), rather than the unrestricted level of $A$. By contrast, \citet{fox2018estimating} is nonparametric in unobserved heterogeneity and identifies preferences only up to scale via equilibrium stability inequalities. Hence, affinity matrices here are not comparable across stages or over time in levels; interpretation is based on sign robustness and within-stage relative importance. Accordingly, Dupuy--Galichon is used for cross-sectional levels and long-run Proposal-stage trends, while Fox is used to identify when assortative patterns become robust along the pipeline.%\textcolor{black}{[To be rigorously discussed, TBA after EALE, Otani]}

%\if0
\section{Estimation Results}\label{sec:results}

This section presents the empirical results in three steps. We first characterize the structure of assortative matching at the Proposal stage using the multidimensional affinity matrix estimated following \citet{dupuy2014personality}. We then document how the strength of assortative matching along key dimensions has evolved over the past decade at the Proposal stage, as in \cite{ciscato2020role}. Finally, we use a many-to-many framework in the spirit of \citet{fox2018estimating} to examine when assortative patterns begin to emerge along the matching pipeline.

\subsection{Affinity Matrix at the Proposal Stage in 2024}

We begin by examining the cross-sectional structure of preferences at the Proposal stage in the most recent year. Table \ref{tb:Dupuy_Galichon_affinity_matrix} reports the estimated affinity matrix for 2024 based on the multidimensional matching framework of \citet{dupuy2014personality}, where rows correspond to male attributes and columns to female attributes. It yields three notable insights into sorting patterns and preference structures in the marriage market.

\paragraph{Sociodemographic and Anthropometric Attributes.}
The affinity matrix reveals positive assortative matching along sociodemographic and anthropometric attributes.
Diagonal elements for age, education, income, height, and weight are all positive, indicating that individuals tend to match with partners who are similar along these dimensions. 
Because all attributes are measured prior to matching, these patterns reflect sorting behavior rather than post-marital coordination.

Among these attributes, age exhibits the largest diagonal coefficient (3.55), substantially exceeding those of other characteristics. 
Other sociodemographic and anthropometric variables also display positive assortative matching, with diagonal coefficients ranging between 0.17 and 0.21.

Cross-attribute interactions among sociodemographic and anthropometric traits are present and, in some cases, asymmetric. 
For example, the interaction between education and age is negative, indicating that, conditional on own age, higher education is associated with matching to younger partners.
The interaction between education and income is positive, implying that more educated individuals tend to match with higher-income partners conditional on their own income. 
This interaction exhibits modest asymmetry: the coefficient for male education–female income ($0.11$) exceeds that for male income–female education ($0.07$).
Interactions involving physical attributes also display asymmetry. 
The interaction between male height and female age is negative ($–0.23$), whereas the corresponding interaction between female height and male age is small ($–0.03$) and statistically indistinguishable from zero. 
This pattern indicates that the contribution of sociodemographic and anthropometric attributes to match surplus differs by gender, consistent with gender-specific roles, expectations, or signaling in the marriage market.

\paragraph{Occupational Flexibility.}
Occupational flexibility, constructed following \citet{goldin2014grand}, exhibits positive assortative matching, although the magnitude of the diagonal coefficient is small ($0.03$). 
Relative to sociodemographic and anthropometric attributes, occupational flexibility plays a more limited role in the affinity matrix.

Ex-ante theoretical predictions regarding assortative matching in occupational flexibility are ambiguous.
If flexibility facilitates within-household specialization, it may function as a substitutable characteristic across spouses, reducing incentives for positive sorting.
By contrast, positive assortative matching may arise if spouses' inputs to home production are complementary, a mechanism supported by recent evidence showing that such complementarities have strengthened over time \citep{calvo2024marriage}.
In a related vein, recent work by \citet{almar2025educational} documents positive assortative matching along educational ambition, measured using expected initial wage levels and wage growth. Because educational ambition is negatively associated with hours flexibility, their findings suggest assortative matching in occupational characteristics related to flexibility.\footnote{
Assortative matching may also reflect preferences for partners in similar occupational environments.
While empirical studies have reported occupational similarities between spouses \citep[e.g.,][]{kalmijn_assortative_1994}, some recent work finds that preferences for similar occupations play a limited role in partner selection \citep{belot2013dating,lee2016effect,mansour_same-occupation_2018}.
}

Interactions involving occupational flexibility are also limited. 
In particular, nearly all interaction coefficients between flexibility and other attributes are statistically indistinguishable from zero, with standard errors ranging between $0.01$ and $0.03$. 
As a result, occupational flexibility contributes relatively little to overall sorting patterns at the Proposal stage.

\paragraph{Household Preferences.}

Household-related preferences play a quantitatively important role in sorting at the Proposal stage. 
In particular, the diagonal coefficient on desire for children is large ($0.25$), making it the second-largest diagonal element in the affinity matrix after age. 
This magnitude exceeds that of education, income, and all other anthropometric and preference-related attributes, indicating that alignment in fertility preferences is a major source of match surplus. 
Diagonal coefficients for childcare and housework preferences are also positive, though smaller in magnitude, confirming assortative matching along multiple dimensions of household orientation.
These results suggest that early alignment in household expectations plays an important role in match formation, even before such preferences are behaviorally realized in marriage.

Interactions among household preferences further clarify the structure of this alignment.
For example, interactions between childcare preferences and desire for children are positive ($0.05$ and $0.06$ across gender pairs), whereas interactions between housework preferences and desire for children are negative ($-0.06$ and $-0.04$).
These patterns indicate that household preferences interact in a structured way: individuals with strong fertility preferences tend to match with partners willing to assume childcare responsibilities, while preferences over housework exhibit a more differentiated relationship with fertility preferences. 
More generally, these coefficients show that alignment across household-related attributes does not necessarily imply similarity, but rather reflects systematic complementarities and trade-offs in household roles that contribute to match surplus.

By contrast, interactions between preferences for children and most sociodemographic or anthropometric attributes are small, with age as a notable exception. 
In particular, the interaction between male child preference and female age is negative, while interactions with education, income, height, and weight are otherwise limited. This implies that fertility preferences constitute a distinct margin of sorting that operates largely independently of education, income, and most physical characteristics. 
Taken together, these patterns show that preferences for children are not only strongly assortative but also form a separate dimension of matching, rather than serving as proxies for traditional sociodemographic attributes.

\paragraph{Factor Decomposition.}
We further assess whether the estimated sorting patterns can be summarized by a low-dimensional structure using the saliency decomposition of \citet{dupuy2014personality}. 
\autoref{tb:Dupuy_Galichon_saliency} reports the relative importance and factor loadings of the leading indices implied by the estimated affinity matrix. 
The hypothesis that a single index fully captures sorting patterns is rejected. 
Nevertheless, three indices together account for nearly 90\% of the observable matching surplus, indicating that sorting operates along a small number of economically meaningful dimensions.

The first index is the dominant source of variation, accounting for more than 70\% of total saliency for both men and women. 
This index loads almost exclusively on age, with loading coefficients of $–0.97$ for men and $–0.99$ for women. 
The magnitude and isolation of these loadings indicate that age-based sorting remains the most salient dimension of marriage formation.
The strong and isolated weight on age suggests that partners tend to form matches within narrowly defined birth cohorts \citep[e.g.,][]{chiappori2024analyzing}.

The second index explains approximately 9\% of total saliency and captures variation in sociodemographic and anthropometric attributes. 
It loads positively on education ($0.26$ for men, $0.20$ for women), income ($0.65$ and $0.40$), and height ($0.49$ and $0.47$), while loading negatively on weight ($–0.43$ and $–0.71$), with notable gender differences in magnitudes. 
Income ($0.65$) and height ($0.49$) are most salient for men, while weight ($-0.71$) and height ($0.47$) play a more prominent role for women. 
This index therefore reflects sorting along a bundle of characteristics associated with sociodemographic status and physical attractiveness, beyond pure age similarity.

The third index is primarily driven by preferences for children for both men ($-0.89$) and women ($-0.91$). 
Although it accounts for only 6\% of total saliency, this dimension represents a distinct margin of sorting that is not subsumed by sociodemographic or anthropometric characteristics. 
Moreover, because individuals search for partners only within their cohort, the contribution of this index to joint surplus is amplified when conditioning on age, amounting to more than 20\% of the remaining surplus.
This finding underscores that preferences for children constitute an economically important and independent dimension of assortative matching, despite their comparatively smaller weight in the unconditional decomposition.

\begin{landscape}
\begin{table}[!htbp]
\footnotesize
\caption{Estimated Affinity Matrix}
\label{tb:Dupuy_Galichon_affinity_matrix}
\begin{center}
  \begin{tabular}{ccccccccccccc}
  \input{figuretable/labor_family_economics_project/dupuy_galichon_affinity_matrix_2024.txt}

  \end{tabular}
\end{center}
\footnotesize
Note: We use 6,592 couples to estimate the affinity matrix. All variables are standardized to have unit variance. Income represents the upper limit of the income category. Standard errors, reported in parentheses, are obtained from 2,000 bootstrap replications. Estimates in bold are statistically significant at the 5\% level. Rows correspond to male attributes and columns to female attributes. 
\end{table}
\end{landscape}

\if0
\begin{landscape}
\begin{table}[!htbp]
\caption{Estimated Affinity Matrix (Rescaled)}\footnotesize
\label{tb:Dupuy_Galichon_affinity_matrix_rescale}
    \centering
    
  \begin{tabular}{ccccccccccccc}
  \input{figuretable/labor_family_economics_project/dupuy_galichon_affinity_matrix_2024_rescale.txt}
  \end{tabular}
\end{table}
\footnotesize
Note: Table \ref{tb:affinity_matrix_2015_2024} reports historical transition of diagonals in affinity matrix of \cite{dupuy2014personality}. 
\end{landscape}
\fi

\begin{table}[!htbp]
\caption{Saliency Analysis}
\label{tb:Dupuy_Galichon_saliency}
\begin{center}
  \begin{tabular}{ccccc|cccc}
  \input{figuretable/labor_family_economics_project/dupuy_Galichon_saliency_2024.txt}

  \end{tabular}
\end{center}
\footnotesize
Note: The table presents the singular vectors associated with men and women, denoted by $U$ and $V$, respectively, along with the singular values contained in $\mathrm{diag}(\Lambda)$, obtained from the singular value decomposition of the affinity matrix $A = U' \Lambda V$. The final row reports the elements of $\mathrm{diag}(\Lambda)$, which measure the relative contribution of each underlying sorting dimension. We use 6,592 couples for the saliency analysis. All variables are standardized to have unit variance. Standard errors, reported in parentheses, are obtained from 2,000 bootstrap replications. Estimates in bold are statistically significant at the 5\% level.
\end{table}

\subsection{Assortative Matching Estimates Under Unidimensional and Multidimensional Specifications}

Another important question in the multidimensional matching literature is whether conclusions drawn from parsimonious specifications—often focusing on a single attribute such as education or age—are sensitive to the exclusion of other relevant characteristics. Many empirical studies estimate assortative matching using a limited set of variables, implicitly assuming that omitted attributes neither interact strongly with included ones nor materially affect measured assortativeness. \autoref{tb:affinity_matrix_different_attributes_2024} examines this issue by progressively enriching the attribute space within the Dupuy–Galichon framework and comparing diagonal coefficients across specifications.

The results indicate that estimates based on parsimonious specifications can differ by economically large magnitudes from those obtained in the fully multidimensional model. When education alone is included, as in \cite{chiappori2017partner} and \cite{eika2019educational}, the estimated diagonal coefficient is $0.25$, compared with $0.17$ in the full specification, implying an estimate about 47\% larger.\footnote{Unidimensional models do not typically adopt quadratic specifications as in \citet{dupuy2014personality}; therefore, the magnitude reported here does not necessarily imply that existing estimates in the literature overstate assortativeness by 47\%. For an extensive exercise, Appendix \ref{sec:unidimensional_assortativeness} estimates unidimensional models for each characteristic as in \cite{choo2006marries}.}
This pattern suggests that part of the assortativeness commonly attributed to education reflects its correlation and interaction with other characteristics. 
A different pattern arises for age. When age alone is used, as in \cite{choo2006marries}, the estimated diagonal coefficient is $2.98$, compared with $3.55$ in the full specification, so the unidimensional estimate is about 16\% smaller than the multidimensional benchmark.

By contrast, the specification that includes both age and income produces coefficients much closer to those in the fully specified model: the age coefficient is $3.61$ versus $3.55$ in the full model, and the income coefficient is $0.22$ versus $0.20$. 
This finding suggests that, when the objective is to measure assortativeness along the age and income dimensions, including both variables jointly yields estimates that are broadly consistent with those from the multidimensional benchmark.

At the same time, incorporating additional attributes such as occupational flexibility and household preferences does not materially alter the education–age–income coefficients, indicating that these core sociodemographic dimensions remain relatively stable as the model is further enriched. 
Nevertheless, several of the added traits—most notably height and preferences for children—exhibit sizable own-trait assortativeness in the fully specified model, in some cases exceeding that of income. 
These results underscore that preference-based and anthropometric characteristics represent quantitatively distinct and important margins in their own right, with direct implications for the fertility discussion in Section \ref{sec:discussion}.

\begin{table}[!htbp]
  \begin{center}
      \caption{Sensitivity to Attribute Selection}
      \label{tb:affinity_matrix_different_attributes_2024}
      \begin{tabular}{cccccccccc}
      \input{figuretable/labor_family_economics_project/Dupuy_Galichon_affinity_matrix_different_set_of_attributes_2024.tex}

      \end{tabular}
  \end{center}\footnotesize
  \textit{Note}: The table displays the diagonal elements of the marital preference parameter matrix $A$, capturing own-trait assortative matching, estimated under progressively richer specifications. Column (1) includes only education; column (2) includes only age; column (3) includes only income; column (4) includes education and age; column (5) includes education and income; column (6) includes age and income; column (7) includes education, age, and income; column (8) further includes flexibility, height, weight, drinking, smoking, and marital history; column (9) presents the full specification with all twelve attributes. Income represents the upper limit of the income category. We use 6,592 couples for these analyses. Standard errors are in parentheses.

\end{table}

\subsection{Transition of Affinity Matrices at the Proposal Stage: 2015--2024}

Having established the structure of assortative matching at the Proposal stage in 2024, we next examine how the strength of assortative matching along each dimension has evolved over time.
To enable comparisons across years, we apply the normalization procedure proposed by \citet{ciscato2020role} explained in Section \ref{subsec:dupuy_galichon}.
\footnote{The method proposed by \citet{dupuy2014personality} is robust to changes in the marginal distributions and therefore identifies the underlying structure of the surplus function. Accordingly, provided that appropriate normalization is imposed on the parameters, as in \citet{ciscato2020like} and \citet{ciscato2020role}, it is possible to analyze variation in affinity matrices over time. Building on this approach, \citet{ciscato2020like} compare sorting patterns across different-sex and same-sex marriage markets, while \citet{ciscato2020role} examine temporal changes in the affinity matrix.}

\autoref{tb:affinity_matrix_2015_2024} reports the rescaled diagonal elements of the affinity matrix at the Proposal stage from 2015 to 2024, allowing for a direct comparison of assortative matching patterns over time.\footnote{The contribution of unobserved heterogeneity increases during the early years of the sample period but returns to its initial level by 2024; see \autoref{tb:affinity_matrix_sigma_2015_2024}.}
Focusing on the endpoints of the sample period, assortative matching by education declines substantially, decreasing from $0.09$ in 2015 to $0.05$ in 2024 (a decline of $44.4\%$), indicating a gradual weakening of educational sorting at the final matching stage. 
In contrast, assortativeness by age remains high throughout the period and increases slightly over the decade, rising from $0.93$ to $0.95$. 
This pattern suggests a persistent, and marginally strengthened, role of age similarity in partner selection.
Assortative matching by marital history also declines markedly, falling from $0.09$ in 2015 to $0.03$ in 2024 (a decline of $66.7\%$). 
While marital history is the second most salient dimension of assortative matching in 2015, it plays a much more limited role by 2024.
By contrast, assortativeness with respect to income, height, weight, and preferences for children remains largely stable between 2015 and 2024 and continues to play an important role in sorting after age.
Overall, these patterns point to a modest reallocation of assortative matching at the Proposal stage away from education and marital history toward age, while most other dimensions exhibit stable sorting patterns over the past decade.

\begin{table}[!htbp]
  \begin{center}
      \caption{Transition of Diagonal Elements of Affinity Matrices}
      \label{tb:affinity_matrix_2015_2024}
      % \subfloat[Original]{\input{figuretable/labor_family_economics_project/affinity_matrix_2015_2024}}\\
      %\subfloat[Rescaled]{
      \input{figuretable/labor_family_economics_project/affinity_matrix_2015_2024_rescale_CW}
      %}
  \end{center}\footnotesize
  \textit{Note}: The table displays the estimated trend of the diagonal elements of the marital preference parameter matrix $A$ capturing the interaction between husband's and wife's characteristics, rescaled following \cite{ciscato2020role} to make estimates comparable across years. Income represents the upper limit of the income category.

  \normalsize
  \begin{center}
      \caption{Transition of the Contribution of Unobserved Heterogeneity in Affinity Matrices}
      \label{tb:affinity_matrix_sigma_2015_2024}
      % \subfloat[Original]{\input{figuretable/labor_family_economics_project/affinity_matrix_2015_2024}}\\
      %\subfloat[Rescaled]{
      \input{figuretable/labor_family_economics_project/affinity_matrix_sigma_2015_2024_rescale_CW}
      %}
  \end{center}\footnotesize
  \textit{Note}: The table displays the estimated trend of the unobserved heterogeneity $\sigma$, following \cite{ciscato2020role} to make estimates comparable across years.
\end{table}

\subsection{When Does Assortative Matching Emerge? Evidence from Dating Process Data}\label{sec:fox_results}

While the preceding analysis documents both the cross-sectional structure and the long-run evolution of assortative matching at the Proposal stage, it remains unclear at which point in the matching process these patterns begin to form. We therefore utilize our dating process data and turn to a many-to-many matching framework following \citet{fox2018estimating} to trace the emergence of assortative matching across stages of relationship formation.

Table \ref{tb:maximum_score_diagonal_all_stage} reports assortative matching estimates from the many-to-many matching framework, with education assortativeness normalized to one within each stage. Given the scale normalization inherent in the maximum score estimator, coefficient magnitudes are interpreted only relative to other covariates within the same stage. Under this interpretation, a limited subset of variables exhibits robust assortative patterns with confidence intervals that exclude zero. Across all stages, assortativeness with respect to age is robustly positive and large relative to education, indicating that age consistently ranks as one of the most salient dimensions of sorting within each stage. Income assortativeness is also robustly positive across stages and smaller than age but clearly larger than education. Family-related characteristics play an important role: assortativeness in marital history and in preferences for children is robustly positive at later stages, indicating that these dimensions are systematically associated with sorting when compared to education within the same stage. By contrast, flexibility, physical attributes, lifestyle variables, housework, and childcare have confidence intervals that include zero in all stages and therefore do not display robust assortative signs under the many-to-many specification.

Because parameters are normalized separately by stage, comparisons of coefficient magnitudes across stages are not meaningful. Instead, the multi-stage estimates are informative about when assortative matching along a given dimension first becomes robustly identified along the matching pipeline. Under this criterion, assortative matching with respect to age and income is already robust at the Application stage and remains robust throughout subsequent stages. In contrast, assortativeness with respect to marital history is not robustly identified at early stages but emerges at later stages of relationship formation, remaining robust through the Proposal stage. These patterns indicate that while some dimensions shape sorting from the outset of search, others become relevant only after earlier screening has occurred. Overall, the many-to-many estimates suggest that assortative matching at the Proposal stage reflects selective continuation along a small set of dimensions whose relevance becomes apparent at different points in the matching process, rather than a monotonic strengthening of sorting intensity across stages.

\begin{table}[!htbp]
  \begin{center}
      \caption{Matching Maximum Score Estimation in 2024}\footnotesize
      \label{tb:maximum_score_diagonal_all_stage}
      \input{figuretable/labor_family_economics_project/maximum_score_diagonal_all_stage}
  \end{center}\footnotesize
  \textit{Note}: The objective function was numerically maximized using the differential evolution (DE) algorithm in the \texttt{BlackBoxOptim.jl} package. For the DE algorithm, we require setting the domain of parameters and the number of population seeds so that we fix the former to [-10, 10]. For estimation, 100 runs of 1000 seeds were performed for all specifications. The numbers in parentheses are the lower and upper bounds of the set of maximizers of the maximum rank estimator. Parameters that can take on only a finite number of values (here 1) converge at an arbitrarily fast rate, so they are superconsistent. All variables are normalized by subtracting their sample means and dividing by their sample standard deviations. Income represents the upper limit of the income category. The full interaction model is shown in Appendix \ref{sec:maximum_score_full_interaction}.
\end{table}

\section{Discussion: Implications for Household Economics}
\label{sec:discussion}

A distinctive feature of our data is that they record individuals' preferences for children at the time of marriage.
In Section \ref{sec:results}, we document that individuals sort into marriages along these pre-marital preferences.
In this section, we discuss how standard frameworks in household economics naturally predict positive assortative matching (PAM) on fertility preferences, and how the predicted strength of sorting depends on the assumed decision-making protocol.
The full model and derivations are provided in \autoref{sec:theoretical_analysis} .

%We consider a static model in which a husband and a wife each have a preference parameter governing the strength of their desire for children, 
We consider a static model in which a husband and a wife each have a preference parameter — taking one of two values, low or high, drawn from a common set — governing the strength of their desire for children, and choose private consumption and the number of children together.
Children enter both spouses' utility functions but entail monetary costs as well as time costs borne by the wife.
Within this environment, we contrast two decision-making protocols: a \emph{unitary} model in which the household jointly maximizes the sum of spousal utilities, and a \emph{veto} model in which children require the consent of both spouses, so that fertility is set by the spouse who prefers fewer, following \citet{doepke2019bargaining}.

Under standard assumptions on the utility from children --- in particular, that the marginal utility of an additional child is increasing in own fertility preferences --- both protocols predict PAM on fertility preferences.
The intuition under joint maximization is that children are effectively a public good within the household: both spouses enjoy them at a shared cost, and two partners with similarly strong preferences generate surplus by jointly choosing a fertility level both prefer, with neither pulling against the other.
A spouse with strong preferences matched with an equally high-preference partner enjoys more children at no extra cost relative to being matched with a low-preference partner.
The veto model produces the same qualitative prediction through a different mechanism: fertility is determined by the spouse who prefers fewer children, so a person with strong preferences benefits from a partner with similarly strong preferences, because the binding choice will then sit closer to both spouses' preferred levels.

While both models predict PAM, they differ sharply in the predicted strength of sorting. 
We show in \autoref{sec:theoretical_analysis} that the veto model generates a strictly larger surplus premium from assortative matching — as measured by the supermodular core of the surplus matrix \citep{chiappori2017partner} — than the unitary model, and this holds without any auxiliary conditions on preferences. 
The mechanism is transparent: in an assortative match, both partners share the same preference type and therefore agree on the optimal number of children, so the veto protocol never binds and no surplus is foregone. 
In a cross-match, by contrast, the lower-preference spouse's announced fertility constrains the pair below the jointly optimal level, and only the unitary protocol can recover this foregone surplus. 
Joint maximization therefore selectively reduces the cost of mismatching while leaving assortative surplus unchanged, narrowing the premium from sorting on fertility preferences. 
The veto institution therefore predicts not only PAM but a measurably stronger degree of sorting on fertility preferences than joint maximization.

% old continuous version
\if0
While both models predict PAM, they differ sharply in the predicted strength of sorting.
Embedded in a frictional matching framework with idiosyncratic compatibility shocks \citep{choo2006marries}, the curvature of the household surplus in the two preference parameters governs how tightly observed matches concentrate around couples with similar fertility preferences: the greater this curvature, the more a high-preference spouse gains specifically from a high-preference partner, and the larger idiosyncratic shocks must be to overturn the systematic preference for similarity.
We show in \autoref{sec:theoretical_analysis} that the veto model generates a strictly larger such cross-partial than the unitary model, under mild regularity conditions.
Two reinforcing forces drive the result.
First, under veto only the binding spouse's preferences resist fertility adjustment, whereas under joint maximization the diminishing marginal utility of children for \emph{both} spouses dampens the response of fertility to either preference --- so the veto protocol makes optimal fertility more responsive to the binding spouse's preferences and amplifies the gains from matching with a similarly preferred partner.
Second, the veto protocol leads to a strictly lower equilibrium fertility level, where the complementarity between fertility and preferences is stronger.
The veto institution therefore predicts not only PAM but a measurably stronger degree of sorting on fertility preferences than joint maximization.
\fi

These predictions provide a lens through which to interpret our empirical findings.
We document that the degree of PAM with respect to fertility preferences is remarkably strong: it is the second largest among all attributes considered in our analysis, and exceeds even the degree of sorting on education --- an attribute for which PAM is typically regarded as one of the most robust empirical regularities in the marriage market \citep[e.g.,][]{chiappori2017partner}.
While both models predict PAM and this observation alone cannot formally discriminate between them, the strong sorting we observe is more naturally consistent with the veto model, which predicts a strictly stronger degree of PAM on fertility preferences than joint maximization.

This interpretation aligns with a growing body of theoretical and empirical work that adopts the veto protocol as the preferred framework for modeling fertility within the household.
In particular, \citet{doepke2019bargaining} provide evidence that disagreement between spouses is a quantitatively important determinant of fertility outcomes, which suggests that a veto model is better suited to describing fertility decisions within the household.
Their findings have prompted a broader shift in the household economics literature toward bargaining and veto-based models of fertility, reflecting a recognition that intra-household conflict over the number of children is empirically relevant and theoretically consequential.
Our finding of exceptionally strong PAM on fertility preferences may be viewed as additional empirical support for the veto model as a description of fertility decision-making within the household.

%%%%%%%%%%%%%%%%%%%%%%%%%%%%%%%%%%%%%%%%%%%%

\section{Conclusion}\label{sec:conclusion}

This paper shows that pre-marital preferences—especially preferences for children—are a first-order determinant of marriage market sorting. While much of the existing literature emphasizes sorting on education, income, or other observable characteristics, far less is known about sorting on the preferences that directly govern household behavior, largely because such preferences are rarely observed prior to marriage and are often measured only after household decisions have already been made.

We address this gap using unique data from a structured marriage matching platform in Japan that record a rich set of attributes and preferences prior to matching and verify objectively measurable characteristics using official documents. By focusing on pre-marital information, we isolate sorting patterns that are not confounded by post-marital coordination or household specialization. A multidimensional matching framework reveals pervasive assortative matching across attributes, but also highlights a clear distinction between sociodemographic and anthropometric characteristics—whose interactions are widespread—and preferences, which constitute a separate margin of sorting. In particular, preferences for children emerge as one of the most salient dimensions, second only to age and exceeding education in importance. Exploiting the platform’s staged matching process in a many-to-many framework, we further show that sorting on age and income is already present at the initial Application stage, whereas sorting on preferences—most notably preferences for children—emerges only at later stages through selective continuation.

\if0
These findings have important implications for empirical and theoretical work on household behavior. Models that link marriage market sorting to fertility, labor supply, or intra-household allocation often abstract from heterogeneity in underlying preferences, implicitly assuming that households with similar observables face similar trade-offs. Our theoretical exercise illustrates that this abstraction can be misleading: when individuals sort on preferences prior to marriage, ignoring such heterogeneity leads to biased predictions of policy effects on household decisions such as labor supply and fertility.
\fi

%More broadly, 
Our results suggest that observed sorting on standard characteristics may mask deeper sorting on primitives that are typically unobserved in conventional data. Incorporating clean measures of pre-marital preferences into matching models offers a promising avenue for improving the empirical foundations of household economics. An important direction for future research is to link pre-marital preference sorting to post-marital outcomes—such as fertility, labor supply, and child investments—in order to better understand how household formation shapes long-run inequality and the effectiveness of family policies.

\bibliographystyle{ecca}
\bibliography{ibj_project}

\newpage
\appendix

\section{Appendix: Data}

\subsection{IBJ Data Advantage Compared with Related Literature}\label{sec:related_literature_data_comparison}

Table \ref{tab:marriage_matching_comparison} summarizes recent literature. Relative to the existing marriage-matching literature, our IBJ platform data uniquely combine (i) a rich set of verified pre-marital attributes and stated preferences, (ii) observation of both matched and unmatched users, and (iii) high-frequency, stage-by-stage logs that track the full matching pipeline from initial applications to Proposal—features that are rarely available simultaneously in administrative registries, surveys, or census-based studies.

\begin{table}[!htbp]\scriptsize
\centering
\caption{Comparison of Recent Marriage Matching Studies}
\label{tab:marriage_matching_comparison}

\begin{tabular}{|l|l|l|l|l|}
\hline
\multicolumn{5}{|c|}{\textbf{Data and Observation Design}} \\
\hline
\textbf{Paper} & \textbf{Data Source} & \textbf{Match Observation} & \textbf{Unmatched Observed} & \textbf{Timing Resolution} \\
\hline
\cite{chiappori2022assortative} & Dutch Admin. tax & Couples (marriage year) & No & Annual \\
\cite{chiappori2024analyzing} & Italian parent survey & Couples (rich traits) & No & Post-marriage \\
\cite{abdellaoui2023trading} & GB/NO Biobank & Couples (genetic, SES) & Partial & Cross-sectional \\
\cite{almar2024optimal} & Danish registers & Couples (edu programs) & Yes & Panel \\
\cite{hoehn2023changes} & MX marriage registry & New marriages & No & Annual \\
\cite{shiue2022marriage} & CN genealogies & Historical matches & No & Inferred \\
\citet{chiappori2025changes} & US Census (IPUMS) & Education bins & No & Cohort \\
\citet{chiappori2020changes} & UK LFS & Couples by education & Partial & Cross-section \\
\cite{fremeaux2024assortative} & KLIPS panel & Couples + earnings & Partial & Annual \\
This paper (IBJ, JP) & IBJ platform logs & All actions & Yes & Monthly \\
\hline
\end{tabular}

\vspace{0.4cm}

\begin{tabular}{|l|l|l|l|}
\hline
\multicolumn{4}{|c|}{\textbf{Modeling Approach and Sample Coverage}} \\
\hline
\textbf{Paper} & \textbf{Modeling Approach} & \textbf{Sample Size} & \textbf{Time Coverage} \\
\hline
\cite{chiappori2022assortative} & CS extension (SEV) & 140k marriages/yr & 2011--2014 \\
\cite{chiappori2024analyzing} & Dupuy--Galichon (DG) & 276 couples & 2019 \\
\cite{abdellaoui2023trading} & SGAM (SES$\times$genetics) & Tens of thousands & 2000s--2010s \\
\cite{almar2024optimal} & k-means + CS & 1.8M indiv./yr & 1998--2018 \\
\cite{hoehn2023changes} & SEV index & Millions & 1993--2019 \\
\cite{shiue2022marriage} & Sorting indices & 14k+ marriages & 1300--1900 \\
\citet{chiappori2025changes} & Axiomatic indices & 100k+/cohort & 1950s--1970s \\
\citet{chiappori2020changes} & CS-based index & 297k indiv. & 1945--1974 \\
\cite{fremeaux2024assortative} & Earnings corr. + sim. & 6.7k HHs & 1998--2018 \\
This paper (IBJ, JP) & DG + Nonparametric matching & 120k+ users & 2014--2025 \\
\hline
\end{tabular}
\end{table}

\subsection{Summary Statistics (Omitted in the Main Text)}

\paragraph{Unmatched Users}
Table \ref{tb:summary_statistics_unmatched_continuous_2024} also reports the summary statistics of unmatched users.
Compared to their matched counterparts, unmatched individuals differ most notably in educational attainment and household preferences. Unmatched men and women are less educated on average: 66.9\% of unmatched men and 64.6\% of unmatched women hold undergraduate or graduate degrees, compared to 79.7\% and 68.7\%, respectively, among matched individuals. Additionally, unmatched men are more likely to smoke regularly (6.9\% vs. 3.4\%) but less likely to drink regularly (22.3\% vs. 26.2\%), suggesting a distinct lifestyle profile relative to matched men. Differences in household attitudes are also present: unmatched women are somewhat less likely to report taking primary responsibility for housework (“Do It Myself”: 7.8\% unmatched vs. 9.1\% matched), and unmatched men are less likely to report wanting children (62.0\% unmatched vs. 70.7\% matched). These patterns indicate that both observable human capital and alignment in household preferences may play a role in selection into match.

\begin{table}[!htbp]
  \begin{center}\footnotesize
      \caption{Summary Statistics in 2024 by Gender: Unmatched}
      \label{tb:summary_statistics_unmatched_continuous_2024}
      \subfloat[Continuous]{\input{figuretable/labor_family_economics_project/summary_statistics_unmatched_continuous_2024}}\\
      \subfloat[Discrete]{\input{figuretable/labor_family_economics_project/summary_statistics_unmatched_discrete_2024}}
  \end{center}\footnotesize
  %\textit{Note}: \textcolor{black}{[TBA]}
\end{table}

\paragraph{Joint Distribution of Family Preferences}

\begin{figure}[!htbp]
\centering
\caption{Joint Distribution of Housework, Childcare, and Desire for Child (2024 Sample)}
\label{fig:joint_housework_childcare_desire}
\subfloat[Male, Matched]{\includegraphics[width=0.95\textwidth]{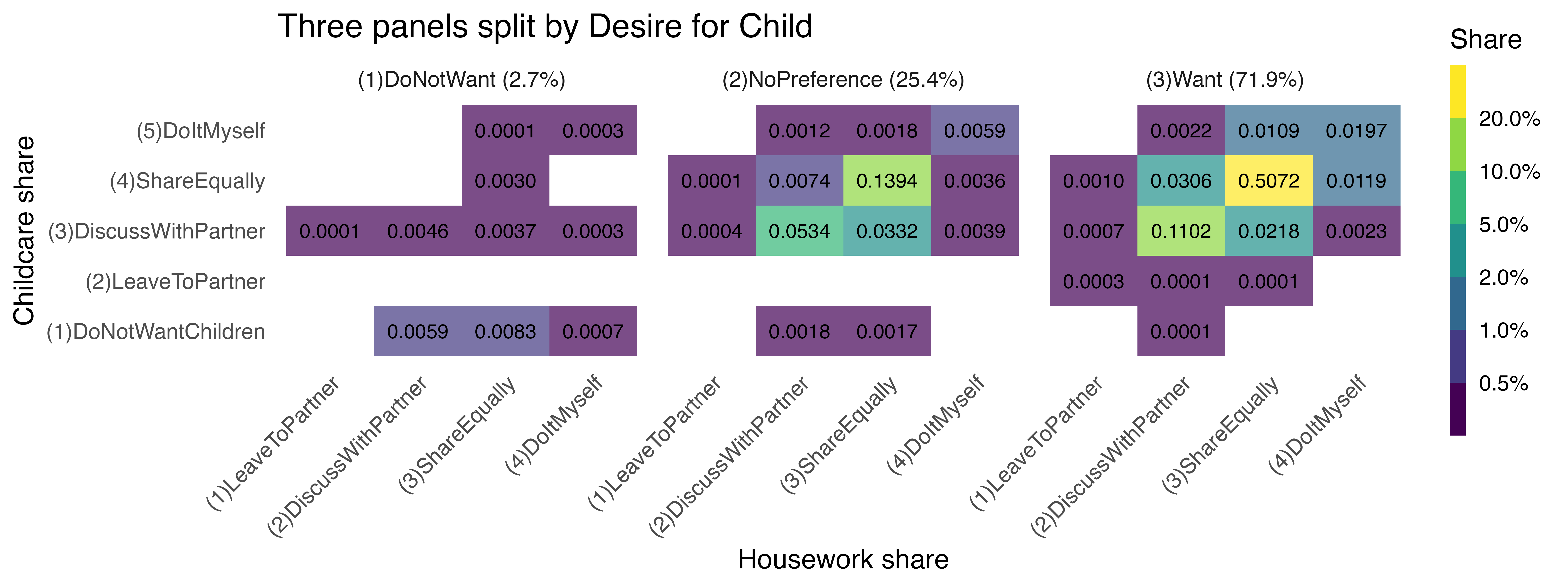}}\\
\subfloat[Female, Matched]{\includegraphics[width=0.95\textwidth]{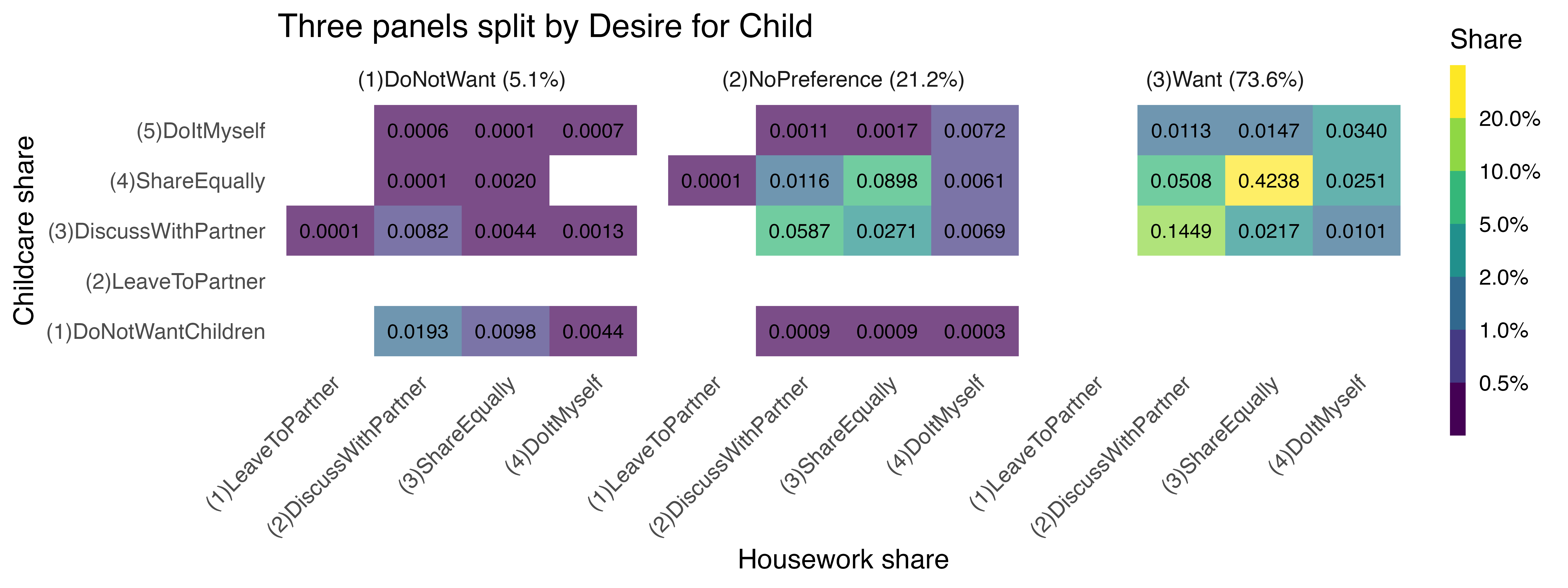}}\\
\subfloat[Male, Matched + Unmatched]{\includegraphics[width=0.95\textwidth]{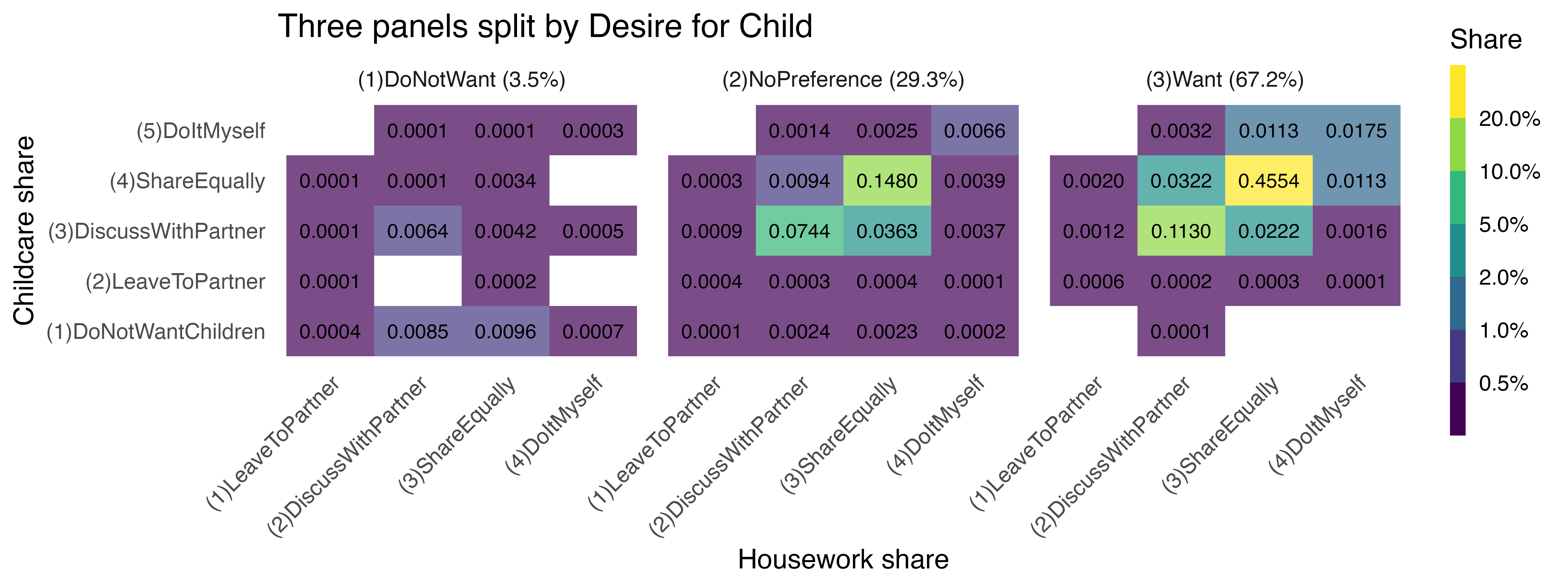}}\\
\subfloat[Female, Matched + Unmatched]{\includegraphics[width=0.95\textwidth]{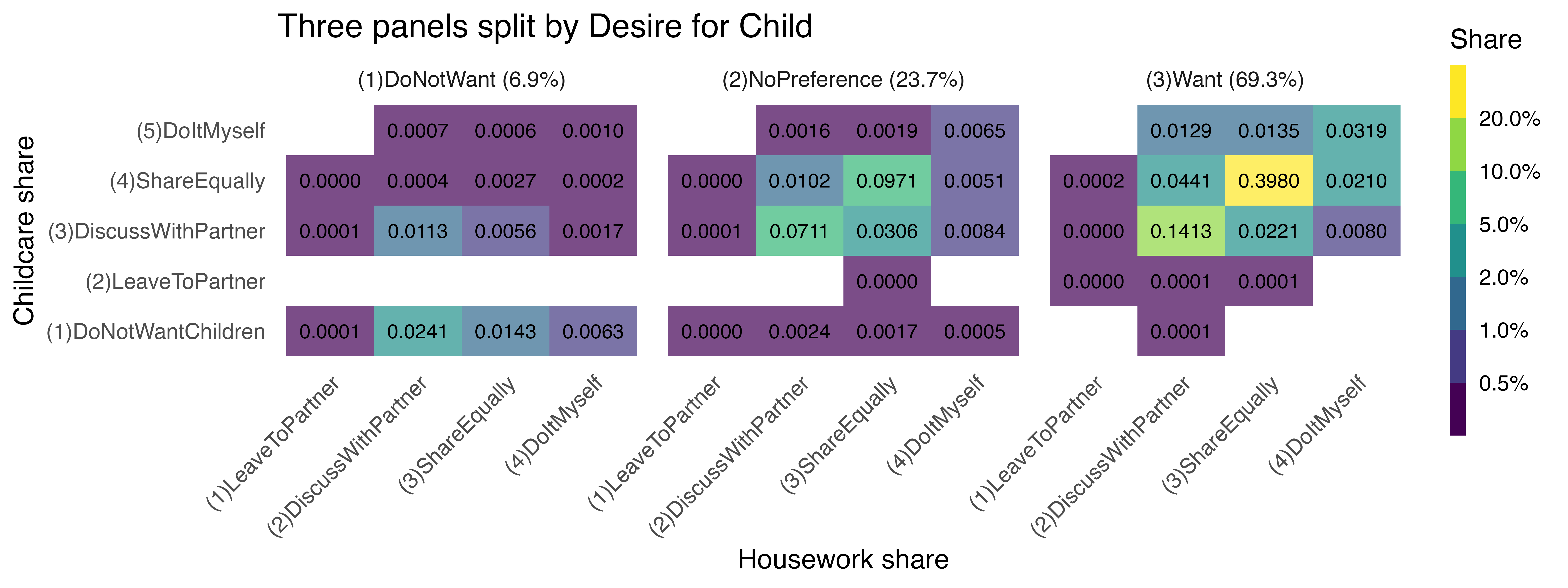}}
\end{figure}

Figure \ref{fig:joint_housework_childcare_desire} plots the joint distribution of (Desire for Child, Housework share, Childcare share) on our 2024 sample, separately for men and women, and for the matched-only and matched-plus-unmatched pools. Cells sum to 100\% across the three panels (the panel title reports the marginal $P(\text{Desire})$). The three preference variables are far from independently drawn: as Desire-for-Child moves from ``Do not want'' through ``No preference'' to ``Want,'' the joint mass concentrates sharply on the cell where both housework and childcare are reported as ``Share equally''---most starkly within the ``Want'' panel, which is also the dominant Desire-for-Child category, so that this single $(\text{Want},\text{Share equally},\text{Share equally})$ cell alone carries the bulk of the entire joint distribution for men. The ``No preference'' panel concentrates on the same cooperative cell at lower mass, and the ``Do not want'' panel is small and comparatively dispersed across cooperative-but-not-equal cells (e.g., ``Discuss with partner'' on either dimension). Note that ``Do not want'' respondents do not uniformly select ``Do not want children'' on the childcare item, reflecting the latter's conditional framing---users signal cooperative flexibility should children eventually arise (e.g., via partner negotiation or step-/adopted-children) rather than absolute commitment. The pattern holds for both sexes, although women exhibit somewhat thicker tails off the equal-share diagonal. Pooling matched and unmatched users (Panels c, d) leaves the qualitative pattern intact, indicating that the joint structure of family preferences is not an artifact of selection into matching but a feature of the platform population. 

\subsection{Additional Data Pattern}\label{subsec:additional_data_pattern}

\subsubsection{Selection at Entry Timing}\label{subsec:selection_entry}

\begin{figure}[!htbp]
\centering
\caption{Income Distribution: IBJ 2023 Entrants vs ESS 2022 Employed (age 25--44)}
\label{fig:selection_entry_income}
\includegraphics[width=0.95\textwidth]{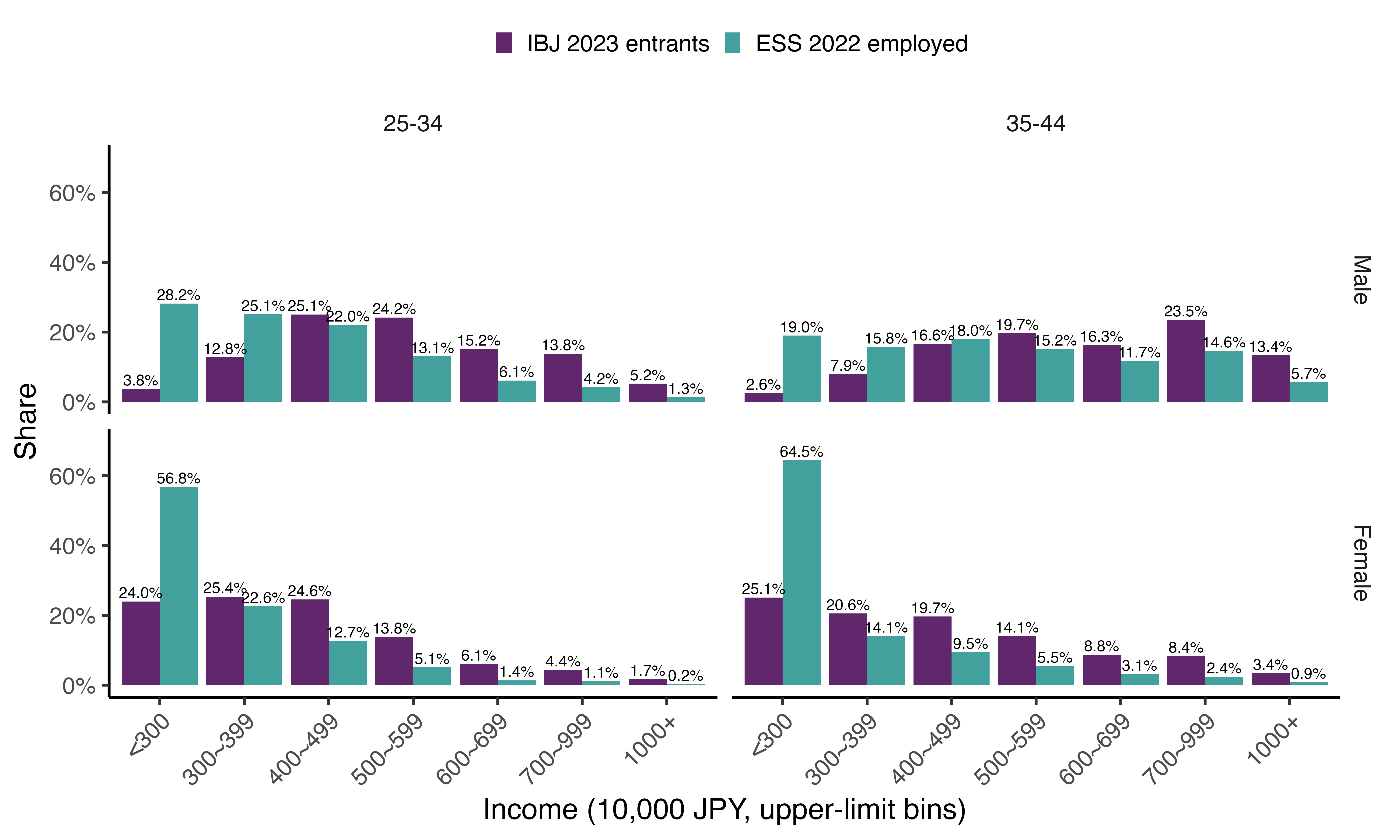}
\end{figure}

\begin{figure}[!htbp]
\centering
\caption{Education Distribution: IBJ 2023 Entrants vs ESS 2022 Employed (age 25--44)}
\label{fig:selection_entry_education}
\includegraphics[width=0.95\textwidth]{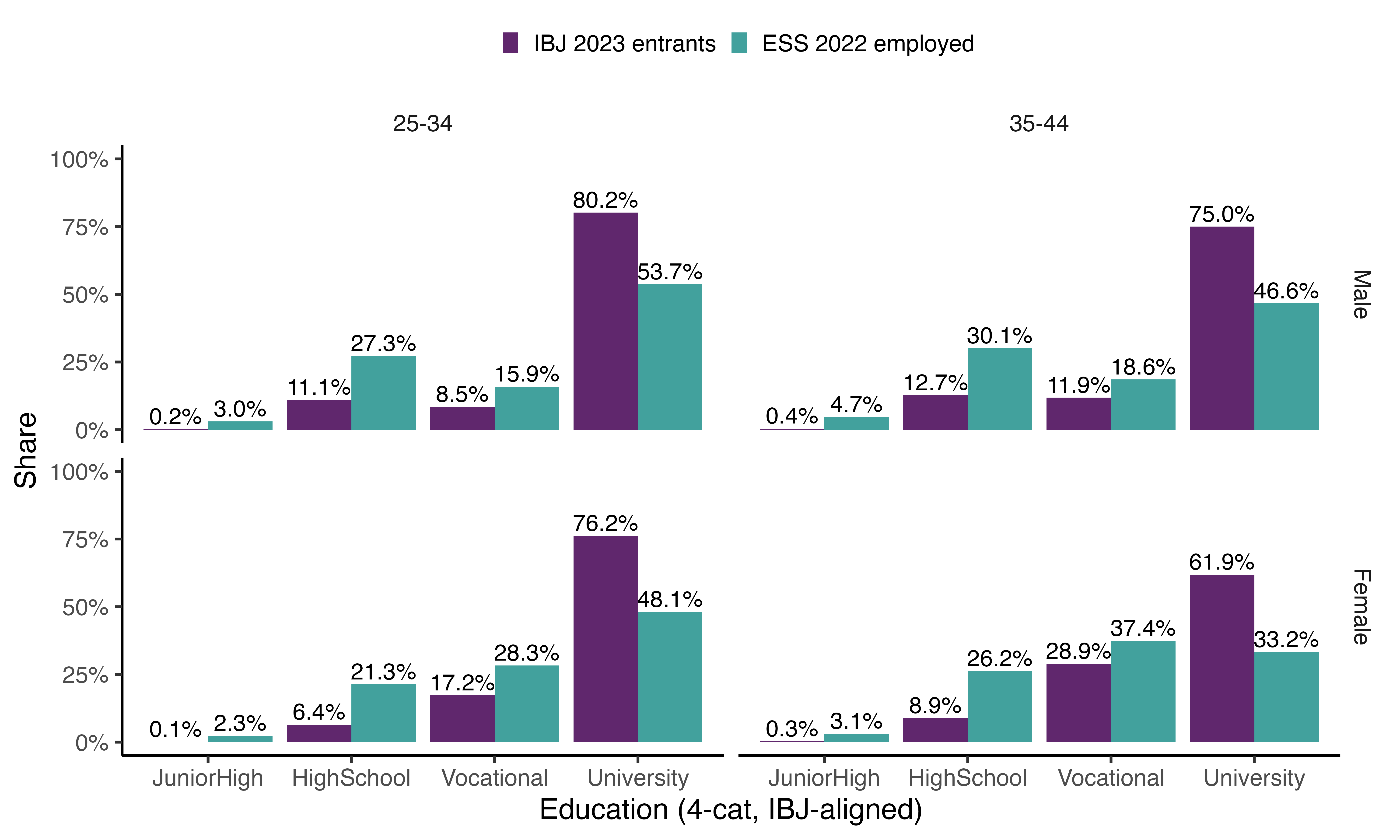}
\end{figure}

Figures \ref{fig:selection_entry_income} and \ref{fig:selection_entry_education} compare the income and education distributions of IBJ users registering on the platform in 2023 (age 25--44 at entry, including users who never subsequently match) with the 2022 Employment Status Survey (ESS) employed population in the same age range. Both panels reveal substantial positive selection on observables, but along different dimensions for each sex. On income, male entrants are sharply right-shifted relative to the national employed benchmark in both 25--34 and 35--44 cohorts: the share earning under 3 million yen collapses while shares above 5 million yen rise markedly, a pattern that is qualitatively present but considerably weaker for female entrants. On education, the asymmetry flips---female entrants are far more concentrated in the university-or-above category than the ESS benchmark, with a particularly pronounced gap in the younger cohort, though the corresponding gap is narrower for male entrants. These patterns are consistent with the platform's institutional design (high upfront fees, document-verified profiles) drawing high-earning men and college-educated women, and provide an entry-time selection benchmark against which subsequent matching outcomes can be interpreted.

\subsubsection{Exit Reason at Exit Timing}\label{subsec:exit_reason}

Table \ref{tab:withdraw_reason_2023} reports the distribution of withdrawal reasons among the 2023 cohort. Of the 17{,}394 male and 23{,}335 female exits, the modal reason is ``Regular withdrawal'' (47.8\% of male and 54.5\% of female exits)---users who left without a successful marriage on the platform. The next-largest category is ``Married within IBJ,'' accounting for 38.6\% of male and 28.8\% of female exits, which constitutes the platform's core success outcome. The gender gap in within-IBJ marriage shares reflects the fact that the matched sample is balanced one-to-one by construction while the unmatched female pool is substantially larger, so the within-IBJ share is mechanically smaller for women. Marriages reported as ``Married outside IBJ'' (covering other matching service federations and friend/personal introductions) account for an additional 13.1\% of male and 16.0\% of female exits, suggesting that a non-trivial fraction of exiting users find partners through alternative channels even after registering on the platform. Other administrative exits (agency-deletion withdrawals and long-inactivity auto-deletions) together represent under 1\% of exits in either gender.

\begin{table}[!htbp]
\centering
\caption{Withdrawal Reason in the 2023 Cohort by Gender (Matched + Unmatched, Exited Only)}
\label{tab:withdraw_reason_2023}
\input{figuretable/labor_family_economics_project/withdraw_reason_2023_male_female}

\vspace{0.2cm}
\flushleft
\textit{Note}: 
``Total exited'' denominator pools matched and unmatched 2023 cohort users with a non-NA withdrawal reason; users with a NA withdrawal reason are interpreted as still active and excluded from these categorization.
% ``Total exited'' denominator pools matched and unmatched 2023 cohort users with a non-NA \texttt{withdraw\_kbn}; users with \texttt{withdraw\_kbn = NA} (interpreted as still active) are excluded.
\end{table}

\section{Additional Results}

\subsection{A Many-to-Many TU Matching Model: Full Interaction Results}

Tables \ref{tb:maximum_score_matrix_application} and \ref{tb:maximum_score_matrix_serious_relation} compare the baseline specification in Table \ref{tb:maximum_score_diagonal_all_stage} that restricts attention to diagonal elements of the affinity matrix with a more flexible full-interaction model that allows for cross-attribute complementarities. Relative to the diagonal specification, the full-interaction model substantially increases the number of parameters to be estimated, and as expected in a semiparametric matching maximum score framework \citep{fox2018estimating,manski1975maximum}, many interaction terms are weakly identified, with wide confidence intervals that frequently include zero. This reflects a general limitation of rank-based estimators \citep{manski1975maximum}: as the parameter dimension grows, point identification becomes increasingly difficult and the identified set expands for many coefficients. Importantly, however, the key diagonal patterns for age, income, and preferences for children remain robustly positive across specifications. Marital-history assortativeness remains positive in point estimates, but it is estimated less precisely in the later stages under the full-interaction model. These results suggest that the main qualitative conclusions from the diagonal specification are not driven by omitted cross-attribute interactions, while also underscoring the precision costs of the richer parameterization. Given the limited additional insight provided by the full-interaction terms and the loss of precision associated with higher-dimensional parameterization, the diagonal specification offers a parsimonious and robust summary of assortative matching in the data.

\begin{landscape}
\begin{table}[!htbp]
  \begin{center}\tiny
      \caption{Matching Maximum Score Estimation: Full Interaction Model}
      \label{tb:maximum_score_matrix_application}
      \subfloat[Application]{\input{figuretable/labor_family_economics_project/maximum_score_matrix_application}}\\
      \subfloat[Pre-relation]{\input{figuretable/labor_family_economics_project/maximum_score_matrix_pre_relation}}
  \end{center}\footnotesize
  \textit{Note}: See the details in the main text.
\end{table} 
\end{landscape}

\begin{landscape}
\begin{table}[!htbp]
  \begin{center}\tiny
      \caption{Matching Maximum Score Estimation: Full Interaction Model}
      \label{tb:maximum_score_matrix_serious_relation}
      \subfloat[Serious Relation]{\input{figuretable/labor_family_economics_project/maximum_score_matrix_serious_relation}}\\
      \subfloat[Proposal]{\input{figuretable/labor_family_economics_project/maximum_score_matrix_proposal}}
  \end{center}\footnotesize
  \textit{Note}: See the details in the main text.
\end{table} 
\end{landscape}

\subsection{Assortativeness in TU One-dimensional Discrete-Type Model \citep{choo2006marries} in 2024}\label{sec:unidimensional_assortativeness}

To highlight the dataset's granularity and novelty, we first apply the well-known framework of \citet{choo2006marries} to quantify assortative matching patterns and compare the contribution of each observed characteristic to match surplus.\footnote{Following \citet{choo2006marries}, we estimate the total systematic surplus $\hat{\Phi}_{ij}$ using the relation $\mu_{ij}^2 = \mu_{i0} \mu_{0j} \exp(\Phi_{ij})$, where $\mu_{ij}$ is the number of matches between type-$i$ men and type-$j$ women, and $\mu_{i0}$, $\mu_{0j}$ denote unmatched counts. Taking logs yields $\hat{\Phi}_{ij} = 2 \log \mu_{ij} - \log \mu_{i0} - \log \mu_{0j}$. We compute $\mu_{ij}$ from observed engagements, and $\mu_{i0}$, $\mu_{0j}$ from unmatched individuals in each type. To avoid undefined values, we replace zero counts with a small constant ($10^{-8}$). See deeper theoretical discussion in \cite{galichon2022cupid}.}

\begin{figure}[!htbp]
  \begin{center}
  \subfloat[Age]{\includegraphics[width = 0.42\textwidth]{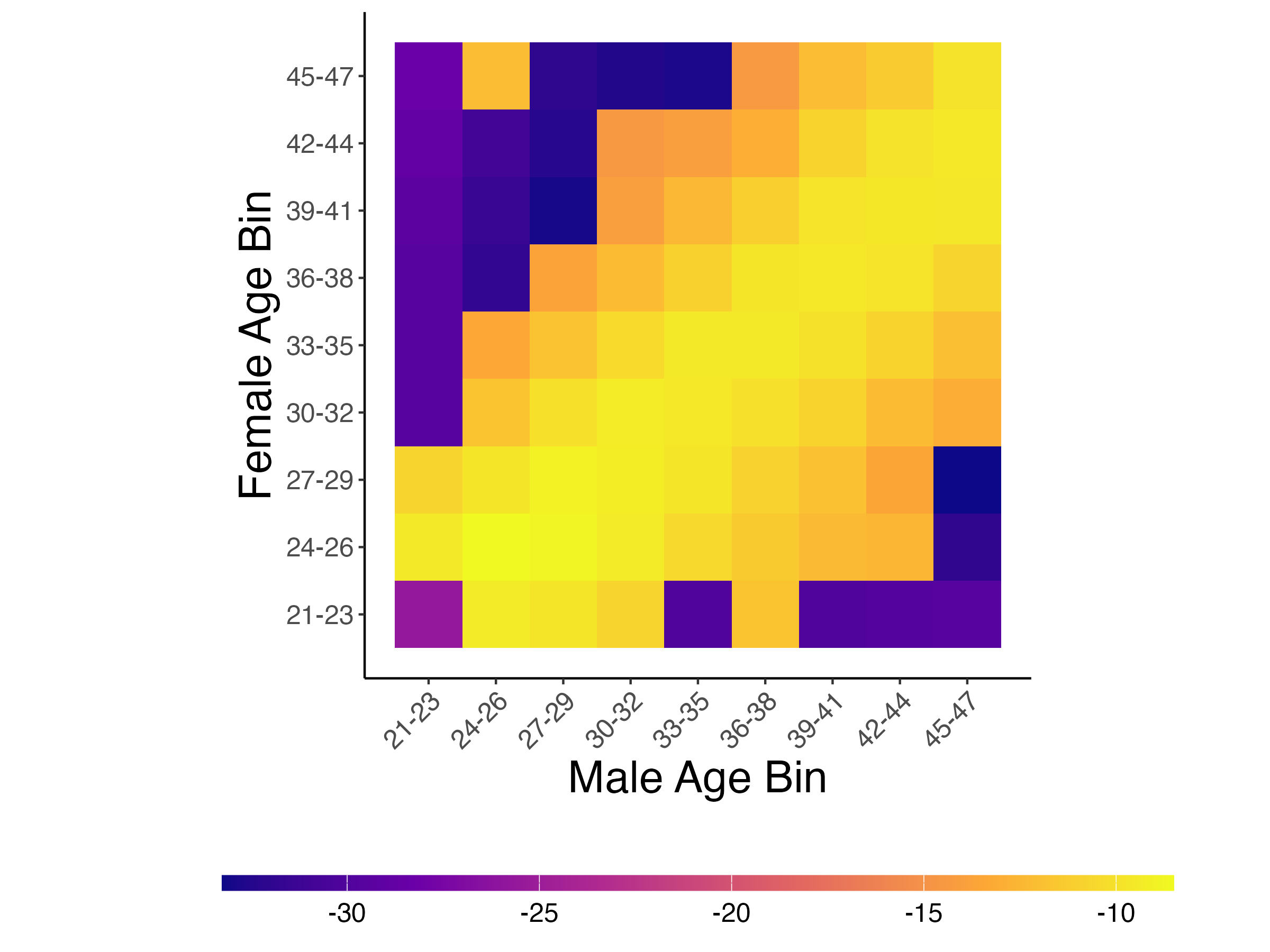}}
  \subfloat[Height]{\includegraphics[width = 0.42\textwidth]{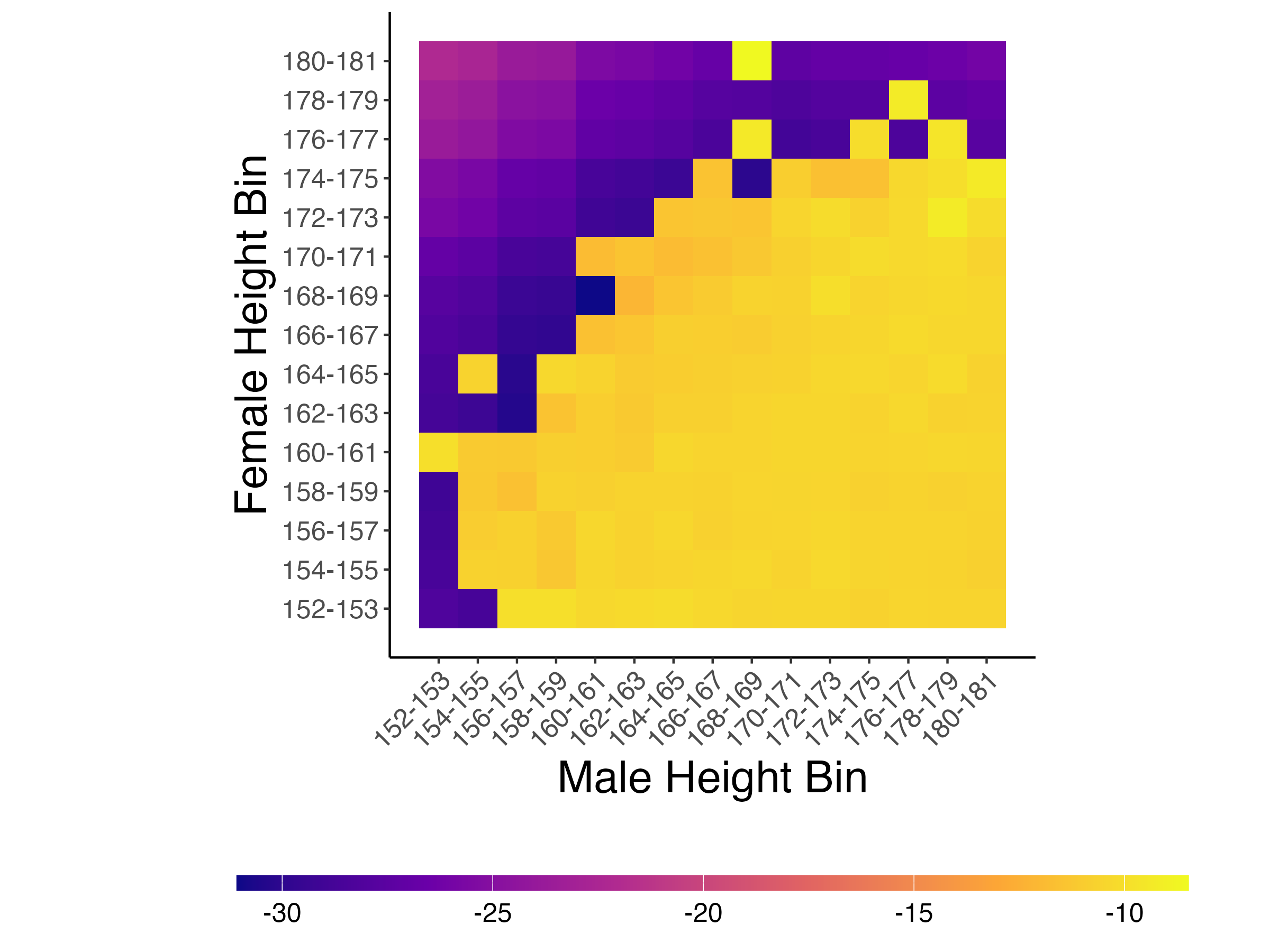}}\\
  \subfloat[Weight]{\includegraphics[width = 0.42\textwidth]{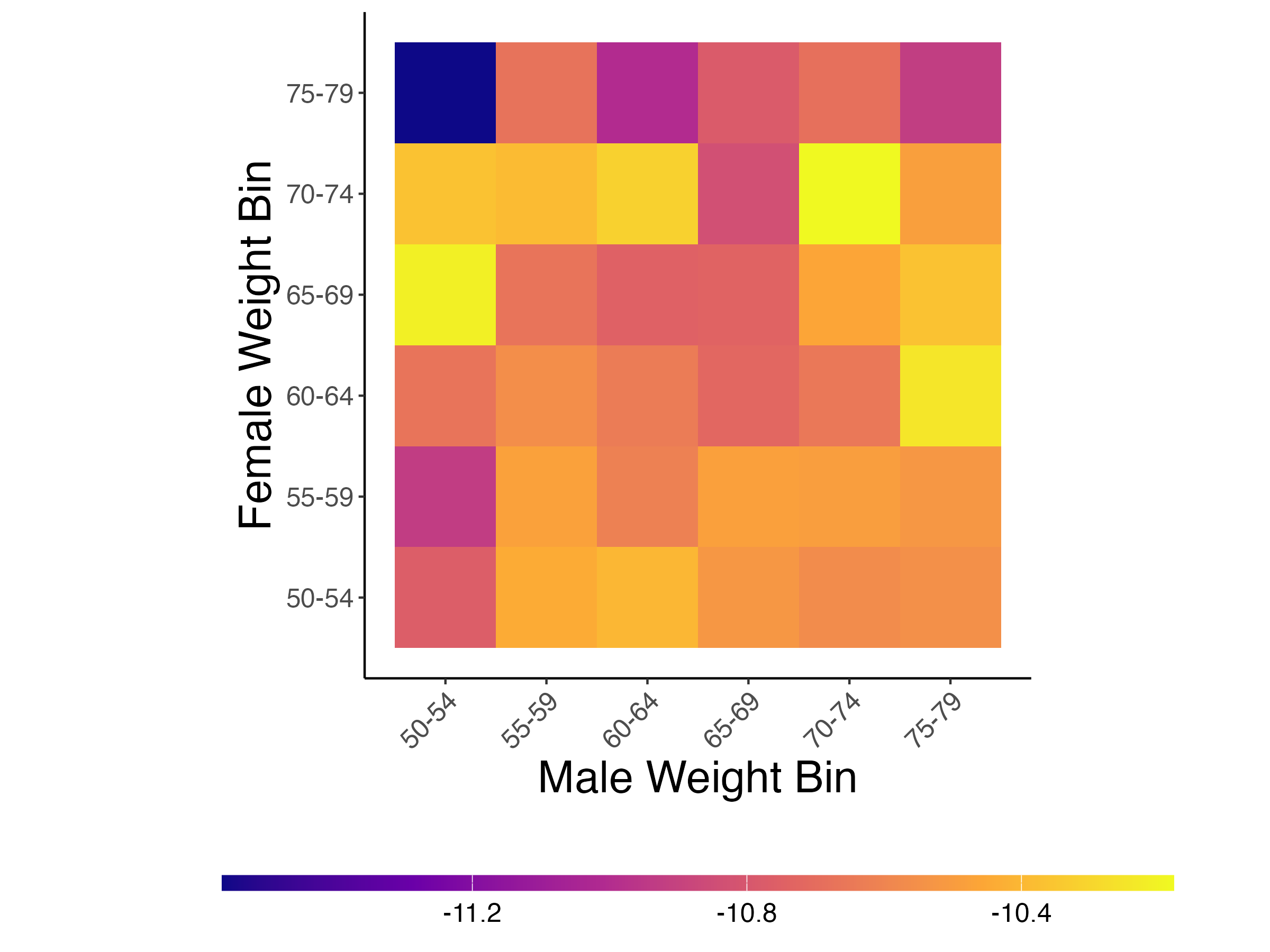}}
  \subfloat[BMI]{\includegraphics[width = 0.42\textwidth]{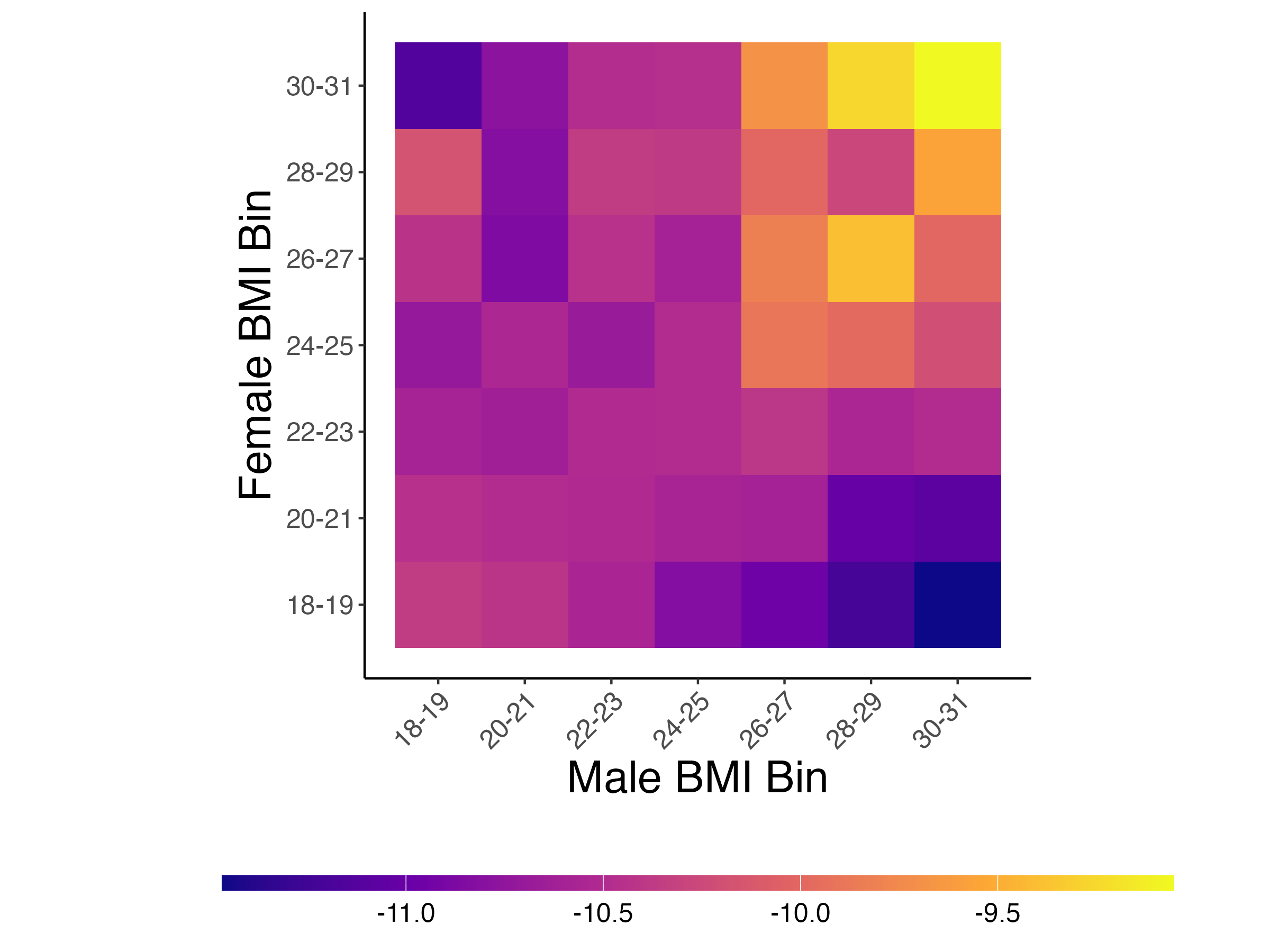}}\\
  \subfloat[Income Category]{\includegraphics[width = 0.42\textwidth]{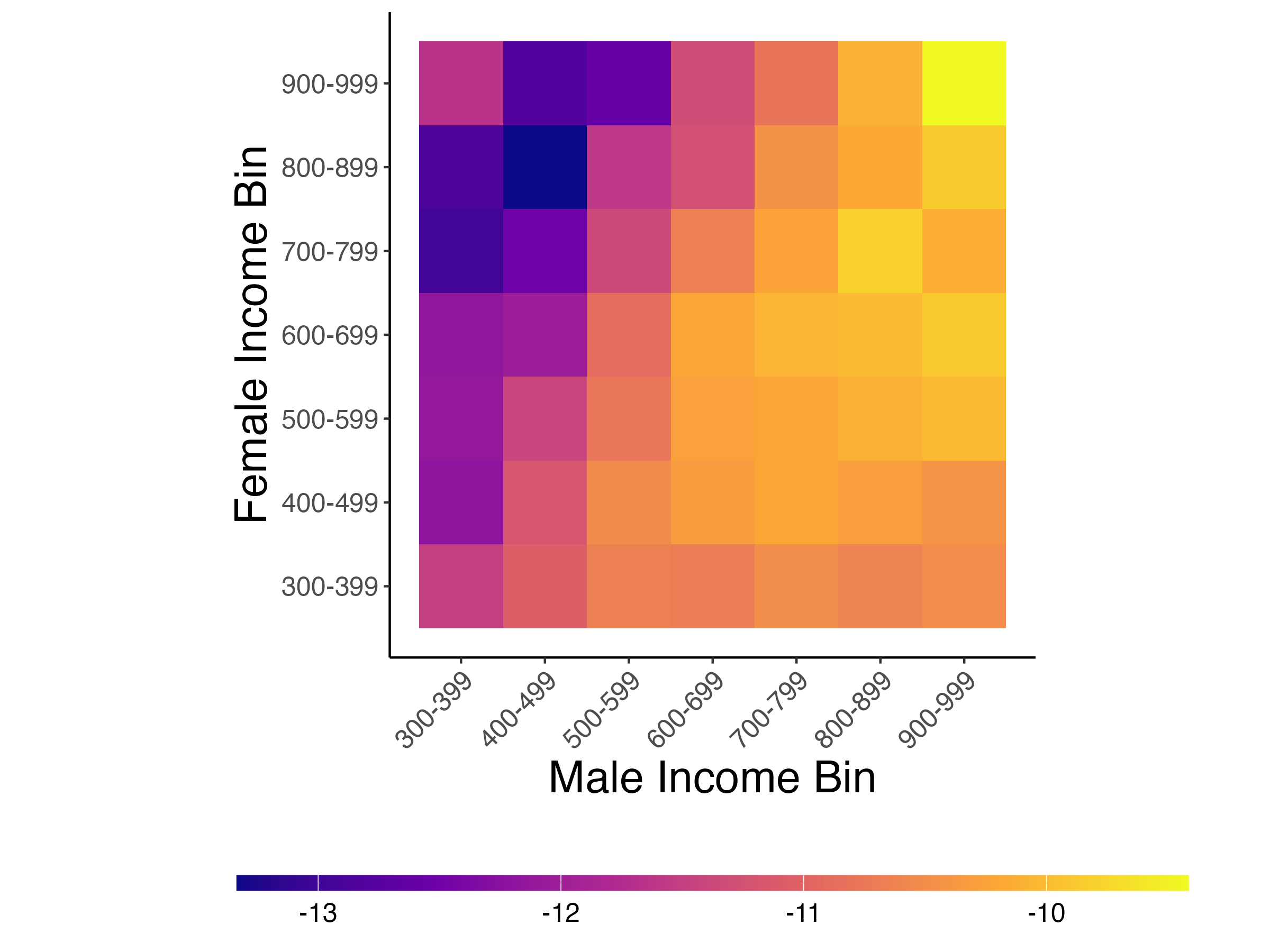}}
  \subfloat[Education]{\includegraphics[width = 0.42\textwidth]{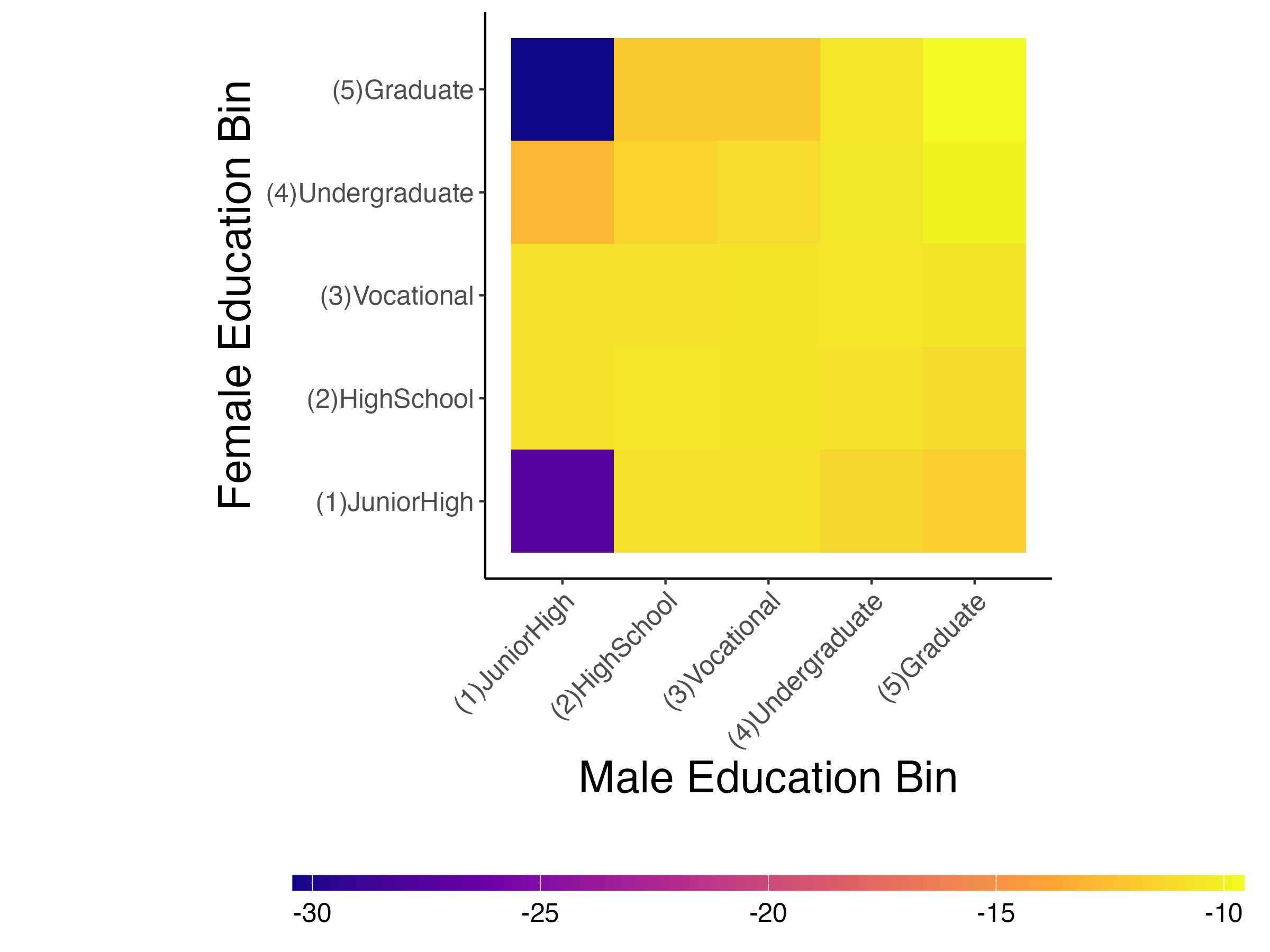}}
  \caption{Systematic Surplus in Anthropometric and Sociodemographic Attractiveness}
  \label{fg:assortativeness_choo_siow_age_2024} 
  \end{center}
  \footnotesize
  Note: Anthropometric physical attractiveness is captured by age, height, weight, and BMI. Sociodemographic attractiveness is captured by income and education. See \cite{chiappori2012fatter} and \cite{dupuy2014personality} for discussion of the use of these characteristics.
\end{figure}

\paragraph{Anthropometric physical attractiveness}
Figure \ref{fg:assortativeness_choo_siow_age_2024} displays the estimated systematic surplus $\hat{\Phi}_{ij}$ across joint bins of male and female attributes. Panel (a), which shows the results for age, reveals strong positive assortative matching: surplus is concentrated in a broad band near the diagonal, with the highest values occurring among women in their late 20s to mid-30s and men in their early 30s to late 30s. The ridge lies slightly below the exact diagonal for much of the support, consistent with a modest tendency for male partners to be somewhat older than female partners. Surplus declines as age differences widen, especially away from the dense central age bins. These results echo common patterns in marriage timing and are consistent with age-related preferences and fertility considerations.

Panels (b) through (d) in Figure \ref{fg:assortativeness_choo_siow_age_2024} explore sorting based on physical characteristics. For height (Panel b), we observe strong positive assortative matching: taller men tend to match with taller women, and the surplus surface peaks along a diagonal ridge. In contrast, weight (Panel c) displays a more diffuse and asymmetric pattern, with several locally high-surplus cells rather than a single sharp diagonal. Panel (d) indicates that BMI-based sorting is more structured than weight-based sorting, with relatively high surplus concentrated in the upper-right portion of the matrix corresponding to higher BMI bins for both sexes. These findings suggest that physical appearance contributes to marital surplus in nuanced ways, with height being positively aligned and weight or BMI subject to more varied patterns.

\paragraph{Sociodemographic attractiveness}
Panels (e) and (f) in Figure \ref{fg:assortativeness_choo_siow_age_2024} depict sorting patterns on income and education, capturing sociodemographic dimensions of match surplus. The income plot reveals a clear positive gradient with respect to male income, while variation across female income is present but less monotone. The education panel shows that surplus is concentrated among pairings involving undergraduate and graduate categories, whereas cells involving junior-high education are consistently weak. These results highlight the salience of both economic capacity and educational alignment in shaping match gains in the marriage market.

\paragraph{Health and Lifestyle Preferences}

\begin{figure}[!htbp]
  \begin{center}
  \subfloat[Drinking]{\includegraphics[width = 0.42\textwidth]{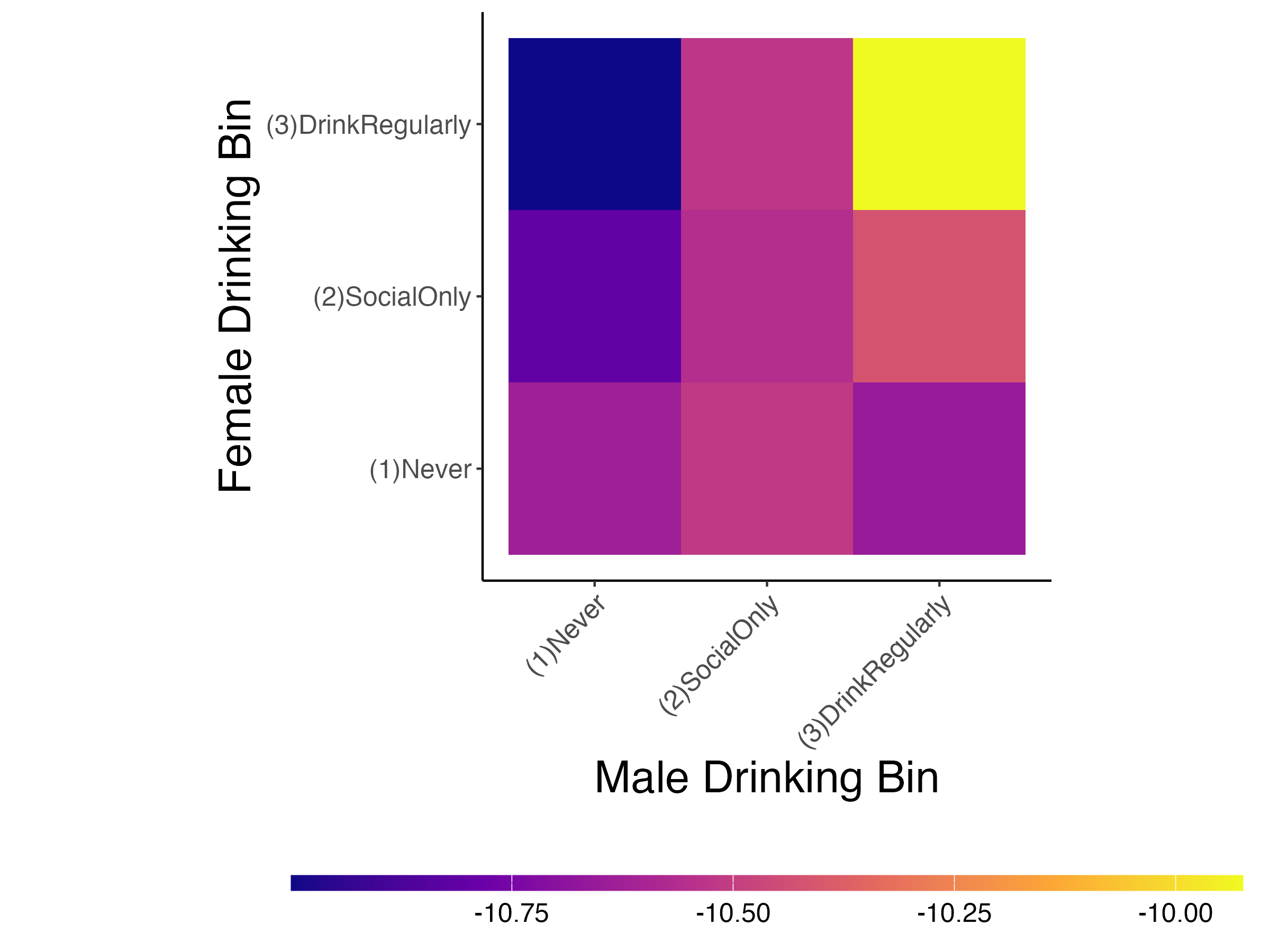}}
  \subfloat[Smoking]{\includegraphics[width = 0.42\textwidth]{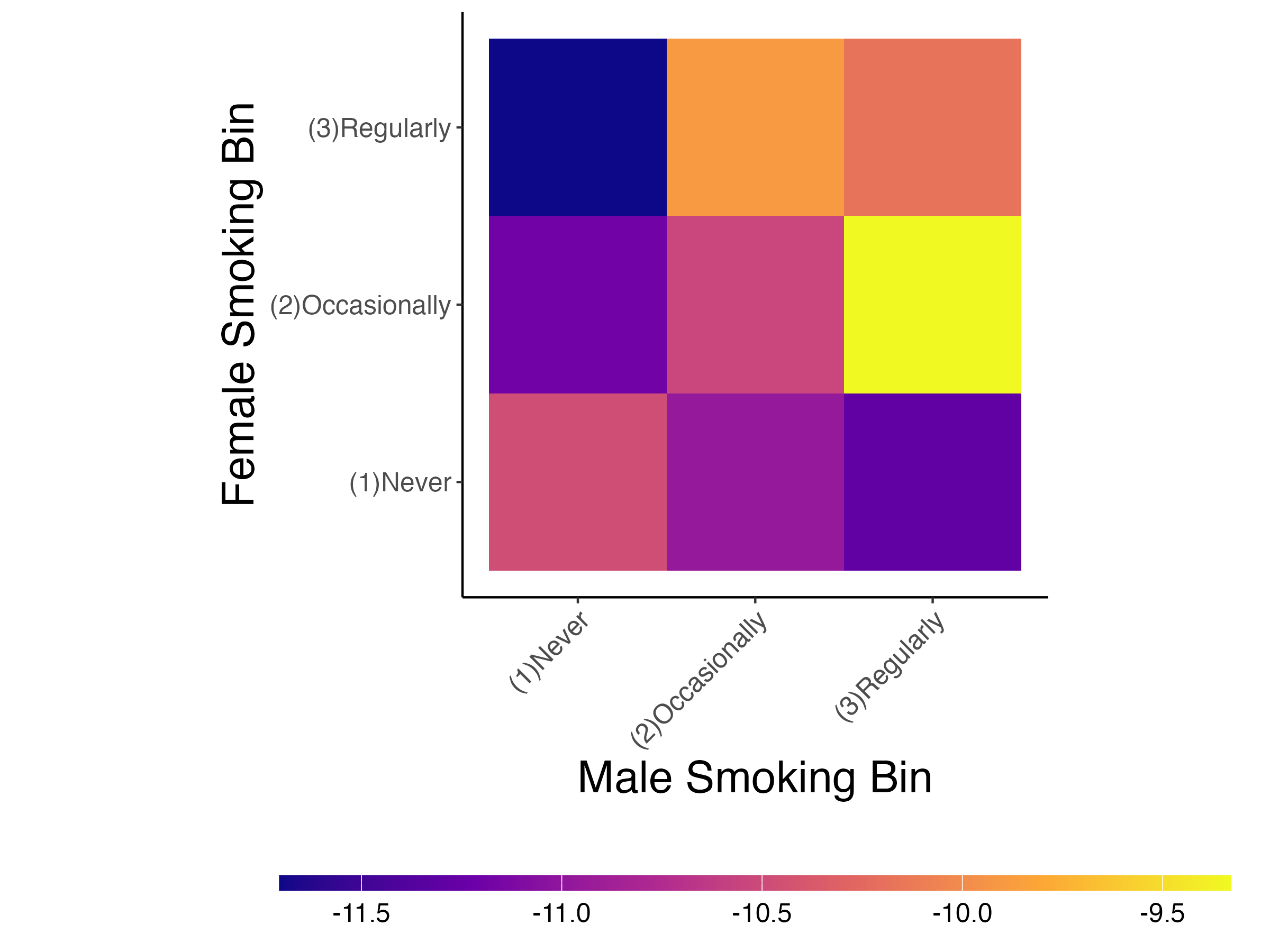}}\\
  \subfloat[Marital History]{\includegraphics[width = 0.42\textwidth]{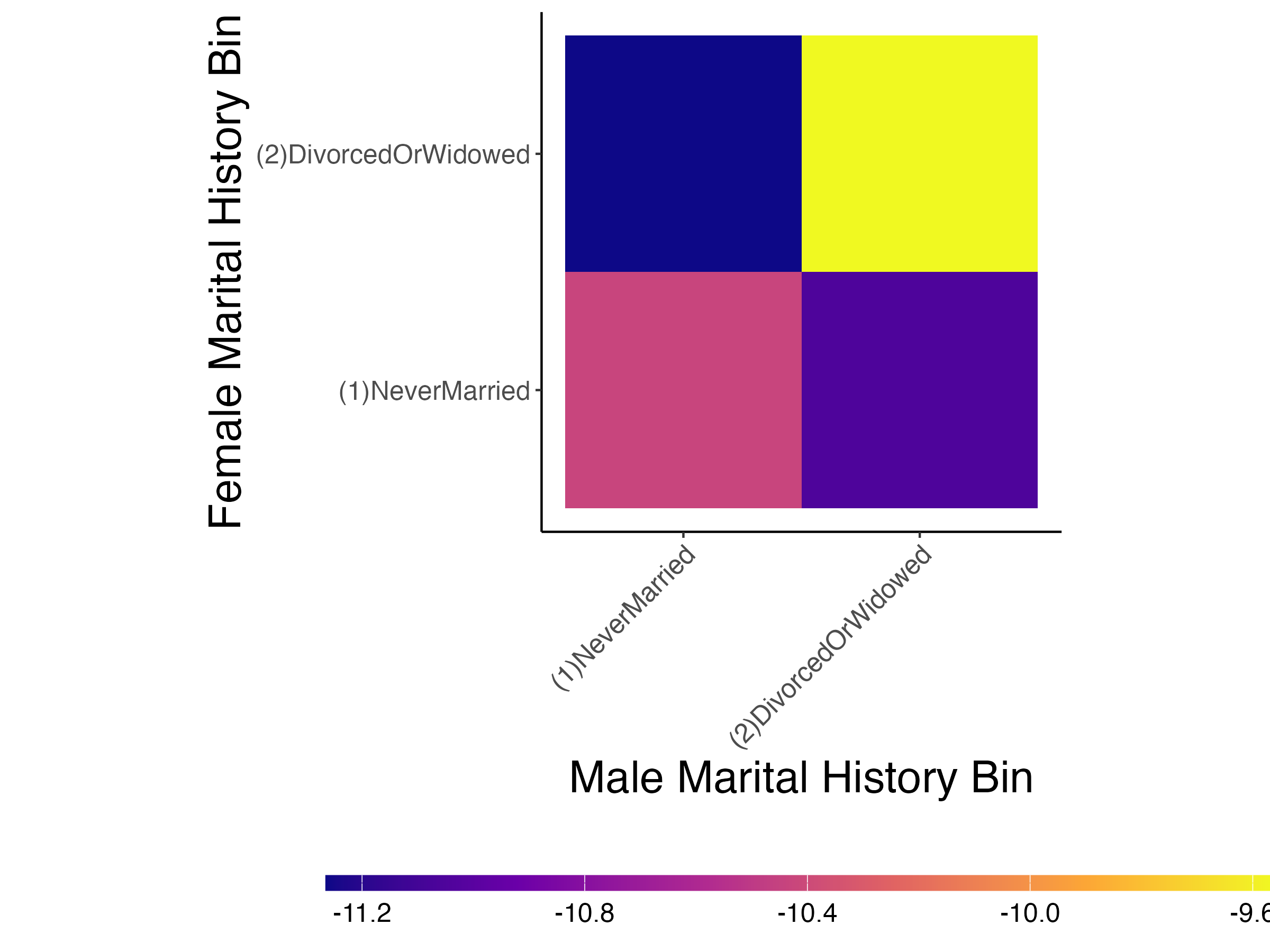}}
  \subfloat[Preference for Children]{\includegraphics[width = 0.42\textwidth]{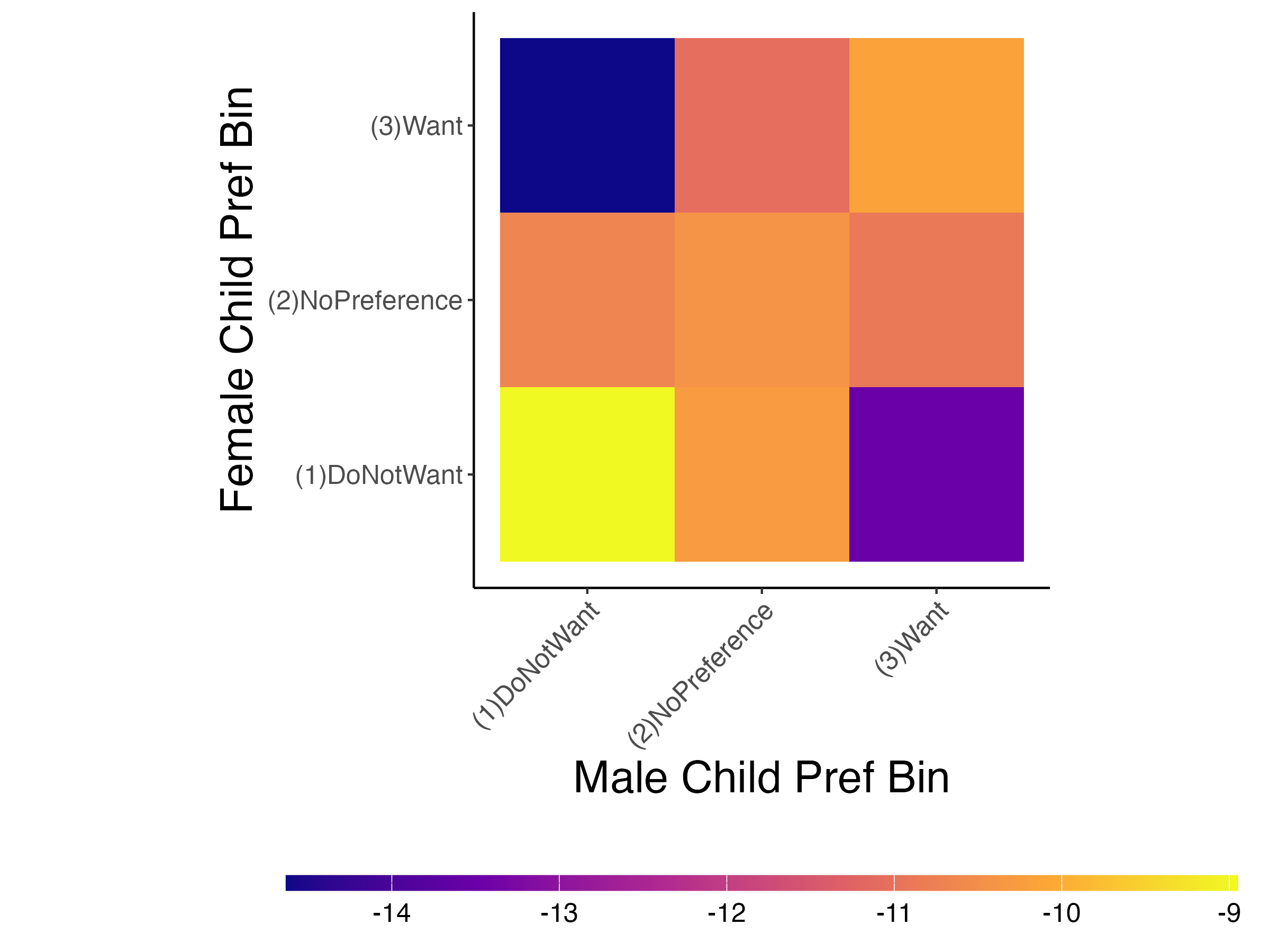}}\\
  \subfloat[Childcare]{\includegraphics[width = 0.42\textwidth]{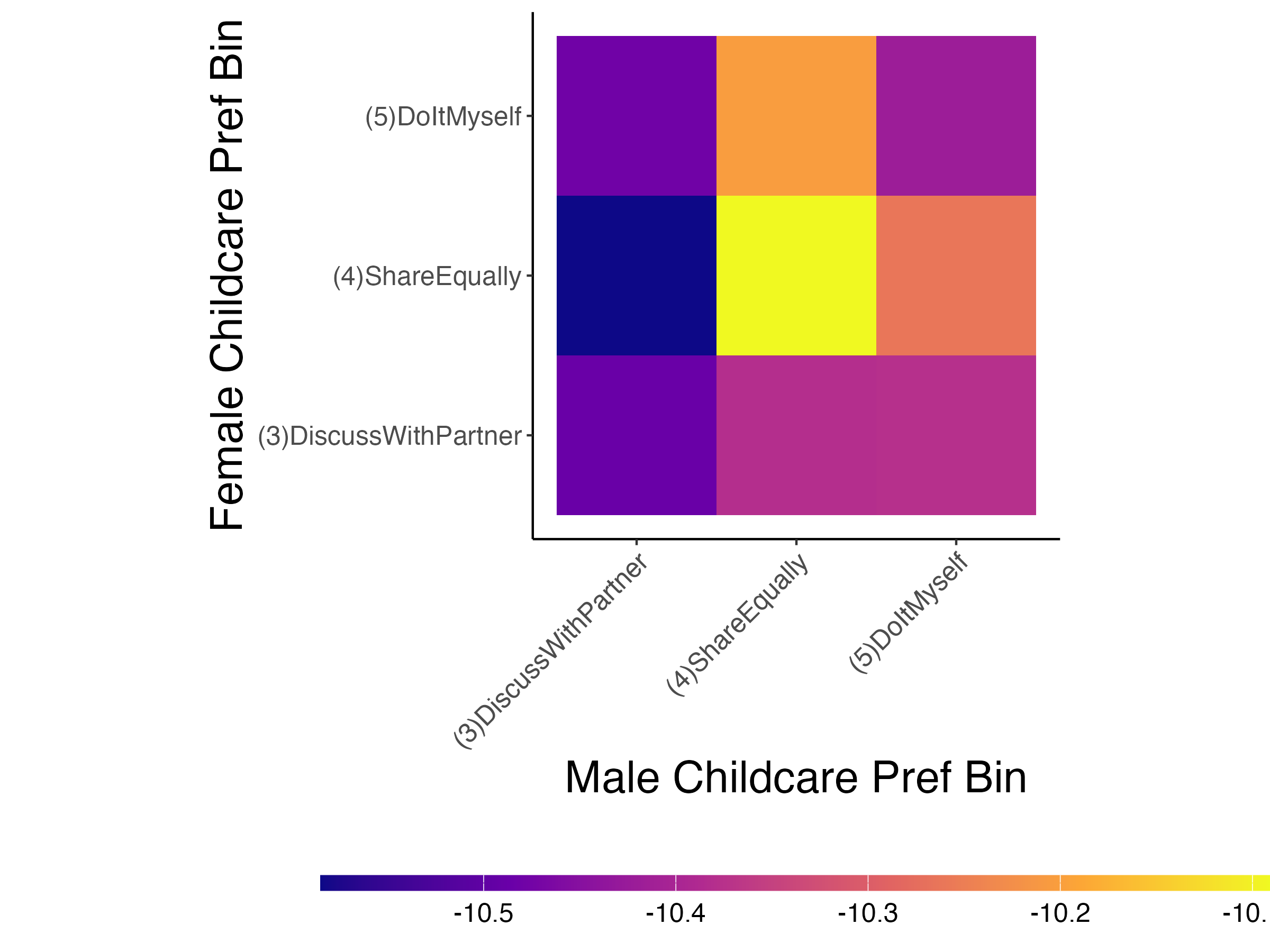}}
  \subfloat[Housework]{\includegraphics[width = 0.42\textwidth]{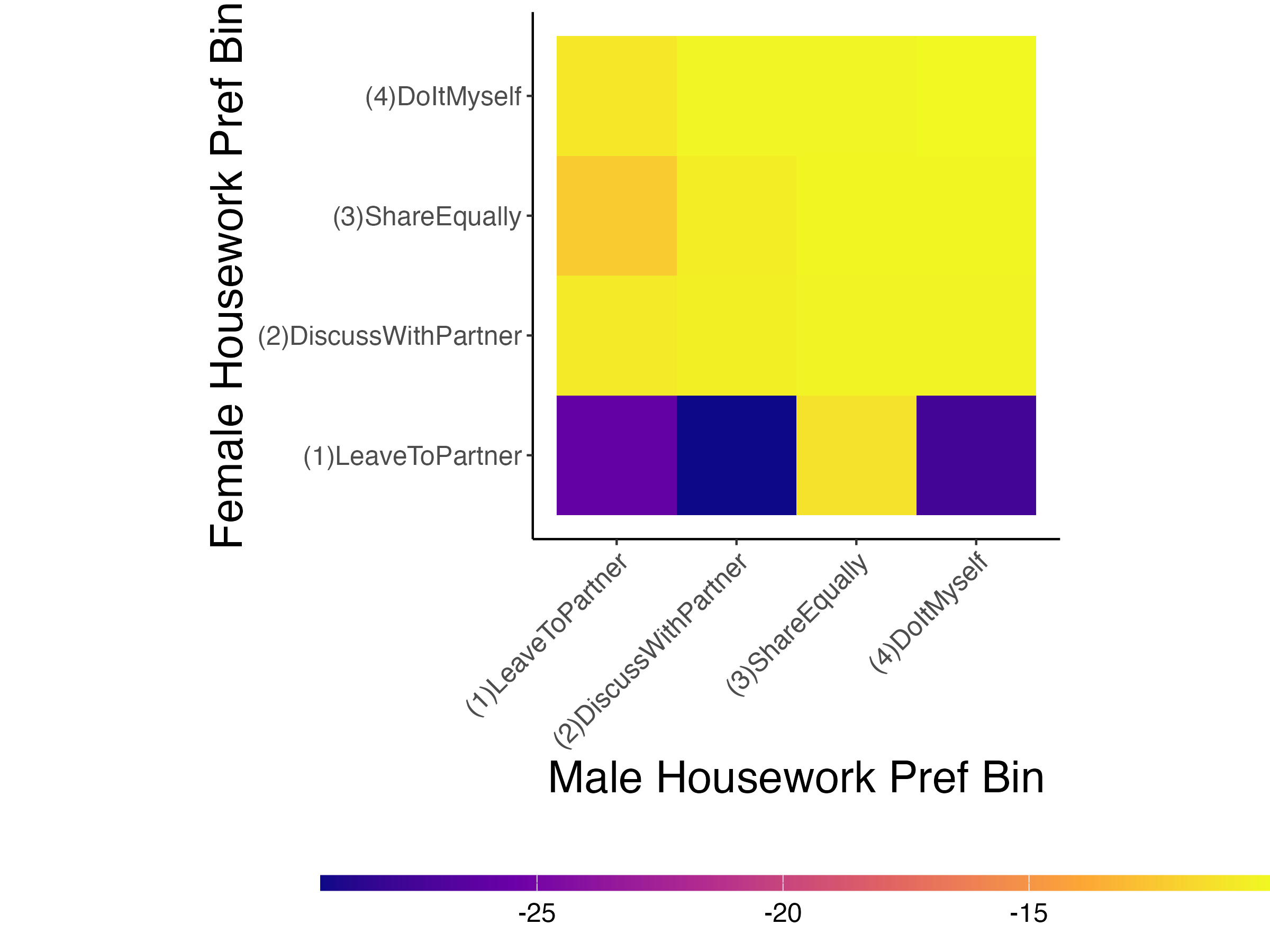}}
  \caption{Systematic Surplus in Lifestyle, Marital History, and Family Preferences}
  \label{fg:assortativeness_choo_siow_housework_2024} 
  \end{center}
  \footnotesize
  Note: 
  See \cite{chiappori2018bidimensional} and \cite{chiappori2024analyzing} for discussion of the use of smoking variables. 
\end{figure}

Panels (a) and (b) in Figure \ref{fg:assortativeness_choo_siow_housework_2024} present surplus estimates by drinking and smoking habits. These lifestyle traits exhibit modest but structured matching patterns. For drinking, surplus tends to be higher among pairings involving regular drinkers than among pairs of non-drinkers. For smoking, the surface is not cleanly diagonal, and some off-diagonal combinations display relatively high surplus. While these factors do not dominate match surplus overall, they reflect nontrivial lifestyle-related complementarities or selection patterns that may influence long-term compatibility.

\paragraph{Marital History}

Panel (c) in Figure \ref{fg:assortativeness_choo_siow_housework_2024} focuses on marital history, distinguishing between never-married individuals and those who are divorced or widowed. The strongest surplus appears when both partners are previously married, while mixed-history pairings are weaker. This pattern suggests that shared life course trajectories along the remarriage margin may play an important role in perceived compatibility and in the structure of surplus.

\paragraph{Preferences over Family Formation}

Panels (d) through (f) in Figure \ref{fg:assortativeness_choo_siow_housework_2024} explore preferences related to family formation, including desire for children, childcare attitudes, and housework division. These surfaces show substantial structure, but the patterns differ across attributes. For desire for children, surplus is high at both ends of the alignment spectrum, with the strongest cell among couples where both report not wanting children and relatively high surplus also among couples where both report wanting children. Childcare attitudes display the clearest assortative pattern, with equal-sharing preferences generating the highest surplus. Housework preferences are more heterogeneous and less cleanly diagonal. Taken together, these figures suggest that family-related values matter for match surplus, but the relevant margin is not always simple like-with-like sorting.

%%%%%%%%%%%%%%%%%%%%%%%%%%%%%%%%%%%%%%%%%%%
\section{Theoretical Analysis}\label{sec:theoretical_analysis}

This appendix provides the formal model and derivations supporting the discussion in Section~\ref{sec:discussion}.

\subsection{Model Setup}

We consider a static model of fertility and female labor supply.\footnote{See \citet{doepke2023economics} for a review of the model of fertility decisions.}
A household consists of a husband and a wife whose fertility preference parameters are drawn from a common discrete set $\{L, H\}$ with $H > L > 0$.
Because the type set is the same for both genders, the utility function $v(n, d_g)$ has the same interpretation for husband and wife.
The utility of spouse $g \in \{m, f\}$ is
\begin{align}
    U_g = c_g + v(n, d_g),
\end{align}
where $c_g$ denotes private consumption and $v(n, d_g)$ captures utility from children.
The quasilinear structure implies that utility is fully transferable between spouses, which is consistent with the transferable utility framework imposed in the matching estimation of Section~\ref{sec:model}.

We impose the following assumptions on $v$.
We assume $v_n > 0$, $v_{nn} < 0$, and standard boundary conditions ensuring an interior solution for all types and matches.\footnote{Formally, we require $\lim_{n \to 0^+} v_n(n, d) = +\infty$ and $\lim_{n \to \infty} v_n(n, d) = 0$ for all $d \in \{L, H\}$. Together with $v_{nn} < 0$, these Inada-type conditions guarantee that the first-order conditions $v_n(n_d, d) = \frac{q}{2}$ and $v_n(n^u, d_m) + v_n(n^u, d_f) = q$ each have a unique interior solution.}
The condition $v_{nd} > 0$ says that the marginal utility of an additional child is increasing in $d_g$: spouses who care more about children benefit more from each additional child.
Together with $v_{nn} < 0$, this implies that the individually optimal number of children is increasing in $d$, which is consistent with our interpretation of $d$ as governing the degree of preferences for children.
These assumptions are standard in the fertility literature and admit the commonly used specification $v(n,d) = d\log n$ as a special case.

Each child entails a monetary cost $\psi > 0$ and a time cost $\phi > 0$, and a fraction $s \in (0,1)$ of total childcare time is provided through public childcare.
The husband works full time at wage $w_m$, while the wife allocates her remaining time after childcare to market work at wage $w_f$.
The household budget constraint is
\begin{align}
    c_m + c_f + \left[\psi + (1-s)w_f\phi\right] n = w_m + w_f.
\end{align}
Let $q \equiv \psi + (1-s)w_f\phi$ denote the full cost per child, and assume wages are independent of preference type so that $q$ is common to all matches.
We define the \emph{individually optimal fertility} $n_d$ as the solution to $\max_n \left[-\frac{qn}{2} + v(n, d)\right]$, where each spouse bears an equal share $\frac{q}{2}$ of the child cost, yielding the first-order condition $v_n(n_d, d) = \frac{q}{2}$; our assumptions imply $n_H > n_L > 0$.

\subsection{Household Joint Maximization}
\label{sec:appendix_unitary}

Since utility is fully transferable, the household maximizes the sum of spousal utilities subject to the budget constraint.
For a match $(d_m, d_f)$, define
\begin{align}
    A(n,\, d_m,\, d_f) \;\equiv\; -qn + v(n, d_m) + v(n, d_f),
\end{align}
and let $B^U(d_m, d_f) \equiv \max_n A(n, d_m, d_f)$ be the joint utility from children net of cost at the jointly optimal fertility.
The interior optimum $n^u(d_m, d_f)$ satisfies
\begin{align}
    v_n\!\left(n^u,\, d_m\right) + v_n\!\left(n^u,\, d_f\right) = q.
    \label{eq:unitary_foc}
\end{align}
In a same-type $(d,d)$ match, \eqref{eq:unitary_foc} reduces to $v_n(n^u, d) = \frac{q}{2}$, which coincides with the individual optimality condition, so $n^u(d,d) = n_d$.
For a cross-match $(H,L)$, evaluating \eqref{eq:unitary_foc} at $n_L$ gives $v_n(n_L,H) + v_n(n_L,L) > q$ (since $v_n(n_L,L) = \frac{q}{2}$ and $v_n(n_L,H) > \frac{q}{2}$ by $v_{nd}>0$ and $H>L$), so $n_L$ lies strictly below the joint optimum: $n^u(H,L) > n_L$.
By the symmetric argument, $n^u(H,L) < n_H$, giving $n^u(H,L) \in (n_L, n_H)$.

In the transferable utility framework, the stable matching exhibits positive assortative matching (PAM) --- whereby high-preference men tend to match with high-preference women --- if and only if the household surplus function is supermodular \citep{becker1973theory}.
In the two-type discrete setting, following \citet{chiappori2017partner}, supermodularity of the surplus corresponds to positivity of the supermodular core of the surplus matrix, which in the two-type case reduces to the single quantity $\Delta \equiv B(H,H) + B(L,L) - 2B(H,L)$: PAM holds if and only if $\Delta > 0$, and a larger $\Delta$ reflects a stronger degree of PAM as the assortative surplus premium is larger and cross-matching more costly in equilibrium.

The unitary model predicts PAM: $\Delta^U \equiv B^U(H,H) + B^U(L,L) - 2B^U(H,L) > 0$.
By the optimality of $n_H$ for the $(H,H)$ match and $n_L$ for the $(L,L)$ match, each evaluated at $n^u(H,L)$:
\begin{align}
    B^U(H,H) + B^U(L,L)
    &\geq A\bigl(n^u(H,L),\, H,\, H\bigr) + A\bigl(n^u(H,L),\, L,\, L\bigr) \nonumber\\
    &= 2\,A\bigl(n^u(H,L),\, H,\, L\bigr) = 2\,B^U(H,L),
    \label{eq:delta_u}
\end{align}
where the equality uses $A(n,H,H) + A(n,L,L) = 2A(n,H,L)$ for any $n$.
The inequality is strict since $n_H \neq n^u(H,L)$ and $n_L \neq n^u(H,L)$, confirming $\Delta^U > 0$.
The cost $q$ governs the level of optimal fertility through \eqref{eq:unitary_foc} but plays no role in determining the direction of PAM: the argument holds for any $q > 0$ and depends only on the shape of $v$.

The intuition is straightforward: children are a public good within the household, enjoyed by both spouses at a shared cost.
Matching two spouses with similarly strong preferences generates a surplus, because both partners benefit from the jointly chosen fertility level and neither is pulling against the other's preferred quantity.
A high-$d$ spouse matched with an equally high-$d$ partner enjoys more children at no extra cost relative to being matched with a low-$d$ partner, making same-preference matches strictly more valuable.

\subsection{Veto Model}
\label{sec:appendix_veto}

We now consider an alternative decision protocol in which children require the consent of both spouses, following the empirical findings of \citet{doepke2019bargaining}.
The timing is as follows.
In the first stage, each spouse independently announces their preferred number of children, and the household adopts the minimum of the two: $n^* = \min(n_m^*, n_f^*)$.
In the second stage, the spouses bargain over the allocation of consumption given $n^*$.
This veto structure captures the idea that a child is born only when both parents agree, which is a natural description of fertility decisions within marriage.

\paragraph{Consumption allocation.}
We assume that spouses can commit ex ante to the consumption allocation that will obtain after fertility is determined.
\footnote{An alternative assumption often made is limited commitment \citep{doepke2019bargaining}, where the outside option may bind after fertility choices are made. However, as we show below, the PAM prediction does not depend on this assumption: the cross-partial of the household value function is determined by the fertility choice of the binding spouse, which is independent of how cost is allocated upon divorce. We therefore adopt full commitment for simplicity, noting that the key sorting result carries through under limited commitment.}
Given $n^*$, spouses bargain over $(c_m, c_f)$ via symmetric Nash bargaining.
The outside option of each spouse $g$ is
\begin{align}
    \bar{U}_g = w_g - \frac{q\,n^*}{2} + v(n^*, d_g).
\end{align}
Since total consumption $w_m + w_f - qn^*$ equals the sum of outside options net of child utility, the cooperative surplus is zero and each spouse receives their outside option payoff.
The equilibrium consumption is therefore $c_g = w_g - \frac{qn^*}{2}$, giving total utility
\begin{align}
    U_g = w_g - \frac{q\,n^*}{2} + v(n^*, d_g). \label{eq:veto_utility}
\end{align}

\paragraph{Fertility choice.}
Anticipating \eqref{eq:veto_utility}, each spouse $g$ chooses their preferred fertility to maximize $-\frac{qn}{2} + v(n, d_g)$, yielding
\begin{align}
    v_n(n_g^*, d_g) = \frac{q}{2}. \label{eq:veto_foc}
\end{align}
Since \eqref{eq:veto_foc} coincides with the individual optimality condition, we have $n_g^* = n_{d_g}$.
In a same-type $(d,d)$ match, both spouses announce $n_d$ and the veto gives $n^* = n_d$.
In a cross-match, the lower-preference spouse is binding: in either $(H,L)$ or $(L,H)$ the $L$-type spouse announces $n_L < n_H$, so $n^* = n_L$, regardless of which gender draws type $L$.\footnote{This is consistent with the empirical pattern that, conditional on disagreement, the lower-preference spouse effectively vetoes higher fertility \citep{doepke2019bargaining}.}

\paragraph{Sorting.}
Let $B^V(d_m, d_f) \equiv A(n^*, d_m, d_f)$ denote the joint utility from children net of cost under the veto.
The four match values are
\begin{align}
    B^V(H,H) &= -qn_H + 2v(n_H,H), \label{eq:sv_hh}\\
    B^V(L,L) &= -qn_L + 2v(n_L,L), \label{eq:sv_ll}\\
    B^V(H,L) &= B^V(L,H) = -qn_L + v(n_L,H) + v(n_L,L). \label{eq:sv_hl}
\end{align}
The veto model predicts PAM: $\Delta^V \equiv B^V(H,H) + B^V(L,L) - 2B^V(H,L) > 0$.
Using \eqref{eq:sv_hh}--\eqref{eq:sv_hl}:
\begin{align}
    2\Delta^V = q(n_L - n_H) + 2\bigl[v(n_H,H) - v(n_L,H)\bigr] = 2\int_{n_L}^{n_H}\!\bigl[v_n(n,H) - \tfrac{q}{2}\bigr]\,dn \;>\; 0.
    \label{eq:delta_v}
\end{align}
Since $v_{nn} < 0$, $v_n(n,H)$ is strictly decreasing in $n$.
Because $v_n(n_H, H) = \frac{q}{2}$, it follows that $v_n(n,H) > \frac{q}{2}$ for all $n < n_H$, so the integrand is strictly positive on $(n_L, n_H)$.

\subsection{Relative Strength of Sorting}
\label{sec:appendix_strength}

Both models predict PAM.
We now show that the veto model predicts \emph{strictly stronger} PAM, with no conditions on $v$ beyond the maintained assumptions.
Define the veto loss of match $(d_m, d_f)$ as
\begin{align}
    \varepsilon(d_m, d_f) \;\equiv\; B^U(d_m,d_f) - B^V(d_m,d_f) \;\geq\; 0,
\end{align}
so that the gap between the supermodular cores of the two surplus matrices satisfies, following \citet{chiappori2017partner},
\begin{align}
    2(\Delta^V - \Delta^U) = \bigl[\varepsilon(H,L) + \varepsilon(L,H)\bigr] - \bigl[\varepsilon(H,H) + \varepsilon(L,L)\bigr].
    \label{eq:gap}
\end{align}
Veto predicts strictly stronger PAM if and only if cross-matches incur larger total veto losses than assortative matches.

The key observation is that assortative matches carry zero veto loss: $\varepsilon(H,H) = \varepsilon(L,L) = 0$.
In a same-type $(d,d)$ match, both spouses announce the same individually optimal fertility $n_d$, so the veto gives $n^* = n_d$ --- the same outcome as joint maximization.
There is no within-couple disagreement, the veto protocol never binds, and no utility is foregone.
Substituting into \eqref{eq:gap}:
\begin{align}
    2(\Delta^V - \Delta^U) = \varepsilon(H,L) + \varepsilon(L,H) = 2\,\varepsilon(H,L) > 0,
    \label{eq:main}
\end{align}
where $\varepsilon(H,L) = B^U(H,L) - A(n_L, H, L) > 0$ since $n^u(H,L) > n_L$, so the cross-match strictly benefits from joint optimization.
The magnitude of the gap equals
\begin{align}
    \Delta^V - \Delta^U = \varepsilon(H,L) = B^U(H,L) - A\!\left(n_L,\, H,\, L\right),
    \label{eq:gap_explicit}
\end{align}
the gain from moving from the veto fertility $n_L$ to the jointly optimal fertility $n^u(H,L)$ in a cross-match.

The result reflects a fundamental asymmetry between assortative and cross-matches under the veto.
Joint maximization raises the joint utility from children net of cost exclusively in cross-matches, leaving the already-optimal assortative outcomes unchanged, and thereby selectively reduces the cost of mismatching.
The veto institution therefore predicts not only PAM but a measurably stronger degree of sorting on fertility preferences than the unitary model, with the magnitude of the difference given explicitly by $\varepsilon(H,L)$ in \eqref{eq:gap_explicit}.

\end{document}

%% file: figuretable/labor_family_economics_project/summary_statistics_action_male_female_2024.tex
\begin{tabular}[t]{llrrrrrrrr}
\toprule
Gender &   & N & Mean & SD & Min & P25 & Median & P75 & Max\\
\midrule
Male & Application & 56915 & 119.93 & 210.58 & 1.00 & 16.00 & 55.00 & 145.00 & 15788.00\\
 & Pre-relation & 56915 & 2.28 & 3.25 & 0.00 & 0.00 & 1.00 & 3.00 & 50.00\\
 & Serious-relation & 56915 & 0.21 & 0.43 & 0.00 & 0.00 & 0.00 & 0.00 & 3.00\\
 & Proposal & 56915 & 0.10 & 0.31 & 0.00 & 0.00 & 0.00 & 0.00 & 1.00\\
Female & Application & 60064 & 113.64 & 134.29 & 1.00 & 29.00 & 71.00 & 149.00 & 2376.00\\
 & Pre-relation & 60064 & 2.16 & 2.92 & 0.00 & 0.00 & 1.00 & 3.00 & 39.00\\
 & Serious-relation & 60064 & 0.20 & 0.42 & 0.00 & 0.00 & 0.00 & 0.00 & 3.00\\
 & Proposal & 60064 & 0.10 & 0.30 & 0.00 & 0.00 & 0.00 & 0.00 & 1.00\\
\bottomrule
\end{tabular}

%% file: figuretable/labor_family_economics_project/summary_statistics_matched_continuous_2024.tex
\begin{tabular}[t]{llrrrrrr}
\toprule
Gender &   & N & mean & median & sd & min & max\\
\midrule
Female & Age & 7562 & 36.01 & 35.00 & 7.15 & 23.00 & 75.00\\
 & Income (upper limit) & 7511 & 503.62 & 500.00 & 208.45 & 300.00 & 2100.00\\
 & Height & 7562 & 158.67 & 158.00 & 5.36 & 142.00 & 180.00\\
 & Weight & 7562 & 49.40 & 50.00 & 7.38 & 35.00 & 95.00\\
 & Flexibility & 7554 & -0.21 & -0.24 & 0.49 & -1.45 & 1.12\\
Male & Age & 7561 & 39.06 & 38.00 & 8.02 & 23.00 & 79.00\\
 & Income (upper limit) & 7544 & 782.91 & 700.00 & 347.35 & 300.00 & 2100.00\\
 & Height & 7561 & 171.65 & 172.00 & 5.75 & 152.00 & 196.00\\
 & Weight & 7561 & 65.19 & 65.00 & 9.88 & 40.00 & 95.00\\
 & Flexibility & 7544 & -0.21 & -0.24 & 0.35 & -1.45 & 0.90\\
\bottomrule
\end{tabular}

%% file: figuretable/labor_family_economics_project/summary_statistics_matched_discrete_2024.tex
\begin{tabular}[t]{llrrrr}
\toprule
\multicolumn{2}{c}{ } & \multicolumn{2}{c}{Female} & \multicolumn{2}{c}{Male} \\
\cmidrule(l{3pt}r{3pt}){3-4} \cmidrule(l{3pt}r{3pt}){5-6}
  &    & N & Pct. & N & Pct.\\
\midrule
Educational level & (1)JuniorHigh & 11 & 0.1 & 21 & 0.3\\
 & (2)HighSchool & 604 & 8.0 & 763 & 10.1\\
 & (3)Vocational & 1709 & 22.6 & 619 & 8.2\\
 & (4)Undergraduate & 4698 & 62.0 & 4207 & 55.5\\
 & (5)Graduate & 504 & 6.7 & 1834 & 24.2\\
 & NA & 51 & 0.7 & 133 & 1.8\\
Drink Alcohol level & (1)Never & 1549 & 20.4 & 912 & 12.0\\
 & (2)SocialOnly & 4838 & 63.9 & 4661 & 61.5\\
 & (3)DrinkRegularly & 1175 & 15.5 & 1988 & 26.2\\
 & NA & 15 & 0.2 & 16 & 0.2\\
Smoking level & (1)Never & 7504 & 99.0 & 7092 & 93.6\\
 & (2)Occasionally & 35 & 0.5 & 210 & 2.8\\
 & (3)Regularly & 23 & 0.3 & 259 & 3.4\\
 & NA & 15 & 0.2 & 16 & 0.2\\
Marital History Dummy & (1)NeverMarried & 6695 & 88.4 & 6426 & 84.8\\
 & (2)DivorcedOrWidowed & 867 & 11.4 & 1135 & 15.0\\
 & NA & 15 & 0.2 & 16 & 0.2\\
Housework Share Level & (1)LeaveToPartner & 2 & 0.0 & 20 & 0.3\\
 & (2)DiscussWithPartner & 2227 & 29.4 & 1596 & 21.1\\
 & (3)ShareEqually & 4247 & 56.1 & 5326 & 70.3\\
 & (4)DoItMyself & 693 & 9.1 & 352 & 4.6\\
 & NA & 408 & 5.4 & 283 & 3.7\\
Child Care Share Level & (1)DoNotWantChildren & 250 & 3.3 & 134 & 1.8\\
 & (2)LeaveToPartner & 0 & 0.0 & 4 & 0.1\\
 & (3)DiscussWithPartner & 2000 & 26.4 & 1702 & 22.5\\
 & (4)ShareEqually & 4297 & 56.7 & 5107 & 67.4\\
 & (5)DoItMyself & 505 & 6.7 & 306 & 4.0\\
 & NA & 525 & 6.9 & 324 & 4.3\\
Desired Child Dummy & (1)DoNotWant & 493 & 6.5 & 256 & 3.4\\
 & (2)NoPreference & 1652 & 21.8 & 1950 & 25.7\\
 & (3)Want & 5417 & 71.5 & 5355 & 70.7\\
 & NA & 15 & 0.2 & 16 & 0.2\\
\bottomrule
\end{tabular}

%% file: figuretable/labor_family_economics_project/correlation_matrix_matched_male.tex
\begin{tabular}[t]{lrrrrrrrrrrrr}
\toprule
  & Educ & Age & Income & Flex & Height & Weight & Drink & Smoke & M Hist & House & C Care & C Pref\\
\midrule
Educ & 1.000 &  &  &  &  &  &  &  &  &  &  & \\
Age & -0.185 & 1.000 &  &  &  &  &  &  &  &  &  & \\
Income & 0.145 & 0.298 & 1.000 &  &  &  &  &  &  &  &  & \\
Flex & -0.028 & -0.008 & -0.137 & 1.000 &  &  &  &  &  &  &  & \\
Height & 0.017 & -0.032 & 0.052 & -0.011 & 1.000 &  &  &  &  &  &  & \\
Weight & -0.050 & 0.081 & 0.054 & -0.017 & 0.476 & 1.000 &  &  &  &  &  & \\
Drink & 0.031 & 0.027 & 0.093 & 0.000 & 0.036 & 0.073 & 1.000 &  &  &  &  & \\
Smoke & -0.135 & 0.078 & 0.014 & 0.006 & 0.019 & 0.063 & 0.073 & 1.000 &  &  &  & \\
M Hist & -0.104 & 0.439 & 0.231 & -0.029 & 0.025 & 0.012 & 0.037 & 0.066 & 1.000 &  &  & \\
House & 0.044 & -0.131 & -0.132 & 0.045 & -0.003 & -0.054 & -0.004 & -0.059 & -0.075 & 1.000 &  & \\
C Care & 0.101 & -0.319 & -0.094 & 0.012 & 0.021 & -0.029 & -0.013 & -0.067 & -0.184 & 0.567 & 1.000 & \\
C Pref & 0.174 & -0.524 & -0.057 & -0.032 & 0.040 & -0.015 & -0.013 & -0.065 & -0.345 & 0.070 & 0.357 & 1.000\\
\bottomrule
\end{tabular}

%% file: figuretable/labor_family_economics_project/correlation_matrix_matched_female.tex
\begin{tabular}[t]{lrrrrrrrrrrrr}
\toprule
  & Educ & Age & Income & Flex & Height & Weight & Drink & Smoke & M Hist & House & C Care & C Pref\\
\midrule
Educ & 1.000 &  &  &  &  &  &  &  &  &  &  & \\
Age & -0.239 & 1.000 &  &  &  &  &  &  &  &  &  & \\
Income & 0.271 & 0.071 & 1.000 &  &  &  &  &  &  &  &  & \\
Flex & -0.095 & 0.093 & -0.214 & 1.000 &  &  &  &  &  &  &  & \\
Height & 0.038 & 0.018 & 0.067 & -0.006 & 1.000 &  &  &  &  &  &  & \\
Weight & -0.064 & 0.049 & -0.002 & 0.006 & 0.422 & 1.000 &  &  &  &  &  & \\
Drink & 0.052 & -0.005 & 0.127 & -0.032 & 0.053 & 0.062 & 1.000 &  &  &  &  & \\
Smoke & -0.082 & 0.059 & -0.020 & 0.014 & -0.001 & 0.024 & 0.027 & 1.000 &  &  &  & \\
M Hist & -0.165 & 0.424 & -0.014 & 0.058 & -0.007 & -0.039 & -0.012 & 0.053 & 1.000 &  &  & \\
House & 0.024 & -0.109 & -0.013 & 0.029 & -0.003 & -0.011 & -0.007 & -0.027 & -0.059 & 1.000 &  & \\
C Care & 0.134 & -0.395 & 0.014 & -0.043 & -0.008 & -0.012 & 0.025 & -0.041 & -0.246 & 0.457 & 1.000 & \\
C Pref & 0.212 & -0.606 & 0.013 & -0.100 & 0.008 & -0.023 & 0.014 & -0.066 & -0.397 & 0.095 & 0.470 & 1.000\\
\bottomrule
\end{tabular}

%% file: figuretable/labor_family_economics_project/dupuy_galichon_affinity_matrix_2024.txt
	& Education	& Age	& Income	& Flexibility	& Height	& Weight	& Drink	& Smoke	& Marital History	& Housework	& Childcare	& Child	\\\midrule

Education	& $\mathbf{0.17}$	& $\mathbf{-0.10}$	& $\mathbf{0.11}$	& $0.01$	& $0.03$	& $\mathbf{-0.05}$	& $0.02$	& $-0.02$	& $\mathbf{-0.03}$	& $0.02$	& $0.00$	& $0.00$	\\

	& (0.01) 	& (0.03) 	& (0.02) 	& (0.01) 	& (0.01) 	& (0.01) 	& (0.01) 	& (0.01) 	& (0.01) 	& (0.02) 	& (0.02) 	& (0.02) 	\\

Age	& $\mathbf{-0.08}$	& $\mathbf{3.55}$	& $\mathbf{-0.21}$	& $0.03$	& $-0.03$	& $\mathbf{0.11}$	& $\mathbf{-0.06}$	& $\mathbf{0.08}$	& $\mathbf{0.08}$	& $0.03$	& $\mathbf{-0.08}$	& $-0.03$	\\

	& (0.03) 	& (0.07) 	& (0.03) 	& (0.03) 	& (0.03) 	& (0.03) 	& (0.03) 	& (0.03) 	& (0.03) 	& (0.03) 	& (0.03) 	& (0.03) 	\\

Income	& $\mathbf{0.07}$	& $\mathbf{-0.74}$	& $\mathbf{0.20}$	& $-0.01$	& $\mathbf{0.12}$	& $\mathbf{-0.27}$	& $\mathbf{0.04}$	& $-0.01$	& $\mathbf{0.06}$	& $-0.01$	& $0.03$	& $0.02$	\\

	& (0.02) 	& (0.03) 	& (0.01) 	& (0.02) 	& (0.02) 	& (0.02) 	& (0.02) 	& (0.01) 	& (0.01) 	& (0.02) 	& (0.02) 	& (0.02) 	\\

Flexibility	& $0.01$	& $\mathbf{0.10}$	& $0.00$	& $\mathbf{0.03}$	& $0.01$	& $0.00$	& $0.01$	& $-0.01$	& $-0.01$	& $-0.01$	& $-0.00$	& $0.00$	\\

	& (0.01) 	& (0.03) 	& (0.01) 	& (0.01) 	& (0.01) 	& (0.01) 	& (0.01) 	& (0.01) 	& (0.01) 	& (0.01) 	& (0.02) 	& (0.02) 	\\

Height	& $0.03$	& $\mathbf{-0.23}$	& $\mathbf{0.08}$	& $0.01$	& $\mathbf{0.21}$	& $\mathbf{-0.13}$	& $0.01$	& $\mathbf{-0.03}$	& $0.02$	& $-0.00$	& $-0.01$	& $\mathbf{0.06}$	\\

	& (0.02) 	& (0.03) 	& (0.02) 	& (0.01) 	& (0.02) 	& (0.02) 	& (0.01) 	& (0.01) 	& (0.02) 	& (0.02) 	& (0.02) 	& (0.02) 	\\

Weight	& $-0.03$	& $\mathbf{0.24}$	& $\mathbf{-0.06}$	& $-0.00$	& $\mathbf{-0.07}$	& $\mathbf{0.20}$	& $0.01$	& $0.02$	& $\mathbf{-0.04}$	& $-0.00$	& $0.03$	& $-0.03$	\\

	& (0.02) 	& (0.03) 	& (0.02) 	& (0.01) 	& (0.02) 	& (0.02) 	& (0.01) 	& (0.01) 	& (0.02) 	& (0.02) 	& (0.02) 	& (0.02) 	\\

Drink	& $0.02$	& $-0.03$	& $0.03$	& $0.02$	& $0.01$	& $-0.03$	& $\mathbf{0.10}$	& $-0.01$	& $0.01$	& $-0.00$	& $0.02$	& $0.01$	\\

	& (0.01) 	& (0.03) 	& (0.01) 	& (0.01) 	& (0.01) 	& (0.01) 	& (0.01) 	& (0.01) 	& (0.01) 	& (0.01) 	& (0.02) 	& (0.02) 	\\

Smoke	& $\mathbf{-0.03}$	& $\mathbf{0.12}$	& $-0.02$	& $0.01$	& $0.01$	& $-0.02$	& $0.01$	& $\mathbf{0.03}$	& $\mathbf{0.03}$	& $-0.01$	& $0.03$	& $-0.00$	\\

	& (0.01) 	& (0.03) 	& (0.02) 	& (0.01) 	& (0.01) 	& (0.01) 	& (0.01) 	& (0.01) 	& (0.01) 	& (0.01) 	& (0.02) 	& (0.02) 	\\

Marital History	& $\mathbf{-0.06}$	& $\mathbf{0.11}$	& $0.02$	& $-0.02$	& $0.02$	& $\mathbf{-0.04}$	& $\mathbf{0.04}$	& $\mathbf{0.02}$	& $\mathbf{0.12}$	& $0.02$	& $-0.02$	& $\mathbf{-0.04}$	\\

	& (0.01) 	& (0.03) 	& (0.01) 	& (0.01) 	& (0.02) 	& (0.02) 	& (0.01) 	& (0.01) 	& (0.01) 	& (0.02) 	& (0.02) 	& (0.02) 	\\

Housework	& $0.01$	& $-0.03$	& $0.03$	& $0.00$	& $0.01$	& $-0.01$	& $-0.01$	& $-0.02$	& $0.03$	& $0.02$	& $0.00$	& $\mathbf{-0.06}$	\\

	& (0.02) 	& (0.03) 	& (0.02) 	& (0.02) 	& (0.02) 	& (0.02) 	& (0.02) 	& (0.01) 	& (0.02) 	& (0.02) 	& (0.02) 	& (0.02) 	\\

Childcare	& $0.00$	& $-0.05$	& $-0.01$	& $0.01$	& $-0.02$	& $0.03$	& $0.01$	& $-0.00$	& $-0.00$	& $-0.03$	& $0.02$	& $\mathbf{0.05}$	\\

	& (0.02) 	& (0.03) 	& (0.02) 	& (0.02) 	& (0.02) 	& (0.02) 	& (0.02) 	& (0.01) 	& (0.02) 	& (0.02) 	& (0.02) 	& (0.02) 	\\

Child	& $-0.01$	& $\mathbf{-0.29}$	& $-0.01$	& $-0.01$	& $0.02$	& $-0.02$	& $\mathbf{-0.04}$	& $-0.01$	& $-0.01$	& $\mathbf{-0.04}$	& $\mathbf{0.06}$	& $\mathbf{0.25}$	\\

	& (0.02) 	& (0.03) 	& (0.02) 	& (0.02) 	& (0.02) 	& (0.02) 	& (0.02) 	& (0.01) 	& (0.02) 	& (0.02) 	& (0.02) 	& (0.02) 	\\

%% file: figuretable/labor_family_economics_project/dupuy_galichon_saliency_2024.txt
	& Index 1	& Index 2	& Index 3	& Index 4	& Index 1	& Index 2	& Index 3	& Index 4	\\\midrule

Education	& $\mathbf{0.03}$	& $\mathbf{0.26}$	& $\mathbf{0.10}$	& $\mathbf{0.79}$	& $\mathbf{0.03}$	& $\mathbf{0.20}$	& $\mathbf{0.09}$	& $\mathbf{0.78}$	\\

	& $(0.00)$	& $(0.02)$	& $(0.03)$	& $(0.02)$	& $(0.01)$	& $(0.02)$	& $(0.03)$	& $(0.03)$	\\

Age	& $\mathbf{-0.97}$	& $\mathbf{0.21}$	& $\mathbf{-0.08}$	& $0.01$	& $\mathbf{-0.99}$	& $\mathbf{0.08}$	& $-0.01$	& $\mathbf{0.04}$	\\

	& $(0.00)$	& $(0.01)$	& $(0.01)$	& $(0.01)$	& $(0.00)$	& $(0.01)$	& $(0.01)$	& $(0.01)$	\\

Income	& $\mathbf{0.21}$	& $\mathbf{0.65}$	& $\mathbf{0.08}$	& $\mathbf{-0.07}$	& $\mathbf{0.07}$	& $\mathbf{0.40}$	& $\mathbf{0.17}$	& $\mathbf{0.26}$	\\

	& $(0.01)$	& $(0.02)$	& $(0.03)$	& $(0.03)$	& $(0.01)$	& $(0.03)$	& $(0.04)$	& $(0.04)$	\\

Flexibility	& $\mathbf{-0.03}$	& $0.03$	& $-0.02$	& $\mathbf{0.13}$	& $-0.01$	& $0.02$	& $-0.01$	& $\mathbf{0.13}$	\\

	& $(0.01)$	& $(0.02)$	& $(0.04)$	& $(0.04)$	& $(0.01)$	& $(0.03)$	& $(0.04)$	& $(0.04)$	\\

Height	& $\mathbf{0.07}$	& $\mathbf{0.49}$	& $\mathbf{-0.19}$	& $\mathbf{-0.06}$	& $\mathbf{0.02}$	& $\mathbf{0.47}$	& $\mathbf{-0.12}$	& $\mathbf{-0.08}$	\\

	& $(0.01)$	& $(0.02)$	& $(0.02)$	& $(0.03)$	& $(0.01)$	& $(0.02)$	& $(0.03)$	& $(0.03)$	\\

Weight	& $\mathbf{-0.07}$	& $\mathbf{-0.43}$	& $0.03$	& $\mathbf{0.14}$	& $\mathbf{-0.05}$	& $\mathbf{-0.71}$	& $-0.01$	& $\mathbf{0.19}$	\\

	& $(0.01)$	& $(0.02)$	& $(0.03)$	& $(0.03)$	& $(0.01)$	& $(0.01)$	& $(0.02)$	& $(0.02)$	\\

Drink	& $\mathbf{0.01}$	& $\mathbf{0.11}$	& $0.03$	& $\mathbf{0.09}$	& $\mathbf{0.02}$	& $\mathbf{0.08}$	& $\mathbf{0.17}$	& $-0.01$	\\

	& $(0.01)$	& $(0.02)$	& $(0.03)$	& $(0.03)$	& $(0.01)$	& $(0.02)$	& $(0.03)$	& $(0.03)$	\\

Smoke	& $\mathbf{-0.03}$	& $0.04$	& $-0.01$	& $\mathbf{-0.21}$	& $\mathbf{-0.02}$	& $-0.04$	& $0.04$	& $\mathbf{-0.14}$	\\

	& $(0.01)$	& $(0.02)$	& $(0.03)$	& $(0.04)$	& $(0.01)$	& $(0.02)$	& $(0.04)$	& $(0.04)$	\\

Marital History	& $\mathbf{-0.03}$	& $\mathbf{0.14}$	& $\mathbf{0.24}$	& $\mathbf{-0.53}$	& $\mathbf{-0.02}$	& $\mathbf{0.20}$	& $\mathbf{0.14}$	& $\mathbf{-0.49}$	\\

	& $(0.01)$	& $(0.02)$	& $(0.03)$	& $(0.03)$	& $(0.01)$	& $(0.02)$	& $(0.03)$	& $(0.03)$	\\

Housework	& $0.01$	& $\mathbf{0.06}$	& $\mathbf{0.22}$	& $0.01$	& $-0.01$	& $0.03$	& $\mathbf{0.19}$	& $0.04$	\\

	& $(0.01)$	& $(0.02)$	& $(0.04)$	& $(0.04)$	& $(0.01)$	& $(0.03)$	& $(0.04)$	& $(0.05)$	\\

Childcare	& $\mathbf{0.01}$	& $\mathbf{-0.08}$	& $\mathbf{-0.20}$	& $0.05$	& $\mathbf{0.02}$	& $-0.03$	& $\mathbf{-0.19}$	& $0.04$	\\

	& $(0.01)$	& $(0.03)$	& $(0.04)$	& $(0.05)$	& $(0.01)$	& $(0.03)$	& $(0.04)$	& $(0.04)$	\\

Child	& $\mathbf{0.08}$	& $0.02$	& $\mathbf{-0.89}$	& $-0.05$	& $\mathbf{0.02}$	& $\mathbf{0.10}$	& $\mathbf{-0.91}$	& $0.05$	\\

	& $(0.01)$	& $(0.02)$	& $(0.02)$	& $(0.03)$	& $(0.01)$	& $(0.02)$	& $(0.02)$	& $(0.03)$	\\

\midrule Index share	& $\mathbf{0.72}$	& $\mathbf{0.09}$	& $\mathbf{0.06}$	& $\mathbf{0.04}$	& $\mathbf{0.72}$	& $\mathbf{0.09}$	& $\mathbf{0.06}$	& $\mathbf{0.04}$	\\

	& $(0.01)$	& $(0.00)$	& $(0.00)$	& $(0.00)$	& $(0.01)$	& $(0.00)$	& $(0.00)$	& $(0.00)$	\\

%% file: figuretable/labor_family_economics_project/Dupuy_Galichon_affinity_matrix_different_set_of_attributes_2024.tex
	& (1)	& (2)	& (3)	& (4)	& (5)	& (6)	& (7)	& (8)	& (9)	\\\midrule

Education	& $0.25$	& 	& 	& $0.22$	& $0.22$	& 	& $0.18$	& $0.17$	& $0.17$	\\

	& (0.01) 	& 	& 	& (0.01) 	& (0.01) 	& 	& (0.01) 	& (0.01) 	& (0.01) 	\\

Age	& 	& $2.98$	& 	& $2.98$	& 	& $3.61$	& $3.58$	& $3.58$	& $3.55$	\\

	& 	& (0.06) 	& 	& (0.06) 	& 	& (0.07) 	& (0.07) 	& (0.07) 	& (0.07) 	\\

Income	& 	& 	& $0.17$	& 	& $0.16$	& $0.22$	& $0.20$	& $0.19$	& $0.20$	\\

	& 	& 	& (0.01) 	& 	& (0.01) 	& (0.01) 	& (0.01) 	& (0.01) 	& (0.01) 	\\

Flexibility	& 	& 	& 	& 	& 	& 	& 	& $0.03$	& $0.03$	\\

	& 	& 	& 	& 	& 	& 	& 	& (0.01) 	& (0.01) 	\\

Height	& 	& 	& 	& 	& 	& 	& 	& $0.21$	& $0.21$	\\

	& 	& 	& 	& 	& 	& 	& 	& (0.02) 	& (0.02) 	\\

Weight	& 	& 	& 	& 	& 	& 	& 	& $0.20$	& $0.20$	\\

	& 	& 	& 	& 	& 	& 	& 	& (0.02) 	& (0.02) 	\\

Drink	& 	& 	& 	& 	& 	& 	& 	& $0.10$	& $0.10$	\\

	& 	& 	& 	& 	& 	& 	& 	& (0.01) 	& (0.01) 	\\

Smoke	& 	& 	& 	& 	& 	& 	& 	& $0.03$	& $0.03$	\\

	& 	& 	& 	& 	& 	& 	& 	& (0.01) 	& (0.01) 	\\

Marital History	& 	& 	& 	& 	& 	& 	& 	& $0.13$	& $0.12$	\\

	& 	& 	& 	& 	& 	& 	& 	& (0.01) 	& (0.01) 	\\

Housework	& 	& 	& 	& 	& 	& 	& 	& 	& $0.02$	\\

	& 	& 	& 	& 	& 	& 	& 	& 	& (0.02) 	\\

Childcare	& 	& 	& 	& 	& 	& 	& 	& 	& $0.02$	\\

	& 	& 	& 	& 	& 	& 	& 	& 	& (0.02) 	\\

Child	& 	& 	& 	& 	& 	& 	& 	& 	& $0.25$	\\

	& 	& 	& 	& 	& 	& 	& 	& 	& (0.02) 	\\\bottomrule

%% file: figuretable/labor_family_economics_project/affinity_matrix_2015_2024_rescale_CW.tex
\begin{tabular}{llclclclclc}
\toprule
  & 2015 & 2016 & 2017 & 2018 & 2019 & 2020 & 2021 & 2022 & 2023 & 2024\\
\midrule
Education & 0.09 & 0.11 & 0.07 & 0.08 & 0.07 & 0.07 & 0.06 & 0.06 & 0.05 & 0.05\\
Age & 0.93 & 0.92 & 0.92 & 0.93 & 0.93 & 0.94 & 0.95 & 0.95 & 0.95 & 0.95\\
Income & 0.04 & 0.05 & 0.04 & 0.04 & 0.05 & 0.06 & 0.05 & 0.04 & 0.04 & 0.05\\
Flexibility & 0.01 & 0.00 & 0.01 & 0.02 & 0.01 & 0.02 & 0.00 & 0.01 & 0.01 & 0.01\\
Height & 0.05 & 0.07 & 0.08 & 0.07 & 0.06 & 0.06 & 0.05 & 0.05 & 0.06 & 0.06\\
Weight & 0.06 & 0.05 & 0.05 & 0.05 & 0.06 & 0.07 & 0.05 & 0.05 & 0.05 & 0.05\\
Drink & 0.01 & 0.03 & 0.02 & 0.02 & 0.03 & 0.02 & 0.02 & 0.03 & 0.02 & 0.03\\
Smoke & 0.01 & 0.03 & 0.01 & 0.01 & 0.01 & 0.01 & 0.01 & 0.01 & 0.01 & 0.01\\
Marital History & 0.09 & 0.07 & 0.07 & 0.05 & 0.06 & 0.05 & 0.04 & 0.05 & 0.04 & 0.03\\
Child & 0.07 & 0.07 & 0.07 & 0.07 & 0.06 & 0.07 & 0.08 & 0.07 & 0.07 & 0.07\\
\bottomrule
\end{tabular}

%% file: figuretable/labor_family_economics_project/affinity_matrix_sigma_2015_2024_rescale_CW.tex
\begin{tabular}{llclclclclc}
\toprule
  & 2015 & 2016 & 2017 & 2018 & 2019 & 2020 & 2021 & 2022 & 2023 & 2024\\
\midrule
Sigma & 0.27 & 0.32 & 0.34 & 0.32 & 0.33 & 0.3 & 0.27 & 0.26 & 0.25 & 0.27\\
\bottomrule
\end{tabular}

%% file: figuretable/labor_family_economics_project/maximum_score_diagonal_all_stage.tex
\begin{tabular}[t]{lllllllll}
\toprule
\multicolumn{1}{c}{ } & \multicolumn{2}{c}{Application} & \multicolumn{2}{c}{Pre-Relation} & \multicolumn{2}{c}{Serious-Relation} & \multicolumn{2}{c}{Proposal} \\
\cmidrule(l{3pt}r{3pt}){2-3} \cmidrule(l{3pt}r{3pt}){4-5} \cmidrule(l{3pt}r{3pt}){6-7} \cmidrule(l{3pt}r{3pt}){8-9}
Variable & Mean & 95\% CI & Mean & 95\% CI & Mean & 95\% CI & Mean & 95\% CI\\
\midrule
Education & 1.00 & {}[1.00, 1.00] & 1.00 & {}[1.00, 1.00] & 1.00 & {}[1.00, 1.00] & 1.00 & {}[1.00, 1.00]\\
Age & 9.24 & {}[7.64, 9.97] & 9.36 & {}[7.86, 9.99] & 9.31 & {}[7.92, 9.97] & 9.27 & {}[8.24, 9.99]\\
Income & 4.24 & {}[1.38, 7.46] & 4.54 & {}[1.12, 8.49] & 4.08 & {}[0.33, 7.74] & 4.42 & {}[1.79, 7.46]\\
Flexibility & -0.24 & {}[-2.22, 1.85] & 1.04 & {}[-0.72, 3.02] & 0.41 & {}[-1.10, 2.46] & -0.34 & {}[-1.99, 1.04]\\
Height & 1.56 & {}[-0.03, 3.93] & 1.14 & {}[-0.82, 2.97] & 0.69 & {}[-1.06, 2.52] & 1.12 & {}[-0.70, 2.79]\\
Weight & 1.60 & {}[-0.72, 4.00] & 0.94 & {}[-0.56, 2.73] & 1.02 & {}[-0.15, 3.40] & 1.68 & {}[-0.08, 3.53]\\
Drink & 0.35 & {}[-1.49, 2.11] & 1.12 & {}[-0.54, 3.67] & 0.61 & {}[-0.91, 3.01] & 1.25 & {}[-0.61, 3.89]\\
Smoke & 0.83 & {}[-8.88, 9.75] & -0.62 & {}[-8.03, 9.07] & 3.24 & {}[-5.10, 9.50] & 3.45 & {}[-4.35, 9.14]\\
Marital History & 3.79 & {}[-1.15, 9.01] & 4.57 & {}[0.49, 8.52] & 5.46 & {}[1.92, 9.59] & 5.24 & {}[0.93, 9.53]\\
Housework & 0.06 & {}[-1.72, 1.90] & -0.26 & {}[-2.12, 1.51] & 0.71 & {}[-1.78, 3.51] & -0.65 & {}[-3.60, 1.59]\\
Childcare & 0.89 & {}[-1.98, 2.90] & 1.00 & {}[-1.85, 3.64] & -0.23 & {}[-3.95, 2.96] & 0.86 & {}[-2.50, 4.83]\\
Child & 3.48 & {}[-0.74, 8.90] & 4.86 & {}[1.17, 9.05] & 4.01 & {}[0.80, 9.13] & 4.86 & {}[0.53, 9.50]\\
\bottomrule
\end{tabular}

%% file: figuretable/labor_family_economics_project/summary_statistics_unmatched_continuous_2024.tex
\begin{tabular}[t]{llrrrrrr}
\toprule
Gender &   & N & mean & median & sd & min & max\\
\midrule
Female & Age & 19885 & 37.39 & 36.00 & 8.42 & 20.00 & 85.00\\
 & Income (upper limit) & 19631 & 499.08 & 500.00 & 220.23 & 300.00 & 2100.00\\
 & Height & 19883 & 158.88 & 158.00 & 5.28 & 142.00 & 182.00\\
 & Weight & 19881 & 49.42 & 50.00 & 7.13 & 35.00 & 95.00\\
 & Flexibility & 19824 & -0.18 & -0.19 & 0.50 & -1.45 & 1.12\\
Male & Age & 14307 & 41.45 & 40.00 & 9.68 & 20.00 & 91.00\\
 & Income (upper limit) & 14263 & 700.05 & 600.00 & 345.59 & 300.00 & 2100.00\\
 & Height & 14306 & 171.17 & 170.00 & 5.84 & 142.00 & 196.00\\
 & Weight & 14306 & 66.33 & 65.00 & 10.65 & 35.00 & 95.00\\
 & Flexibility & 14237 & -0.19 & -0.21 & 0.35 & -1.45 & 0.90\\
\bottomrule
\end{tabular}

%% file: figuretable/labor_family_economics_project/summary_statistics_unmatched_discrete_2024.tex
\begin{tabular}[t]{llrrrr}
\toprule
\multicolumn{2}{c}{ } & \multicolumn{2}{c}{Female} & \multicolumn{2}{c}{Male} \\
\cmidrule(l{3pt}r{3pt}){3-4} \cmidrule(l{3pt}r{3pt}){5-6}
  &    & N & Pct. & N & Pct.\\
\midrule
Educational level & (1)JuniorHigh & 60 & 0.3 & 125 & 0.9\\
 & (2)HighSchool & 1973 & 9.9 & 2661 & 18.5\\
 & (3)Vocational & 4845 & 24.3 & 1719 & 12.0\\
 & (4)Undergraduate & 11612 & 58.2 & 7514 & 52.3\\
 & (5)Graduate & 1285 & 6.4 & 2105 & 14.6\\
 & NA & 194 & 1.0 & 247 & 1.7\\
Drink Alcohol level & (1)Never & 4176 & 20.9 & 2223 & 15.5\\
 & (2)SocialOnly & 12983 & 65.0 & 8876 & 61.8\\
 & (3)DrinkRegularly & 2715 & 13.6 & 3200 & 22.3\\
 & NA & 95 & 0.5 & 72 & 0.5\\
Smoking level & (1)Never & 19615 & 98.2 & 12716 & 88.5\\
 & (2)Occasionally & 126 & 0.6 & 597 & 4.2\\
 & (3)Regularly & 133 & 0.7 & 986 & 6.9\\
 & NA & 95 & 0.5 & 72 & 0.5\\
Marital History Dummy & (1)NeverMarried & 17220 & 86.2 & 11844 & 82.4\\
 & (2)DivorcedOrWidowed & 2657 & 13.3 & 2456 & 17.1\\
 & NA & 92 & 0.5 & 71 & 0.5\\
Housework Share Level & (1)LeaveToPartner & 17 & 0.1 & 102 & 0.7\\
 & (2)DiscussWithPartner & 5742 & 28.8 & 3364 & 23.4\\
 & (3)ShareEqually & 10174 & 50.9 & 8316 & 57.9\\
 & (4)DoItMyself & 1553 & 7.8 & 553 & 3.8\\
 & NA & 2483 & 12.4 & 2036 & 14.2\\
Child Care Share Level & (1)DoNotWantChildren & 953 & 4.8 & 340 & 2.4\\
 & (2)LeaveToPartner & 6 & 0.0 & 43 & 0.3\\
 & (3)DiscussWithPartner & 5257 & 26.3 & 3445 & 24.0\\
 & (4)ShareEqually & 9679 & 48.5 & 7846 & 54.6\\
 & (5)DoItMyself & 1197 & 6.0 & 532 & 3.7\\
 & NA & 2877 & 14.4 & 2165 & 15.1\\
Desired Child Dummy & (1)DoNotWant & 1887 & 9.4 & 695 & 4.8\\
 & (2)NoPreference & 5047 & 25.3 & 4659 & 32.4\\
 & (3)Want & 12923 & 64.7 & 8912 & 62.0\\
 & NA & 112 & 0.6 & 105 & 0.7\\
\bottomrule
\end{tabular}

%% file: figuretable/labor_family_economics_project/withdraw_reason_2023_male_female.tex
\begin{tabular}{lrr}
\toprule
Withdrawal reason & Male & Female\\
\midrule
Married within IBJ & 6707 (38.6\%) & 6722 (28.8\%)\\
Married outside IBJ & 2281 (13.1\%) & 3724 (16.0\%)\\
Regular withdrawal & 8318 (47.8\%) & 12724 (54.5\%)\\
Withdrawn on agency deletion & 28 (0.2\%) & 20 (0.1\%)\\
Auto-deleted (long inactivity) & 60 (0.3\%) & 145 (0.6\%)\\
Total exited & 17394 (100.0\%) & 23335 (100.0\%)\\
\bottomrule
\end{tabular}

%% file: figuretable/labor_family_economics_project/maximum_score_matrix_application.tex
\begin{tabular}[t]{lllllllllllll}
\toprule
  & Education & Age & Income & Flexibility & Height & Weight & Drink & Smoke & Marital History & Housework & Childcare & Child\\
\midrule
Education & 1.00 & -3.59 & 5.63 & 3.78 & 0.93 & 0.53 & 3.92 & -2.35 & -1.91 & 2.35 & 1.52 & 4.12\\
 & {}[1.00, 1.00] & {}[-9.35, 4.19] & {}[-0.72, 9.78] & {}[-6.50, 9.67] & {}[-6.81, 8.11] & {}[-6.63, 8.70] & {}[-5.98, 9.83] & {}[-9.42, 6.34] & {}[-8.30, 4.90] & {}[-5.38, 9.64] & {}[-8.89, 9.64] & {}[-3.81, 9.56]\\
Age & -5.08 & 9.28 & 5.42 & 1.96 & 3.76 & 3.89 & -4.81 & -0.75 & 6.36 & 1.55 & -5.04 & -4.89\\
 & {}[-9.77, 1.04] & {}[8.06, 10.00] & {}[-0.65, 9.88] & {}[-6.62, 7.99] & {}[-3.41, 9.54] & {}[-2.64, 9.17] & {}[-9.78, 1.85] & {}[-9.78, 9.44] & {}[-1.01, 9.87] & {}[-7.03, 8.46] & {}[-9.92, 1.13] & {}[-9.98, 4.18]\\
Income & 5.05 & 2.57 & 7.74 & -4.30 & -3.72 & -5.69 & -0.57 & 0.51 & 3.32 & 0.49 & 0.29 & 4.39\\
 & {}[-1.26, 9.74] & {}[-6.25, 9.65] & {}[3.31, 9.89] & {}[-9.38, 4.81] & {}[-9.40, 2.81] & {}[-9.96, 1.91] & {}[-9.92, 7.57] & {}[-8.65, 9.22] & {}[-6.48, 9.66] & {}[-7.57, 9.32] & {}[-8.31, 8.35] & {}[-7.00, 9.93]\\
Flexibility & 0.83 & 3.88 & -2.07 & 2.04 & -3.13 & 1.27 & 4.10 & 1.12 & -2.24 & -1.39 & 3.42 & 1.79\\
 & {}[-8.27, 8.74] & {}[-3.66, 9.49] & {}[-8.33, 5.21] & {}[-8.95, 9.52] & {}[-9.69, 6.54] & {}[-6.04, 8.91] & {}[-0.97, 8.97] & {}[-8.10, 9.46] & {}[-9.65, 6.97] & {}[-8.91, 6.25] & {}[-4.37, 9.42] & {}[-8.58, 9.52]\\
Height & -0.39 & -5.21 & 6.09 & -2.55 & 3.33 & 0.76 & -1.19 & -2.57 & 3.25 & 3.30 & 1.13 & -0.87\\
 & {}[-8.98, 6.61] & {}[-9.88, 1.42] & {}[-1.52, 9.82] & {}[-8.43, 8.35] & {}[-3.76, 9.15] & {}[-8.56, 9.04] & {}[-8.23, 7.43] & {}[-9.43, 6.50] & {}[-5.73, 9.62] & {}[-3.46, 9.04] & {}[-6.86, 9.54] & {}[-9.51, 7.28]\\
Weight & -5.02 & 1.09 & -0.52 & 1.20 & 4.83 & 2.70 & 0.47 & -0.42 & -2.48 & 1.63 & -2.23 & 1.22\\
 & {}[-9.70, 1.29] & {}[-8.03, 8.93] & {}[-8.14, 7.33] & {}[-7.40, 6.44] & {}[-4.25, 9.90] & {}[-6.43, 9.16] & {}[-7.78, 8.36] & {}[-8.16, 9.64] & {}[-9.33, 7.25] & {}[-7.56, 9.26] & {}[-9.59, 7.80] & {}[-9.75, 9.24]\\
Drink & -2.73 & -0.71 & 2.44 & 1.76 & -3.66 & 0.92 & 7.71 & 1.48 & 2.35 & -1.89 & 0.54 & 0.78\\
 & {}[-9.13, 6.05] & {}[-9.15, 9.19] & {}[-6.46, 9.62] & {}[-7.24, 8.20] & {}[-9.37, 4.55] & {}[-8.31, 8.89] & {}[3.94, 9.62] & {}[-8.07, 9.57] & {}[-6.50, 8.68] & {}[-8.46, 7.76] & {}[-7.95, 8.43] & {}[-7.56, 8.51]\\
Smoke & -1.60 & 1.06 & -3.16 & -0.86 & -0.99 & 0.11 & 2.44 & 2.04 & 1.57 & -2.91 & -2.06 & -1.65\\
 & {}[-9.60, 9.64] & {}[-8.43, 7.95] & {}[-9.68, 6.90] & {}[-8.16, 7.78] & {}[-8.55, 9.33] & {}[-7.91, 9.05] & {}[-9.32, 8.57] & {}[-8.64, 9.21] & {}[-7.63, 9.40] & {}[-9.19, 7.20] & {}[-8.99, 5.98] & {}[-8.90, 7.65]\\
Marital History & -1.12 & 5.11 & 2.45 & 0.70 & -1.46 & -2.50 & -2.30 & -0.58 & 6.28 & -0.85 & -1.38 & -2.45\\
 & {}[-9.35, 8.33] & {}[-3.68, 9.73] & {}[-4.32, 9.44] & {}[-8.52, 8.71] & {}[-9.00, 6.63] & {}[-9.35, 6.31] & {}[-9.03, 7.08] & {}[-8.88, 9.06] & {}[0.70, 9.80] & {}[-8.62, 9.25] & {}[-8.82, 7.83] & {}[-9.47, 5.52]\\
Housework & 1.25 & -1.30 & -0.21 & -1.47 & -1.44 & 1.47 & 1.14 & 0.32 & 0.97 & 3.17 & 0.15 & -0.82\\
 & {}[-7.62, 9.68] & {}[-9.54, 8.79] & {}[-8.65, 9.15] & {}[-9.34, 6.79] & {}[-9.10, 6.54] & {}[-7.75, 9.23] & {}[-7.75, 8.18] & {}[-9.60, 9.06] & {}[-7.43, 8.53] & {}[-4.87, 9.47] & {}[-8.49, 7.51] & {}[-7.96, 6.20]\\
Childcare & 0.39 & -4.91 & -0.28 & -2.47 & -3.73 & -0.50 & 1.26 & -1.49 & -1.29 & -0.24 & 1.33 & 2.84\\
 & {}[-8.58, 7.22] & {}[-9.78, 1.74] & {}[-7.26, 8.48] & {}[-9.56, 6.29] & {}[-9.60, 6.08] & {}[-7.85, 6.93] & {}[-8.82, 8.01] & {}[-8.47, 7.25] & {}[-8.90, 9.05] & {}[-8.51, 5.85] & {}[-8.59, 9.27] & {}[-7.23, 9.76]\\
Child & 1.63 & -7.45 & 2.86 & 1.18 & 0.08 & 0.29 & -1.12 & -1.53 & -0.85 & -0.87 & 3.23 & 6.97\\
 & {}[-7.63, 9.19] & {}[-9.76, -2.27] & {}[-8.26, 9.83] & {}[-8.15, 9.60] & {}[-8.04, 8.90] & {}[-8.18, 7.38] & {}[-8.84, 8.41] & {}[-9.81, 9.45] & {}[-9.86, 9.18] & {}[-8.58, 6.96] & {}[-5.07, 9.08] & {}[2.88, 9.75]\\
\bottomrule
\end{tabular}

%% file: figuretable/labor_family_economics_project/maximum_score_matrix_pre_relation.tex
\begin{tabular}[t]{lllllllllllll}
\toprule
  & Education & Age & Income & Flexibility & Height & Weight & Drink & Smoke & Marital History & Housework & Childcare & Child\\
\midrule
Education & 1.00 & -5.46 & 6.48 & 3.12 & 3.49 & -2.99 & -0.95 & -2.76 & -1.30 & -2.32 & -0.43 & 0.88\\
 & {}[1.00, 1.00] & {}[-9.74, -0.32] & {}[1.82, 9.94] & {}[-4.45, 9.54] & {}[-3.66, 9.17] & {}[-9.67, 7.80] & {}[-9.08, 9.10] & {}[-9.69, 8.44] & {}[-9.40, 8.11] & {}[-8.77, 5.90] & {}[-8.44, 7.95] & {}[-5.76, 7.86]\\
Age & -4.92 & 9.15 & 2.85 & 1.13 & 1.93 & 2.82 & -2.08 & 1.95 & 4.56 & -4.01 & -4.69 & -5.81\\
 & {}[-9.57, 2.62] & {}[8.21, 9.97] & {}[-6.96, 9.85] & {}[-7.10, 9.22] & {}[-8.40, 8.51] & {}[-4.65, 9.59] & {}[-9.59, 6.45] & {}[-6.54, 9.57] & {}[-3.13, 9.72] & {}[-9.51, 3.74] & {}[-9.88, 4.96] & {}[-9.77, 0.06]\\
Income & 5.98 & 4.25 & 7.56 & -3.49 & 2.77 & -5.18 & 0.74 & -1.73 & -0.44 & -1.61 & 1.71 & 0.90\\
 & {}[-1.91, 9.71] & {}[-6.20, 9.68] & {}[3.51, 9.82] & {}[-9.41, 6.75] & {}[-8.45, 9.80] & {}[-9.31, 3.65] & {}[-8.39, 7.96] & {}[-9.38, 6.59] & {}[-8.23, 8.21] & {}[-8.71, 7.84] & {}[-9.06, 9.58] & {}[-9.20, 9.36]\\
Flexibility & -1.61 & 2.50 & -2.74 & 4.09 & -4.00 & 0.77 & -0.94 & -0.68 & -0.83 & 0.32 & -0.42 & -2.22\\
 & {}[-8.75, 9.27] & {}[-6.82, 9.41] & {}[-9.49, 6.38] & {}[-5.57, 9.37] & {}[-9.70, 4.58] & {}[-7.09, 7.91] & {}[-7.02, 9.17] & {}[-9.50, 6.66] & {}[-9.15, 7.84] & {}[-6.25, 6.78] & {}[-9.12, 7.88] & {}[-9.83, 5.65]\\
Height & -2.64 & -3.33 & 3.52 & -0.92 & 5.34 & -3.67 & 3.87 & -0.14 & 1.59 & -1.93 & -2.40 & 0.05\\
 & {}[-9.11, 5.01] & {}[-9.12, 5.88] & {}[-5.71, 9.37] & {}[-7.85, 5.00] & {}[-1.04, 9.86] & {}[-9.66, 4.73] & {}[-2.65, 8.79] & {}[-9.42, 7.67] & {}[-7.21, 9.42] & {}[-9.24, 7.77] & {}[-9.62, 7.76] & {}[-8.55, 8.24]\\
Weight & -3.04 & 3.13 & -1.48 & 2.13 & 2.86 & 7.21 & 0.19 & 2.24 & 1.12 & 2.36 & 3.57 & -0.64\\
 & {}[-8.72, 3.11] & {}[-4.33, 8.72] & {}[-9.18, 7.89] & {}[-6.18, 8.79] & {}[-8.20, 9.50] & {}[2.50, 9.91] & {}[-9.54, 8.30] & {}[-6.59, 9.43] & {}[-8.73, 8.83] & {}[-8.20, 8.85] & {}[-4.08, 9.70] & {}[-9.32, 8.75]\\
Drink & 1.55 & -0.16 & 4.11 & 4.45 & 0.55 & 1.92 & 5.38 & -0.39 & -1.99 & -2.24 & 3.92 & -0.88\\
 & {}[-8.86, 7.51] & {}[-9.11, 7.94] & {}[-2.25, 9.76] & {}[-3.41, 9.30] & {}[-7.76, 7.65] & {}[-6.91, 9.25] & {}[-0.42, 9.23] & {}[-9.25, 8.66] & {}[-9.79, 4.76] & {}[-9.44, 5.41] & {}[-3.93, 9.40] & {}[-8.34, 8.41]\\
Smoke & -0.79 & 0.51 & -0.65 & 1.78 & 1.14 & -0.24 & 1.24 & -0.68 & -1.50 & 0.18 & 0.21 & -2.26\\
 & {}[-9.29, 8.35] & {}[-9.51, 8.88] & {}[-8.44, 8.24] & {}[-6.18, 9.68] & {}[-6.26, 9.10] & {}[-8.46, 7.21] & {}[-7.95, 8.29] & {}[-9.34, 8.77] & {}[-8.99, 8.75] & {}[-9.17, 7.65] & {}[-8.55, 9.15] & {}[-8.90, 6.59]\\
Marital History & -2.02 & 6.61 & 4.21 & 2.59 & -1.45 & 0.43 & 2.40 & -0.20 & 5.74 & -0.87 & -2.01 & -0.10\\
 & {}[-8.36, 6.89] & {}[0.11, 9.93] & {}[-7.98, 9.33] & {}[-5.87, 8.91] & {}[-8.71, 6.67] & {}[-8.71, 8.48] & {}[-8.02, 8.07] & {}[-9.58, 9.79] & {}[0.46, 9.51] & {}[-9.29, 8.11] & {}[-9.00, 7.45] & {}[-8.14, 8.87]\\
Housework & -1.09 & 1.38 & 3.03 & -1.51 & -0.08 & 3.22 & -4.19 & 0.66 & 3.11 & 2.81 & -0.43 & -2.36\\
 & {}[-9.41, 8.43] & {}[-6.55, 8.25] & {}[-6.77, 9.38] & {}[-9.35, 5.63] & {}[-9.22, 6.64] & {}[-5.75, 8.67] & {}[-9.72, 5.65] & {}[-7.30, 9.59] & {}[-4.49, 9.13] & {}[-5.33, 9.74] & {}[-9.14, 7.99] & {}[-9.35, 7.28]\\
Childcare & -2.08 & -2.41 & 1.03 & -1.09 & 0.42 & -2.25 & 1.11 & -0.56 & 1.19 & -1.02 & 1.88 & 4.45\\
 & {}[-8.73, 6.88] & {}[-9.47, 6.73] & {}[-8.65, 9.50] & {}[-9.59, 8.80] & {}[-8.52, 8.10] & {}[-9.16, 4.60] & {}[-8.59, 9.56] & {}[-9.47, 8.00] & {}[-8.67, 8.53] & {}[-9.19, 8.37] & {}[-7.17, 9.50] & {}[-4.18, 9.62]\\
Child & 1.10 & -7.32 & 2.24 & -0.87 & -0.97 & -1.49 & 2.47 & -1.03 & -3.47 & -1.66 & 3.33 & 7.39\\
 & {}[-4.41, 8.94] & {}[-9.71, -2.32] & {}[-5.46, 9.22] & {}[-9.43, 7.30] & {}[-9.42, 7.79] & {}[-7.14, 5.12] & {}[-6.04, 9.45] & {}[-9.05, 8.91] & {}[-9.64, 4.63] & {}[-9.00, 6.69] & {}[-6.61, 9.42] & {}[4.00, 9.75]\\
\bottomrule
\end{tabular}

%% file: figuretable/labor_family_economics_project/maximum_score_matrix_serious_relation.tex
\begin{tabular}[t]{lllllllllllll}
\toprule
  & Education & Age & Income & Flexibility & Height & Weight & Drink & Smoke & Marital History & Housework & Childcare & Child\\
\midrule
Education & 1.00 & -3.69 & 5.98 & -1.99 & 0.13 & -3.92 & 3.35 & -1.18 & 1.06 & -0.33 & -0.86 & 2.80\\
 & {}[1.00, 1.00] & {}[-8.86, 6.45] & {}[-0.54, 9.90] & {}[-9.12, 5.86] & {}[-8.08, 9.62] & {}[-9.39, 2.17] & {}[-6.40, 9.16] & {}[-9.70, 8.74] & {}[-6.93, 9.06] & {}[-8.33, 8.75] & {}[-9.39, 7.95] & {}[-6.44, 9.54]\\
Age & -4.89 & 9.50 & 4.01 & 2.58 & 2.52 & 1.87 & -0.33 & -1.18 & 5.37 & -2.07 & -5.07 & -6.10\\
 & {}[-9.53, 3.09] & {}[8.20, 9.99] & {}[-3.61, 9.78] & {}[-8.28, 9.24] & {}[-4.73, 9.21] & {}[-8.40, 9.20] & {}[-9.02, 7.88] & {}[-9.15, 7.83] & {}[-1.89, 9.76] & {}[-8.72, 7.19] & {}[-9.73, -0.35] & {}[-9.87, -0.41]\\
Income & 4.39 & 2.80 & 7.22 & -1.20 & 2.66 & -6.25 & 1.29 & 1.34 & 2.81 & -0.86 & 2.26 & -0.49\\
 & {}[-4.82, 9.92] & {}[-7.83, 9.22] & {}[2.13, 9.88] & {}[-9.20, 6.20] & {}[-5.11, 9.68] & {}[-9.70, 0.59] & {}[-7.22, 9.33] & {}[-9.03, 9.28] & {}[-7.28, 8.80] & {}[-8.58, 7.98] & {}[-6.68, 9.07] & {}[-7.49, 7.57]\\
Flexibility & 2.53 & 3.82 & -2.60 & 1.76 & 0.72 & 3.93 & 1.45 & 1.43 & 1.32 & 2.70 & 1.90 & 1.51\\
 & {}[-7.35, 9.89] & {}[-3.15, 9.68] & {}[-8.68, 4.16] & {}[-5.91, 8.64] & {}[-6.61, 8.01] & {}[-3.64, 9.60] & {}[-7.40, 8.61] & {}[-7.88, 8.60] & {}[-7.75, 9.46] & {}[-4.10, 8.83] & {}[-6.80, 8.81] & {}[-8.20, 8.97]\\
Height & 3.69 & -2.10 & 3.78 & -1.50 & 6.33 & -3.44 & 3.41 & 0.58 & -2.01 & -3.33 & 1.02 & 1.89\\
 & {}[-3.31, 9.64] & {}[-7.88, 4.21] & {}[-4.90, 9.48] & {}[-7.75, 5.82] & {}[1.10, 9.76] & {}[-9.71, 4.25] & {}[-3.55, 9.33] & {}[-7.99, 9.22] & {}[-9.71, 7.57] & {}[-9.81, 3.69] & {}[-7.21, 8.83] & {}[-5.93, 9.23]\\
Weight & -1.09 & 4.70 & -3.91 & 0.04 & 0.85 & 6.52 & 1.43 & 1.67 & -0.73 & -2.50 & -1.44 & -0.66\\
 & {}[-8.04, 7.35] & {}[-2.65, 9.52] & {}[-9.68, 3.58] & {}[-9.74, 7.81] & {}[-8.98, 8.06] & {}[-0.33, 9.89] & {}[-6.00, 8.05] & {}[-9.11, 8.10] & {}[-8.59, 8.51] & {}[-9.12, 6.09] & {}[-9.74, 7.74] & {}[-9.17, 8.89]\\
Drink & 1.69 & -0.71 & 3.98 & -1.20 & 1.97 & 1.99 & 7.09 & -0.89 & 2.25 & -1.32 & -1.64 & 1.22\\
 & {}[-7.13, 7.81] & {}[-6.89, 7.90] & {}[-4.45, 9.66] & {}[-8.33, 7.05] & {}[-7.05, 8.07] & {}[-6.30, 8.60] & {}[1.33, 9.80] & {}[-9.15, 8.28] & {}[-7.72, 9.30] & {}[-8.90, 6.30] & {}[-9.07, 6.36] & {}[-7.19, 8.41]\\
Smoke & -3.15 & 3.52 & -2.76 & -0.10 & -1.29 & -3.00 & 0.52 & 3.00 & 1.16 & -0.02 & 0.83 & -1.33\\
 & {}[-9.22, 4.89] & {}[-5.04, 9.87] & {}[-8.94, 6.31] & {}[-9.45, 9.33] & {}[-8.99, 5.32] & {}[-9.64, 7.70] & {}[-8.14, 9.70] & {}[-5.76, 9.44] & {}[-8.30, 8.81] & {}[-8.74, 9.49] & {}[-8.85, 8.47] & {}[-8.56, 7.63]\\
Marital History & -2.91 & 5.61 & 3.65 & 1.90 & 0.59 & 1.36 & -0.22 & 1.58 & 4.74 & 0.37 & -0.28 & -3.30\\
 & {}[-9.27, 5.91] & {}[-2.37, 9.68] & {}[-2.93, 9.71] & {}[-3.80, 9.37] & {}[-7.08, 8.75] & {}[-7.27, 8.90] & {}[-8.33, 8.24] & {}[-9.05, 9.35] & {}[-2.74, 9.17] & {}[-8.50, 9.20] & {}[-9.75, 8.60] & {}[-9.87, 3.40]\\
Housework & -1.58 & 1.77 & 5.02 & -0.17 & 1.09 & 0.08 & 0.24 & -1.65 & 1.92 & 3.68 & 0.68 & -1.63\\
 & {}[-9.07, 5.51] & {}[-8.81, 9.41] & {}[-3.97, 9.33] & {}[-8.24, 8.57] & {}[-8.83, 9.12] & {}[-8.94, 8.51] & {}[-8.04, 8.59] & {}[-9.48, 8.99] & {}[-7.44, 8.35] & {}[-4.91, 9.68] & {}[-8.37, 8.43] & {}[-9.46, 6.81]\\
Childcare & -2.18 & -1.90 & -0.22 & -1.82 & -1.36 & 0.30 & 2.10 & -0.93 & -1.83 & 2.45 & 2.74 & 4.63\\
 & {}[-8.76, 6.18] & {}[-8.99, 5.20] & {}[-8.68, 8.50] & {}[-7.58, 5.12] & {}[-8.79, 7.21] & {}[-8.91, 8.56] & {}[-5.85, 8.05] & {}[-9.21, 9.41] & {}[-9.21, 5.49] & {}[-6.38, 9.31] & {}[-4.99, 9.79] & {}[-3.02, 9.46]\\
Child & 1.49 & -6.50 & 3.93 & 0.47 & 0.33 & -3.00 & 1.19 & 1.60 & -1.53 & 1.69 & 5.22 & 6.92\\
 & {}[-6.88, 9.33] & {}[-9.84, -2.76] & {}[-2.72, 9.43] & {}[-7.29, 8.76] & {}[-8.25, 8.71] & {}[-9.68, 6.34] & {}[-6.65, 7.80] & {}[-7.47, 9.72] & {}[-8.90, 9.00] & {}[-6.10, 9.11] & {}[-3.42, 9.95] & {}[1.77, 9.89]\\
\bottomrule
\end{tabular}

%% file: figuretable/labor_family_economics_project/maximum_score_matrix_proposal.tex
\begin{tabular}[t]{lllllllllllll}
\toprule
  & Education & Age & Income & Flexibility & Height & Weight & Drink & Smoke & Marital History & Housework & Childcare & Child\\
\midrule
Education & 1.00 & -4.58 & 6.75 & 0.54 & 2.60 & -2.63 & 2.67 & -1.61 & -2.39 & 1.92 & 2.76 & 0.93\\
 & {}[1.00, 1.00] & {}[-9.63, 0.50] & {}[2.26, 9.49] & {}[-8.24, 8.27] & {}[-7.87, 9.35] & {}[-9.06, 6.63] & {}[-4.62, 9.61] & {}[-9.51, 7.98] & {}[-9.76, 5.85] & {}[-6.46, 8.60] & {}[-6.84, 9.58] & {}[-6.24, 8.85]\\
Age & -5.65 & 9.11 & 3.07 & 0.93 & 2.25 & -0.32 & -2.68 & 0.26 & 5.62 & -1.88 & -4.68 & -5.42\\
 & {}[-9.81, 0.97] & {}[7.61, 9.92] & {}[-6.79, 9.88] & {}[-7.58, 8.00] & {}[-5.18, 9.22] & {}[-8.17, 8.76] & {}[-9.73, 7.56] & {}[-9.48, 9.25] & {}[-1.93, 9.96] & {}[-9.23, 8.47] & {}[-9.88, 6.96] & {}[-9.89, -0.62]\\
Income & 3.68 & 2.29 & 7.48 & -3.57 & 2.45 & -7.24 & 5.67 & 1.07 & 3.15 & 1.67 & 1.97 & 3.79\\
 & {}[-2.74, 8.89] & {}[-5.83, 9.35] & {}[2.44, 9.91] & {}[-9.08, 7.62] & {}[-7.67, 8.19] & {}[-9.78, -2.50] & {}[-3.49, 9.94] & {}[-8.62, 9.65] & {}[-3.05, 8.53] & {}[-6.78, 9.08] & {}[-7.31, 9.10] & {}[-5.47, 9.58]\\
Flexibility & 2.22 & 3.12 & 1.59 & 1.04 & 0.28 & 0.39 & 2.19 & 0.21 & -2.34 & -1.96 & -0.85 & -0.62\\
 & {}[-7.78, 9.36] & {}[-5.30, 8.77] & {}[-6.98, 9.55] & {}[-8.71, 7.90] & {}[-8.35, 9.74] & {}[-7.94, 7.96] & {}[-5.52, 9.11] & {}[-8.40, 9.20] & {}[-9.27, 7.02] & {}[-9.09, 5.37] & {}[-8.70, 7.91] & {}[-7.60, 9.22]\\
Height & -1.30 & 0.56 & 2.41 & -1.26 & 7.39 & -2.53 & -0.47 & -1.60 & 1.47 & 0.21 & 0.28 & 4.87\\
 & {}[-9.47, 7.70] & {}[-9.44, 8.91] & {}[-6.33, 8.76] & {}[-8.29, 7.79] & {}[1.47, 9.93] & {}[-9.11, 4.66] & {}[-8.39, 7.88] & {}[-9.71, 7.34] & {}[-8.12, 7.88] & {}[-8.39, 7.58] & {}[-9.29, 8.08] & {}[-2.76, 9.51]\\
Weight & -0.20 & 4.89 & 0.73 & -5.06 & 5.05 & 6.88 & -0.63 & -0.42 & -0.43 & 2.79 & 2.28 & 2.51\\
 & {}[-8.33, 6.86] & {}[-3.25, 9.22] & {}[-8.16, 9.42] & {}[-9.37, 1.17] & {}[-1.29, 9.98] & {}[2.03, 9.86] & {}[-8.35, 5.83] & {}[-8.85, 8.30] & {}[-9.42, 9.10] & {}[-4.98, 9.74] & {}[-7.24, 9.06] & {}[-6.77, 9.40]\\
Drink & 1.86 & -2.52 & 2.22 & -1.95 & 0.20 & -4.05 & 7.52 & 1.18 & 2.70 & 1.01 & -2.39 & -0.31\\
 & {}[-9.02, 7.76] & {}[-9.77, 6.90] & {}[-6.89, 9.32] & {}[-9.60, 6.17] & {}[-9.04, 7.75] & {}[-9.85, 3.63] & {}[4.15, 9.83] & {}[-8.33, 9.52] & {}[-2.97, 8.45] & {}[-8.07, 8.95] & {}[-9.24, 5.52] & {}[-7.52, 9.09]\\
Smoke & -2.91 & 4.26 & -0.96 & 2.07 & 1.80 & 1.02 & 4.02 & 2.62 & 1.35 & -3.57 & 0.76 & -2.16\\
 & {}[-9.12, 7.71] & {}[-6.45, 9.60] & {}[-8.96, 8.68] & {}[-5.13, 9.73] & {}[-8.31, 9.83] & {}[-6.76, 8.34] & {}[-6.55, 9.74] & {}[-6.79, 9.76] & {}[-8.93, 8.00] & {}[-9.80, 2.67] & {}[-8.20, 9.12] & {}[-8.88, 5.31]\\
Marital History & -4.07 & 6.32 & 3.60 & -2.41 & 0.45 & -3.01 & 5.11 & 1.08 & 5.62 & -1.28 & -3.40 & -1.65\\
 & {}[-9.77, 2.91] & {}[0.19, 9.83] & {}[-3.28, 9.58] & {}[-9.72, 6.31] & {}[-6.43, 8.66] & {}[-9.85, 8.85] & {}[-2.48, 9.66] & {}[-9.59, 9.24] & {}[-1.47, 9.82] & {}[-9.15, 8.13] & {}[-9.80, 5.25] & {}[-9.29, 8.09]\\
Housework & -0.32 & -2.46 & -0.34 & 0.82 & 1.65 & 0.49 & 1.30 & -0.67 & 3.16 & -2.71 & -0.13 & 1.37\\
 & {}[-8.53, 9.21] & {}[-9.66, 5.29] & {}[-9.12, 9.27] & {}[-6.21, 8.33] & {}[-8.75, 9.54] & {}[-7.63, 8.22] & {}[-8.83, 9.24] & {}[-8.94, 8.98] & {}[-5.49, 9.58] & {}[-9.01, 5.20] & {}[-9.40, 8.10] & {}[-6.99, 9.10]\\
Childcare & -0.83 & -3.84 & -0.51 & -1.50 & -1.43 & 1.35 & -1.73 & -0.19 & 0.02 & 3.48 & 1.47 & 3.23\\
 & {}[-9.67, 9.02] & {}[-9.82, 4.18] & {}[-9.43, 7.91] & {}[-8.15, 7.97] & {}[-9.04, 7.01] & {}[-8.39, 9.06] & {}[-9.18, 9.44] & {}[-8.89, 8.45] & {}[-9.06, 7.43] & {}[-6.22, 9.02] & {}[-4.45, 7.95] & {}[-4.20, 8.92]\\
Child & -0.30 & -6.64 & 0.19 & -0.34 & -2.02 & -1.18 & -1.78 & -2.17 & -4.13 & 2.73 & 5.27 & 6.67\\
 & {}[-9.57, 8.07] & {}[-9.96, -0.14] & {}[-8.71, 6.89] & {}[-8.93, 7.26] & {}[-8.86, 7.65] & {}[-8.92, 6.78] & {}[-9.64, 6.61] & {}[-9.85, 8.01] & {}[-9.84, 6.24] & {}[-7.23, 9.65] & {}[-0.20, 9.49] & {}[0.41, 9.82]\\
\bottomrule
\end{tabular}